\documentclass[%
reprint,
superscriptaddress,
 amsmath,amssymb,
 aps,
prb,
showkeys
]{revtex4-2}

\usepackage{graphicx}
\usepackage{dcolumn}
\usepackage{bm}
\usepackage{placeins}
\usepackage{xcolor}
\usepackage[normalem]{ulem}
\usepackage{physics}

\definecolor{myblue}{RGB}{0, 112, 192}
\usepackage[pagebackref=false]{hyperref}
\newcommand{\mbeq}{\overset{!}{=}}

\begin{document}

\title{The role of antisymmetric orbitals and electron-electron interactions on the two-particle spin and valley blockade in graphene double quantum dots}

\author{S. M\"oller}
\affiliation{JARA-FIT and 2nd Institute of Physics, RWTH Aachen University, 52074 Aachen, Germany,~EU}%
\affiliation{Peter Gr\"unberg Institute  (PGI-9), Forschungszentrum J\"ulich, 52425 J\"ulich,~Germany,~EU}
\author{L. Banszerus}
\thanks{Present address: University of Vienna, Faculty of Physics, Boltzmanngasse 5, 1090 Vienna, Austria}
\affiliation{JARA-FIT and 2nd Institute of Physics, RWTH Aachen University, 52074 Aachen, Germany,~EU} %
\author{K. Hecker}
\author{H. Dulisch}
\affiliation{JARA-FIT and 2nd Institute of Physics, RWTH Aachen University, 52074 Aachen, Germany,~EU}%
\affiliation{Peter Gr\"unberg Institute  (PGI-9), Forschungszentrum J\"ulich, 52425 J\"ulich,~Germany,~EU}
\author{K.~Watanabe}
\affiliation{Research Center for Functional Materials, 
National Institute for Materials Science, 1-1 Namiki, Tsukuba 305-0044, Japan}
\author{T.~Taniguchi}
\affiliation{International Center for Materials Nanoarchitectonics, 
National Institute for Materials Science,  1-1 Namiki, Tsukuba 305-0044, Japan}%
\author{C. Volk}
\email{c.volk@fz-juelich.de}
\author{C. Stampfer}
\email{stampfer@physik.rwth-aachen.de}
\affiliation{JARA-FIT and 2nd Institute of Physics, RWTH Aachen University, 52074 Aachen, Germany,~EU}%
\affiliation{Peter Gr\"unberg Institute  (PGI-9), Forschungszentrum J\"ulich, 52425 J\"ulich,~Germany,~EU}%

\date{\today}


\begin{abstract} 
We report on an experimental study of spin and valley blockade in two-electron bilayer graphene (BLG) double quantum dots (DQDs) and explore the limits set by asymmetric orbitals and electron-electron interactions. The results obtained from magnetotransport measurements on two-electron BLG DQDs, where the resonant tunneling transport involves both orbital symmetric and antisymmetric two-particle states, show a rich level spectrum. 
We observe a magnetic field tunable spin and valley blockade, which is limited by the orbital splitting, the strength of the electron-electron interaction and the difference in the valley g-factors between the symmetric and antisymmetric two-particle orbital states. Our conclusions are supported by simulations based on rate equations, which allow the identification of prominent interdot transitions associated with the transition from single to two-particle states observed in the experiment.
\end{abstract}

\maketitle

Double quantum dots (DQDs) hosting a finite number of charge carriers -- usually one or two -- are a promising  building block for quantum computing~\cite{Loss1998Jan, Vajner2022Jul, Burkard1999Jan}, since the charge, spin and/or valley degree of freedom of those charge carriers can be used to encode a quantum bit (qubit)~\cite{Loss1998Jan, Vajner2022Jul, Burkard1999Jan}. 
This has been achieved in a variety of material systems, e.g. gallium arsenide~\cite{Petta2005Sep, Bluhm2011Feb, Cerfontaine2020Aug, Sala2020Mar}, silicon~\cite{Philips2022Sep, DumoulinStuyck2021Aug, Klemt2023Oct, Huang2024Mar, Cai2023Mar, Maurand2016Nov, Kim2015Oct, Shi2012Apr}, germanium~\cite{Jirovec2021Aug, Watzinger2018Sep, Wang2022Jan, Hendrickx2020Jan}, or carbon nanotubes~\cite{Laird2013Jul}.
A DQD in bilayer graphene (BLG)~\cite{Zebrowski2017Jul,Eich2018Aug,Banszerus2018AugFirstQD,Tong2021JanTunableValleySplitting, Banszerus2023MayNature,Banszerus2021SepSpinOrbit, Banszerus2020MarSingleElectronDQD,Banszerus2021MarTunableInterdot,Tong2022FebBlockade,Hecker2023Nov,Tong2024JanCatalogue,Garreis2024MarLongLived,Knothe2024Jun,Tong2024Jul}  may offer a compelling alternative for hosting a qubit thanks to the BLG's low nuclear spin densities, weak hyperfine coupling and weak spin-orbit interaction~\cite{McCann2013Apr}, promising long spin coherence times~\cite{Trauzettel2007Feb}. 
Additionally, the well tunable valley degree of freedom~\cite{Moller2023SepShellFilling} offers the possibility to create valley based qubits~\cite{Rohling2014Oct, Rohling2012Aug, Wu2013Sep}.
Successful qubit operation requires, among others, the ability to quickly read out the quantum state of a qubit. 
For spin or valley qubits, this requires a spin- or valley-to-charge conversion, since charge states can be read out rapidly by e.g. radiofrequency reflectometry (RF)~\cite{Noiri2020Feb, Volk2019Aug, Cassidy2007Nov}.
In DQDs, this conversion is usually provided by Pauli blockade \cite{ Petta2005Sep, Buitelaar2008Jun, Churchill2009Apr,Churchill2009AprA,Lai2011Oct, Pakkiam2018Nov, Seedhouse2021Jan}, where an energetically allowed interdot charge transition is forbidden due to incompatible quantum numbers of the involved charge carriers.
Therefore, a comprehensive understanding of the limits and the tunablility of spin and valley blockade in BLG DQDs is necessary for evaluating their potential for hosting qubits. 
\begin{figure}[!thb]
\centering
\includegraphics[draft=false,keepaspectratio=true,clip,width=\linewidth]{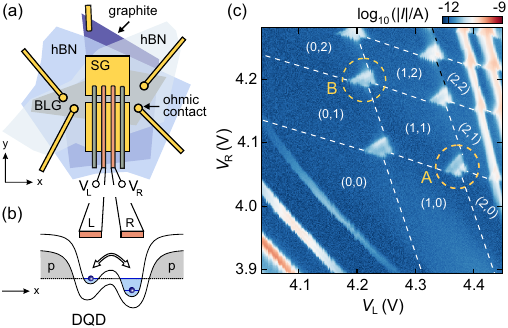}
\caption[Fig01]{\textbf{(a)} Schematic top view of the van-der-Waals heterostructure with metallic gates. The BLG is contacted by etching through the top hBN. Together with the graphite BG, two layers of metallic gates, SGs and FGs, separated by aluminum oxide, electrostatically defines well-controllable QDs. The SG channel is $\approx 100$~nm wide, while FGs have a width of 70~nm and a pitch of 140~nm.
\textbf{(b)}~Schematic of the band edge profile along the channel illustrating the formation of a DQD at the $(1,1) \leftrightarrow (0,2)$ charge transition. 
\textbf{(c)} Charge stability diagram at $V_\mathrm{SD} = -1$~mV as a function of the voltages applied to the left and right FGs. The formation of an electron-electron DQD is visible for $V_\mathrm{L} > 4.2$~V and $V_\mathrm{R} > 4.05$~V. White numbers indicate its occupation when in Coulomb blockade. The yellow dashed circle labeled A and B indicate the  $(1,1) \leftrightarrow (2,0)$ and  $(1,1) \leftrightarrow (0,2)$ bias triangles, respectively. White dashed lines indicate co-tunneling lines.}
\label{f1}
\end{figure}

\begin{figure*}[!thb]
	\centering
\includegraphics[draft=false,keepaspectratio=true,clip,width=0.93\linewidth]{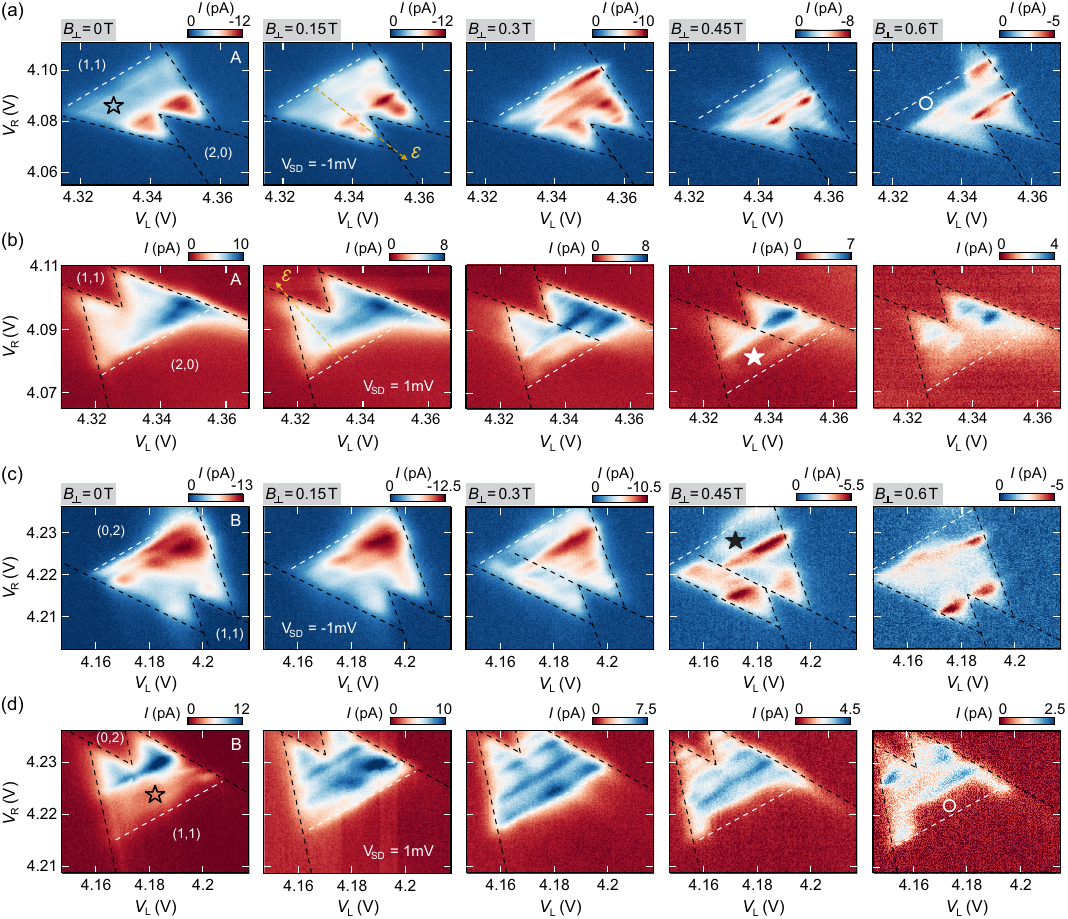}
\caption[Fig02]{\textbf{(a)}  Zoom-in on the $(1,1) \rightarrow (2,0)$ bias triangle for $V_\mathrm{SD} = - 1$ mV and different perpendicular magnetic fields (see labels). At zero magnetic field ($B_\perp=0$~T), a distinct valley blocked region is visible, indicated by the star. For increasing magnetic field, the extent of the valley blockade along the detuning axis ($\varepsilon$) decreases and eventually, a spin blocked region appears, indicated by the circle in the right most panel. The white dashed line indicates the estimated onset of the bias triangle, defining zero detuning, $\varepsilon = 0$.  
\textbf{(b)} Zoom-in on the $(2,0) \rightarrow (1,1)$ bias triangle for $V_\mathrm{SD} = 1$ mV and different perpendicular magnetic fields (see labels in panel a). For increasing magnetic fields, a small valley blocked region appears in the bias triangles, indicated by the white star.
\textbf{(c)} Zoom-in on the $(0,2) \rightarrow (1,1)$ bias triangle for $V_\mathrm{SD} = - 1$ mV and different perpendicular magnetic fields (see labels). \textbf{(d)} Zoom-in on the $(1,1) \rightarrow (0,2)$ bias triangle for $V_\mathrm{SD} = 1$ mV. For more details see text.
}
\label{f2}
\end{figure*}

First investigations of blockade effects in the $(1,1) \leftrightarrow (0,2)$ charge transition of BLG DQDs,
reported by Tong~\textit{et~al.} 
~\cite{Tong2022FebBlockade, Tong2024JanCatalogue}, showed a magnetic field tunable spin and valley blockade at low bias voltage. 
So far, however, only the six energetically lowest -- orbitally symmetric -- two-particle states have been discussed. The role of the energetically higher antisymmetric orbital states and the question of what limits the spin and valley blockade remain largely unexplored. 

In this study, we present feature-rich magnetotransport spectroscopy measurements of two-electron bilayer graphene DQDs, where resonant tunneling involves both orbital symmetric and antisymmetric two-particle states.
We observe a magnetic field tunable spin and valley blockade that is limited by the orbital splitting, and the magnitude of both the electron-electron interaction and the valley magnetic moments, expressed by the valley $g$-factor. 
Our findings are supported by transport simulations of the DQD system following a rate equation approach with detailed knowledge of the single and two-particle states in BLG QDs and with the assumption of a weak interdot tunnel coupling. 
This allows to identify individual transitions that give rise to prominent features in our data originating from both orbital symmetric and antisymmetric two-particle states. 
We do not only demonstrate the viability of treating the two QDs independently and neglecting the influence of the interdot tunnel coupling on the energy spectrum but also provide a profound understanding of the DQD transition spectrum. 
In the future, this will allow to explore different qubit regimes in BLG QDs, as reading out qubit states -- possible by using spin or valley blockade mechanisms -- is crucial for qubit operation. 

To electrostatically define a DQD, we follow previous works\cite{Eich2018Aug, Banszerus2018AugFirstQD, Tong2021JanTunableValleySplitting, Banszerus2023MayNature, Eich2018Jul, Garreis2021AprShellfillingWarping, Banszerus2021SepSpinOrbit}
and fabricate heterostructures consisting of a BLG crystal encapsulated within two layers of hexagonal boron nitride (with a thickness of $\approx 30$~nm each)~\cite{Engels2014Sep,Wang2013Nov}. The heterostructure is placed on a graphite flake which acts as a back gate (BG)~\cite{Icking2022Jul}.
Split gates (SGs) consisting of $\approx 25$~nm Cr/Au are deposited on top, forming a $\approx 100$~nm wide channel. An additional $\approx 75$~nm thick layer of metallic Cr/Au finger gates (FGs) is then fabricated across the channel. The top gate layers are separated by a 15~nm thick layer of atomic layer deposited Al$_2$O$_3$. A schematic top view of a finished sample is shown in Fig.~\ref{f1}(a).
All measurements are perfomed in a dilution refrigerator with a base temperature below 30 mK.
Voltages of opposite sign applied to the SGs and BG create an electric displacement field perpendicular to the plane of the BLG, opening a band gap and allowing the Fermi energy to be tuned into the  gap~\cite{Icking2022Jul, McCann2013Apr}. This leaves a narrow conductive channel, defined by the SGs, connecting the source and drain reservoirs. Voltages applied to two adjacent FGs ($V_\mathrm{L}, V_\mathrm{R}$) are used to locally invert the polarity of the channel, creating a DQD and giving rise to tunnel barriers where the Fermi energy lies within the band gap, as illustrated in Fig.~\ref{f1}(b). 
At a finite bias voltage ($V_\mathrm{SD}$), we measure the conduction through the channel as a function of $V_\mathrm{L}$ and $V_\mathrm{R}$, as shown in Fig.~\ref{f1}(c). For $V_\mathrm{L} > 4.2$~V and $V_\mathrm{R} > 4.05$~V, we observe the formation of an electron-electron DQD, which is incrementally filled for increasingly positive FG voltages. White numbers $(N_\mathrm{L}, N_\mathrm{R})$, indicate the charge occupation in the left and right QD, respectively.
Note that for smaller $V_\mathrm{L}$ and $V_\mathrm{R}$, a hole QD is formed between the two electron QDs, which appears as curved Coulomb resonances~\cite{Banszerus2020OctEHcrossover, Banszerus2023MayNature}. For larger $V_\mathrm{L}$, $V_\mathrm{R}$, both the interdot tunnel coupling and the tunnel coupling to the leads electron-electron DQD increase significantly~\cite{Banszerus2021MarTunableInterdot}. In particular, the tunnel barrier to the right lead is weakened already at $V_\mathrm{L} > 4.4$ V, $V_\mathrm{R} > 4$ V, leading to a strongly pronounced co-tunneling line.

To investigate interdot charge transitions from single- to two-particle states, we focus on the bias triangles highlighted by the yellow dashed circles in Fig.~\ref{f1}(c). 
They require an interdot charge transition from both QDs being occupied by a single electron to a single QD being occupied by two electrons. The position of the two QDs interchanged for the two bias triangles $(1,1) \leftrightarrow (2,0)$ and  $(1,1) \leftrightarrow (0,2)$, respectively. 
Fig.~\ref{f2}(a,b) shows charge stability diagrams of the $(1,1) \leftrightarrow (2,0)$ bias triangle (labeled A in Fig.~\ref{f1}(c)) for (a) negative and (b) positive $V_\mathrm{SD}$ at different perpendicular magnetic fields, $B_\perp$. The sign of the bias determines the direction of the interdot charge transition, i.e. for positive bias,  transitions $(2,0) \rightarrow (1,1)$ are required for a finite tunnel current, while negative bias requires $(1,1) \rightarrow (2,0)$ transitions.
Corresponding measurements of the $(1,1) \leftrightarrow (0,2)$ bias triangle are shown in Fig.~\ref{f2}(c,d) (more details in \textit{Appendix} A). 
Black dashed lines mark the outline of the triple points and the white dashed lines indicated the base line, i.e. the ground state to ground state transition, defining the zero detuning energy, $\varepsilon = 0$.
Note, that the detuning energy can be obtained from the FG voltages using the respective lever arms extracted from the estimated outlines of the bias triangles.
Significant co-tunneling to the right QD increases the current in the bias triangles and complicates the estimation of the position of the base line with an estimated uncertainty of $\sim 15\%$. 
Note that the overall current through the DQD reduces for increasing perpendicular magnetic field, which is probably due to a stronger localization of the QD wavefunction reducing the tunnel rates to the source/drain leads, in agreement with previous work~\cite{Banszerus2020DecPSSB, Banszerus2020MarSingleElectronDQD, Tong2022FebBlockade, Moller2021Dec}.
For $V_\mathrm{SD} = -1$~mV and zero magnetic field, the current within the triple point is suppressed for detuning values $\varepsilon < 0.6$~meV (see arrow for the direction of $\varepsilon$ in the 2nd panel of Fig.~\ref{f2}(a)), marked by the star (see 1st panel of Fig.~\ref{f2}(a)). 
\begin{figure}[!t]
\centering
\includegraphics[draft=false,keepaspectratio=true,clip,width=\linewidth]{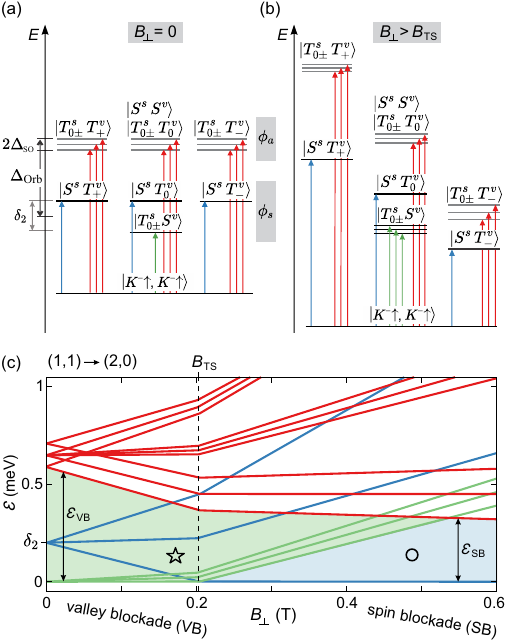}
\caption[Fig01]{\textbf{(a)} Schematic illustrating possible $(1,1) \rightarrow (2,0)$ interdot transitions from the $(1,1)$ ground state (GS) into orbitally symmetric, $\phi_s$, and antisymmetric, $\phi_a$, two-particle states ($B_\perp = 0$~T). 
\textbf{(b)} As in panel (a) but for $B_\perp > B_\mathrm{TS}$. 
\textbf{(c)} The required detuning energy (measured from the base line of the bias triangle) for each transition changes with perpendicular magnetic field, plotted for $g^{(1)}_v = 15, g^s_v = g^a_v = 19, \delta_2 = 0.2$~meV, $\Delta_\mathrm{Orb} = 0.65$~meV. For low magnetic fields a valley blockade is expected (green region and star), which changes to a spin blockade (blue region and circle) at $B_\perp = B_\mathrm{TS} \approx 0.2$~T due to the change in two-particle GS.}
\label{f3}
\end{figure}
\noindent
Increasing the magnetic field to $B_\perp = 0.3$~T makes the blocked region disappear,  while resonant features emerge. 
Exceeding $B_\perp \approx 0.4$~T, an increasingly large region is blocked again, marked by the circle (see rightmost panel of Fig.~\ref{f2}(a)).
For $V_\mathrm{SD} = 1$~mV and at zero magnetic field, the region of strongest conductance lies close to zero detuning and no blockade is observed. For increasing magnetic field resonant features are observed in the bias triangle and a blocked region appears starting from $B_\perp \approx 0.15$~T (see white star in the 4th panel of Fig.~\ref{f2}(b)). 
The data shown in Figs.~\ref{f2}(c) and~\ref{f2}(c) are overall very similar as the one presented in Figs.~\ref{f2}(a) and~\ref{f2}(b) (compare different symbols) but for different current direction as they belong to the $(0,2) \rightarrow (1,1)$ and the $(1,1) \rightarrow (2,0)$ transitions, i.e. triple point B in Fig.~\ref{f1}(c). 

To understand the observed resonant features and the blockade effects, we discuss possible transitions between single- and two-particle states in a DQD when including both symmetric and antisymmetric orbital states.
For the investigated energy range we focus on 16 relevant two-particle states, which we denote as spin ($s$) and valley ($v$) singlet ($S$) and triplet ($T_{0},T_+,T_-$) states. For example, $\ket{T^s_-S^v}$ represents a fully polarized spin triplet and a valley singlet state. The notation $0\pm$ encompasses all triplet states, namely the fully polarized ($+,-$) and unpolarized ($0$) triplet states. The relevant states group into 6 orbitally symmetric and 10 orbitally antisymmetric states, separated by the orbital splitting $\Delta_\mathrm{Orb}$~\cite{Moller2021Dec,Knothe2022Apr}.
The orbitally symmetric states $\phi_s$ are further split by short-range electron-electron interactions into a valley singlet -- spin triplet ($\ket{T^s_{0\pm}S^v}$) ground state (GS) and valley triplet -- spin singlet ($\ket{S^sT^v_{0\pm}}$) excited states (ESs) with a separation of $\delta_2$~\cite{Moller2021Dec, Knothe2020JunQuartetStates, Knothe2022Apr, Lemonik2010NovLifshitzSymmBreak, Lemonik2012JunCompetingBLGOrders}, as shown in Fig.~\ref{f3}(a). 
The next higher orbital state is antisymmetric $\phi_a$, featuring spin triplet - valley triplet states ($\ket{T^s_{0\pm}T^v_{0\pm}}$) and a spin singlet - valley singlet state ($\ket{S^sS^v}$), which are separated by the Kane-Mele spin-orbit coupling $\Delta_\mathrm{SO} \approx 65~\mu$eV~\cite{Kurzmann2021OctKondo, Banszerus2021SepSpinOrbit, Banszerus2023MayNature}. 
For the $(1,1)$ states, we assume a negligible mixing between the two QDs due to a small interdot tunnel coupling and treat them as independent single particle states in each QD.

First, we focus at the $(1,1) \rightarrow (2,0)$ charge transition at $V_\mathrm{SD}  < 0$~V. 
For simplicity, we initially only consider transitions from the $(1,1)$ GS to the 16 different two-particle states.
Fig.~\ref{f3}(a) displays possible transitions from the $(1,1)$ GS, which is the $\ket{K^- \uparrow, \, K^- \uparrow}$ state  for $B_\perp \geq 0$~\cite{Kurzmann2021OctKondo, Banszerus2021SepSpinOrbit, Banszerus2023MayNature}, into spin triplet - valley singlet, $\ket{T^s_{0\pm}S^v}$ (green), spin singlet - valley triplet, $\ket{S^sT^v_{0\pm}}$ (blue) and spin triplet - valley triplet states, $\ket{T^s_{0\pm}T^v_{0\pm}}$ (red) of the symmetric and antisymmetric two-particle states in the $(2,0)$ configuration at $B_\perp = 0$. 
Increasing the perpendicular magnetic field shifts all states according to their spin and valley magnetic moments~\cite{Moller2021Dec}, which is illustrated in Fig.~\ref{f3}(b).
The condition for a charge transition to resonantly occur is that the detuning energy of the DQD needs to compensate the energy difference between the electrochemical potentials of the $(1,1)$ and $(2,0)$ states, given by
\begin{align}
    \mu_{(1,1) \leftarrow (1,0)} + \frac{\Tilde{\varepsilon}}{2} \mbeq  \mu_{(2,0) \leftarrow (1,0)} -  \frac{\Tilde{\varepsilon}}{2}
\end{align}
with the electrochemical potentials of the right and left QD, $\mu_{(1,1) \leftarrow (1,0)}$, and $\mu_{(2,0) \leftarrow (1,0)}$, respectively. 
Note that here $\Tilde{\varepsilon}$ is the absolute detuning energy (in gate space), which is shifted compared to the detuning energy $\varepsilon$, which is always measured relative to the base line of the bias triangle.
From this, we obtain a resonance condition for transport via the $(1,1) \rightarrow (2,0)$ charge transition
\begin{equation}
   \Tilde{\varepsilon} (B_\perp) \mbeq E_{(0,2)}(B_\perp) -  E_{(1,1)}(B_\perp) \,,
    \label{eq:1102resonanceCondition}
\end{equation}
where $E_{(0,2)}$ and $E_{(1,1)}$ are the energies of the DQD in the respective charge configuration.
\begin{figure}[!t]
\centering
\includegraphics[draft=false,keepaspectratio=true,clip,width=\linewidth]{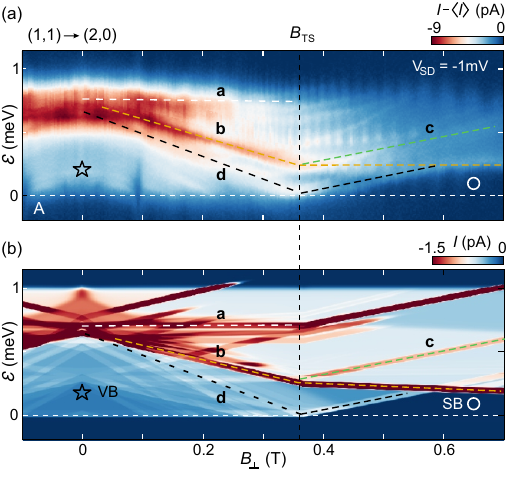}
\caption[Fig01]{\textbf{(a)} Transport through the DQD measured along a detuning cut of the $(1,1) \rightarrow (2,0)$ bias triangle at $V_\mathrm{SD} = -1$~mV, as a function of $B_\perp$. The switch from valley to spin blockade is clearly visible, as indicated by the  star and the circle, respectively. 
Note that the conducting region does not always reach up to $\varepsilon = 1$~mV, since the detuning cut is not always taken perfectly through the tip of the bias triangle. For clarity, we subtracted the average current $\langle I \rangle$ of each detuning trace.
\textbf{(b)} Simulation of the measurement presented in (a) using $\Delta_\mathrm{Orb} = 0.7$~meV, $\delta_2 = 0.34$~meV, $g^{(1)}_v = 15$, $g^s_v = 18$, $g^a_v = 19$. Colored dashed lines highlight prominent similarities between (a) and (b). A spin and valley blocked region is visible both in the experimental data and the simulation, indicated by the circle and the star, respectively.}
\label{f4}
\end{figure}
In order to obtain the detuning dependency of a given transition with respect to the base line of the bias triangle, we need to take into account that the base line also shifts in magnetic field along the absolute detuning axis, as it is given by the GS to GS transition. For low magnetic fields, $B_\perp < B_\mathrm{TS}$ (where $B_\mathrm{TS}$ labels the magnetic field at which the two-particle GS turns from a spin triplet to a spin singlet [REFS]), the GS to GS transition, given by $\ket{K^-\uparrow, \, K^-\uparrow} \rightarrow \ket{T^s_- S^v}$, shifts on the absolute detuning energy axis as
\begin{align}
    \label{eq:GS-GS_transition}
    \Tilde{\varepsilon}_\mathrm{GS-GS} (B_\perp) =& \, E(\ket{T^s_- S^v}) - E(\ket{K^-\uparrow}) - E(\ket{K^-\uparrow}) \nonumber \\
    =& \,g^{(1)}_v \mu_B B_\perp,
\end{align}
with the Bohr magneton, $\mu_B$, and the valley g-factor of the single particle orbital, $g^{(1)}_v$. For $B_\perp > B_\mathrm{TS}$, where the two-particle GS changes to $\ket{S^s T^v_-}$, the GS to GS transition is $\ket{K^-\uparrow,\, K^-\uparrow} \rightarrow \ket{S^s T^v_-}$, which shifts as
\begin{align}
    \Tilde{\varepsilon}_\mathrm{GS-GS}(B_\perp) &= E(\ket{S^s T^v_-}) - E(\ket{K^-\uparrow}) - E(\ket{K^-\uparrow}) \nonumber \\ 
    &= \delta_2 + (g^{(1)}_v + g_s - g^s_v) \mu_B B_\perp \,,
\end{align}
with the spin g-factor $g_s = 2$ and the vally g-factors $g^s_v$ of the symmetric and  $g^a_v$ antisymmetric orbital states.
The detuning dependency of a given transition with respect to the base line of the triple point is then given by 
\begin{equation}
    \varepsilon(B_\perp) = \Tilde{\varepsilon}(B_\perp) -  \Tilde{\varepsilon}_\mathrm{GS-GS}(B_\perp)\, .
\end{equation}

Finally, Fig.~\ref{f3}(c) visualizes the $B_\perp$ dependency of each transition from the $(1,1)$ GS to the 16 different two particle $(2,0)$ states, each potentially giving rise to a resonant feature in the bias triangle, with the color code of the lines corresponding to the colored arrows in Figs.~\ref{f3}(a) and~\ref{f3}(b).
\begin{figure*}[!thb]
	\centering
\includegraphics[draft=false,keepaspectratio=true,clip,width=0.95\linewidth]{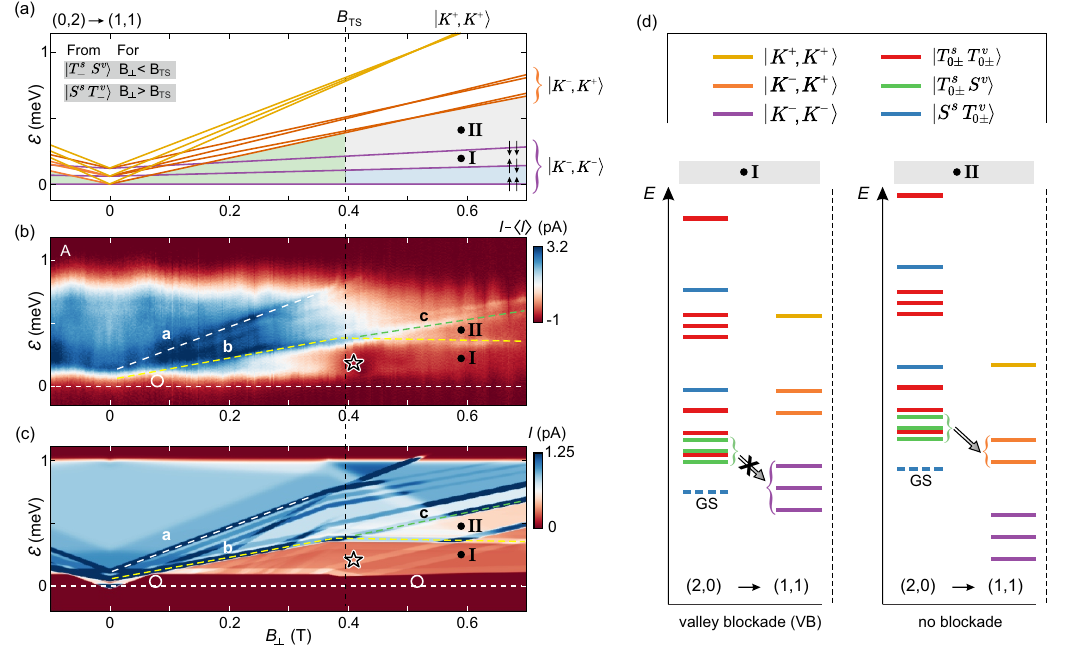}
\caption[Fig03]{ \textbf{(a)}  Schematic illustrating the required detuning energy (measured from the base line of the bias triangle)  for each transition from the $(2,0)$ GS into the 16 $(1,1)$ single particle states, as a function of perpendicular magnetic field.
For low magnetic fields a valley blockade is expected (see green region), which changes to a spin blockade (blue region) at $B_\perp = B_\mathrm{TS} \approx 0.4$~T due to the change in two-particle GS. 
\textbf{(b)} Transport through the DQD measured along a detuning cut of the bias triangle $(2,0) \rightarrow (1,1)$ at $V_\mathrm{SD} = 1$~mV, as a function of perpendicular magnetic field, see dashed arrow in Fig.~\ref{f2}(b). For clarity, we subtracted the average current $\langle I \rangle$ of each detuning trace.
\textbf{(c)} Simulation of the measurement presented in panel (b) using the same parameters as in Fig.~\ref{f4}(b). Colored dashed lines highlight prominent similarities between panel (b) and (c). Both a spin and valley blocked region are predicted by the simulation, indicated by the circles and stars, occurring before and after $B_\mathrm{TS}$, which is in contrast to what is shown panel (a). \textbf{(d)} Illustration of the energetic alignment of the $(1,1)$ and $(2,0)$ states at positions (I) and (II) highlighted in panels (a-c).}
\label{f5}
\end{figure*}
Importantly, Fig.~\ref{f3}(c) illustrates that we expect to observe a magnetic field tunable valley and spin blockade in the $(1,1) \rightarrow (2,0)$ bias triangle, as indicated by the green and blue colored regions, respectively. For $B_\perp < B_\mathrm{TS}$, the $(1,1)$ GS to $(2,0)$ GS transition $\ket{K^-\uparrow,\, K^-\uparrow} \rightarrow \ket{T^s_- S^v}$ (green region) requires a valley flip. 
For increasing detuning energy $\varepsilon$, transitions to the excited states $\ket{S^sT^v_{0\pm}}$ become energetically accessible (blue arrows in Fig.~\ref{f3}(a)). These transitions are spin blocked. However, the valley blocked transition to the ground state remains energetically available. 
Since the valley relaxation rate is expected to be faster than the spin relaxation rate ~\cite{Banszerus2022JunSpinRelax,Banszerus2024FebValleyLive}, valley relaxation is the process limiting the blockade.
Valley blockade (VB) is therefore only lifted if the antisymmetric orbital states are accessible, limiting the detuning energy range of the valley blocked region to
\begin{equation}
    \varepsilon^{(1,1) \rightarrow (2,0)}_\mathrm{VB} (B_\perp)=  \Delta_\mathrm{Orb} - \Delta_\mathrm{SO} - g^a_v \mu_B B_\perp \,. \nonumber
\end{equation}
For $B_\perp > B_\mathrm{TS}$, the $(1,1)$ GS to $(2,0)$ GS transition requires a spin flip (blue region),  $\ket{K^-\uparrow,\, K^-\uparrow} \rightarrow \ket{S^s T^v_-}$, limiting the detuning energy range of the spin blocked (SB) region to 
\begin{equation}
    \varepsilon^{(1,1) \rightarrow (2,0)}_\mathrm{SB} = \Delta_\mathrm{Orb} - \Delta_\mathrm{SO} - \delta_2 + (g^s_v - g_s - g^a_v) \mu_B B_\perp.
\end{equation}
The calculation shows that the valley blocked region is limited only by the properties of the antisymmetric orbital states, while the region of the spin blockade depends on properties of both symmetric and antisymmetric orbital states. In particular, for $g^s_v < g^a_v + g_s$, the spin blockade eventually vanishes at a finite out-of-plane magnetic field $B_\perp := B_\mathrm{Orb}$, if $\ket{S^s T^v_-}$ and $\ket{T^s_- T^v_-}$ become degenerate (in the depicted case this would happen at $\approx 3.35$~T). The spin blockade thus persists in the magnetic field range $B_\mathrm{TS} < B_\perp < B_\mathrm{Orb}$. 

In the following, we investigate in detail the magnetic field dependency of the resonances and blockade effects observed in the bias spectroscopy measurements (Fig.~\ref{f2}(a)) and predicted by theory (Fig.~\ref{f3}(c)). 
For that purpose we measure the current through the DQD as a function of $\varepsilon$ (see dashed arrow in Fig.~\ref{f2}(a)) and $B_\perp$, as shown in Fig.~\ref{f4}(a).
In particular, we aim to reproduce the data by a simulation of the DQD based on the Pauli rate equation (see {\it Appendix} B), which includes all 256 possible transitions instead of only considering transitions from the $(1,1)$ GS as shown in Fig.~\ref{f3}. With this, we aim to understand the origin of the most prominent features and quantify orbital splitting, valley g-factors and short-range interaction strength. The result of the simulation is shown in Fig.~\ref{f4}(b).
For a better comparison of theory and experiment, we subtract the average current $\langle I \rangle$ of each detuning trace of the experimental data in order to compensate the magnetic field induced reduction in tunnel rates.

The position of the zero detuning energy in the gate space changes slightly during the measurement, firstly due to slow electrostatic noise in the sample, and secondly due to the GS to GS transition shifting with the magnetic field. Therefore, using magnetic field dependent measurements of the bias triangles (see Fig.~\ref{f2}), the detuning traces are shifted such that $\varepsilon = 0$ is always at the same position, as indicated by the white dashed line in Fig.~\ref{f4}(a). 
Our simulation is able to reproduce many features in the data using $\Delta_\mathrm{Orb} = 0.7$~meV, $\delta_2 = 0.34$~meV, $g^{(1)}_v = 15$, $g^s_v = 18$, $g^a_v = 19$, which is in agreement with other measurements on QDs of similar dimension~\cite{Moller2023SepShellFilling, Banszerus2023MayNature, Banszerus2021SepSpinOrbit}. 
Both simulation and measurements show the transition from valley blockade (marked by the star) to spin blockade (marked by the circle), as can be seen by comparing Figs.~\ref{f4}(a) and~\ref{f4}(b). 
We identify prominent resonances labeled \textbf{a - d} in the data~\footnote{The software of the simulation allows to interactively inspect any position in gate space, where it illustrates the alignment of the chemical potentials in the left and right QD.}, with resonance \textbf{a} corresponding to a transition from the $(1,1)$ GS to the valley unpolarized triplet state in the antisymmetric orbital, $\ket{K^-\uparrow,\, K^-\uparrow} \rightarrow \ket{ T^s_- T^v_0}$, which requires a valley flip. 
Note that this transition is pronounced distinctly, despite the necessary valley flip. This is likely due to the different distribution of the antisymmetrical orbital state (compared to the symmetric orbital state) in k-space, leading to a non-zero overlap between the $(1,1)$ $\ket{K^- K^-}$ state and the anti-symmetric $\ket{T^v_0}$, which leads to an increased tunnel rate compared to transitions targeting the symmetric $\ket{T^v_0}$. 
Resonance \textbf{b} corresponds to the transition that first lifts the valley and spin blockade, $\ket{K^-\uparrow,\, K^-\uparrow} \rightarrow \ket{ T^s_- T^v_-}$, and is strongly pronounced in both data and simulation, as it is the transition from the (1,1) GS to the antisymmetric orbital state that requires the lowest detuning energy where both spin and valley blockade are lifted.
Resonance \textbf{c} involves a $(1,1)$ ES and requires a valley flip, $\ket{K^-\uparrow,\, K^-\downarrow} \rightarrow \ket{S^s T^v_0}$.
Finally, different valley blocked transitions contribute to resonance \textbf{d}, $\ket{K^-\uparrow,\, K^+\downarrow} \rightarrow \ket{ T^s_0 T^v_-}$, $\ket{K^-\downarrow,\, K^+\downarrow} \rightarrow \ket{ T^s_+ T^v_-}$, $\ket{K^-\downarrow,\, K^+\uparrow} \rightarrow \ket{ T^s_0 T^v_-}$, which is therefore rather broadened. Some of the resonant features do not shift linearly with $B_\perp$, which has been observed before in similar measurements and can be caused by the magnetic field slightly changing the effective confinement potential and thereby changing the wavefunction dependent valley g-factors. Additionally, we observe a dip in conductance at $B_\perp \approx \pm 0.05$ and $0.1~$T. Similar effects can be observed in Fig.~\ref{fig:DetuningsTriplePointB} and in equivalent measurements of BLG DQDs in the $ (1,2)  \leftrightarrow (0,3) $ charge configuration reported in Ref.~\cite{Tong2024Jul}, but have not yet been understood comprehensively.

For positive bias, we probe the $(2,0) \rightarrow (1,1)$ charge transition. When only considering the lowest single particle orbital in each QD, there are four possible states for a single charge carrier, $\ket{K^-\downarrow}, \ket{K^+\downarrow}$, $\ket{K^-\uparrow}, \ket{K^+\uparrow}$ and thus 16 possible $(1,1)$ states, that group into 8 valley polarized states and 8 valley unpolarized states.
We first only consider transitions from the two particle GS to the $(1,1)$ states. The resonance condition is in this case given by
\begin{equation}
   \Tilde{\varepsilon}(B_\perp) \mbeq E_{(1,1)}(B_\perp) - E_{(0,2)}(B_\perp)   \,.
    \label{eq:2011resonanceCondition}
\end{equation}
Following the same procedure as presented in the context of Fig.~\ref{f3}(c), we are able to plot the $B_\perp$ dependency of the expected resonant features as a function of detuning and perpendicular magnetic field, shown in Fig.~\ref{f5}(a). Here, the change in the two particle GS does not affect the detuning dependency of the transitions, because it also affects the onset of the bias triangle (defining $\varepsilon = 0$), see also equation~(\ref{eq:GS-GS_transition}). 
Therefore, Fig.~\ref{f5}(a) basically shows the magnetic field dispersion of the $(1,1)$ states with the energetically lowest state set to zero.
For $B_\mathrm{TS} > B_\perp > 0$,  we expect an increasingly large valley blocked region (green region in Fig.~\ref{f5}(a)) since the two particle GS is a valley singlet, $\ket{T^s_- S^v}$, while the $(1,1)$ GS is valley polarized, $\ket{K^-\uparrow,K^-\uparrow}$, already at small $B_\perp \neq 0~$T. 
The blockade is lifted as soon as the valley unpolarized $(1,1)$ states are available, which occurs at 
\begin{equation}
    \varepsilon^{(1,1) \leftarrow (2,0)}_\mathrm{VB} (B_\perp)= (g_s + g^{(1)}_v) \mu_B B_\perp \, .
\end{equation}
However, transport is still limited by valley blockade since the measured average tunnel current includes many charge cycles through the DQD. Thus, after a few cycles, $\ket{K^-\uparrow,K^-\uparrow}$ will be occupied again by chance, reestablishing the valley blockade.
For $B_\perp > B_\mathrm{TS}$, the two-particle GS becomes a spin singlet - valley triplet, $\ket{S^s T^v_-}$, immediately lifting the valley blockade. Instead, a small spin blockade (blue region in Fig.~\ref{f5}(a)) appears close to zero detuning since the $(1,1)$ GS is spin polarized. Black arrows in Fig.~\ref{f5}(a) indicate the spin orientation of the three lowest $(1,1)$ states. The spin blockade is lifted as soon as the first spin unpolarized state is available
\begin{equation}
    \varepsilon^{(1,1) \leftarrow (2,0)}_\mathrm{SB} = g_s \mu_B B_\perp + \Delta_\mathrm{SO}\, .
\end{equation}

We experimentally probe the $(2,0) \rightarrow (1,1)$ charge transition by measuring magnetic field dependent detuning cuts through the bias triangle, similar to Fig.~\ref{f4}(a), but for positive bias. These are presented in Fig.~\ref{f5}(b). Using the same parameters as for negative bias direction (see Fig.~\ref{f4}(b)), our simulation is able to reproduce the main features of the experimental data. 
Interestingly, both data and simulation show that the valley blocked region (star) extends beyond $B_\mathrm{TS} \approx 0.4$~T, which is in contrast to Fig.~\ref{f5}(a).
Furthermore, the simulation predicts that the spin blocked region (circle) appears along the entire magnetic field range, not only for $B_\perp > B_\mathrm{TS}$. The spin blocked region predicted by the simulation is barely visible in the data, it can only be surmised close to zero detuning for $B_\perp \approx 0.05 - 0.15$ T. However, this can be caused by  difficulties in estimating the onset of the bias triangle due to significant co-tunneling, see white dashed lines in Fig.~\ref{f2}(b).

The extended valley blocked region in Figs.~\ref{f5}(b) and~\ref{f5}(c) can be understood when inspecting the energetic alignment of the $(1,1)$ and $(2,0)$ states at $B_\perp\approx 0.6$~T and two different detuning energies (I) and (II), which are illustrated in Fig.~\ref{f5}(d). At (I) the DQD system is valley blocked, even though both the $(1,1)$ GS (purple line) and the $(2,0)$ GS (dashed blue line) are $K^-$ polarized. 
However, when the left QD is loaded from the lead, i.e. $(1,0) \rightarrow (2,0)$, also $\ket{T^s_{0\pm} S^v}$ ESs can be loaded, as they are energetically close by and thus also in the bias window. As soon as such a state is loaded, the system is valley blocked due to the $(2,0)$ state being a valley singlet. The blockade is lifted at higher detuning energies, when the valley unpolarized $(1,1)$ states are available, which is illustrated at position (II). 
Alternatively, the system could relax to the $(2,0)$ GS lifting the blockade, which would require both a valley and a spin flip. Estimating the interdot tunnel rate to be in the order of $10 $~ns~\footnote{We assume the tunnel current to be limited by the interdot tunnel rate.}, and considering first reported measurements of spin and valley lifetimes~\cite{Banszerus2022JunSpinRelax,Denisov2025Feb, Banszerus2024FebValleyLive}, it is not surprising that relaxation does not play a major role and valley blockade is indeed observed. 
Similarly, the spin blocked region for $B_\perp < B_\mathrm{TS}$ -- predicted by the simulation -- can be explained by loading a $(2,0)$ ES which is not $\uparrow \uparrow$-polarized.
This underlines that it is not sufficient to only consider transitions from the $(2,0)$ GS, but the full picture including all 16 two particle states is indeed required. 
Apart from that, the simulation allows to identify prominent resonances in Fig.~\ref{f5}(b), namely resonance \textbf{a}, which requires a valley flip and corresponds to $\ket{ T^s_- S^v} \rightarrow \ket{K^+\uparrow,\, K^+\uparrow}$. And resonance \textbf{b}, which first lifts the valley blockade, $\ket{T^s_{-} S^v } \rightarrow \ket{K^+\uparrow,\, K^-\uparrow}$, and resonance \textbf{c}, which corresponds to $\ket{ S^s T^v_-} \rightarrow \ket{K^+\downarrow,\, K^-\uparrow}$.

Please note, that we also observe discrepancies between simulated and experimental data presented in Figs.~\ref{f4} and~\ref{f5}, especially in the magnitude of the current (see e.g. resonance \textbf{d} in Fig.~\ref{f4}). Furthermore, the simulation shows more resonances than observed experimentally. We attribute this to different reasons.
First, there is the obvious reason of the interdot tunnel coupling potentially being to large to justify the approximation of completely independent single particle states in the $(1,1)$ configuration. Second, we assume equal tunnel probabilities for all states both for the interdot transition as well as for tunneling from the leads, which does not need to be the case, especially at finite magnetic field. Finally, we do not include a detuning and magnetic field dependency  - apart from the Zeeman effect - of our DQD states, which might be crucial since both the detuning and the magnetic field can change the confinement potential, possibly altering tunnel rates, orbital and short-range splitting, and valley g-factors. 
Nevertheless, we are able to obtain reasonable good agreement between the experimental data and simulation by including antisymmetric orbital states of the QD in the $(2,0)$ charge configuration. 

In conclusion, we investigated transport through a bilayer graphene DQD focusing on the $(1,1) \leftrightarrow (2,0)$ charge transitions. We obtain good agreement between our experimental data and a transport simulation based on rate equations. 
We included antisymmetric orbital states of the QD in the $(0,2)$ charge configuration, which give rise to a manifold of resonant features in magnetotransport measurements and our simulation. 
Crucially, we find that the observed magnetic field dependent spin and valley blockade is limited by the magnitude of the orbital splitting $\Delta_\mathrm{Orb}$, the difference between antisymmetric and symmetric orbital valley g-factor, $|g^a_v - g^s_v|$, and the short-range splitting $\delta_2$. 
Further improvements to our work could be made by taking a finite interdot tunnel coupling into account, which leads to a mixing of the $(1,1)$ and $(0,2)$ states depending on various parameters such as interdot tunnel coupling, QD geometry and screening~\cite{Knothe2024Jun}.
Together with other recent works on BLG QDs~\cite{Moller2023SepShellFilling, Tong2022FebBlockade, Tong2024JanCatalogue, Banszerus2022JunSpinRelax, Garreis2024MarLongLived}, we have obtained a comprehensive understanding of two-electron DQDs in BLG and are in a good position to explore different qubit regimes in the future. 

\textbf{Acknowledgements} The authors thank S.~Trellenkamp, F.~Lentz and M. Otto for their support in device fabrication, as well as F.~Hassler and A.~Knothe for helpful discussions.
This project has received funding from the European Research Council (ERC) under grant agreement No. 820254, the Deutsche Forschungsgemeinschaft (DFG, German Research Foundation) under Germany's Excellence Strategy - Cluster of Excellence Matter and Light for Quantum Computing (ML4Q) EXC 2004/1 - 390534769, and by the Helmholtz Nano Facility~\cite{Albrecht2017May}. K.W. and T.T. acknowledge support from the JSPS KAKENHI (Grant Numbers 21H05233 and 23H02052) and World Premier International Research Center Initiative (WPI), MEXT, Japan.

\section{Appendix}
\subsection{Additional measurements of the $(1,1) \leftrightarrow (0,2)$ charge transition}
\begin{figure*}[]\centering
\includegraphics[draft=false,keepaspectratio=true,clip,width=\linewidth]{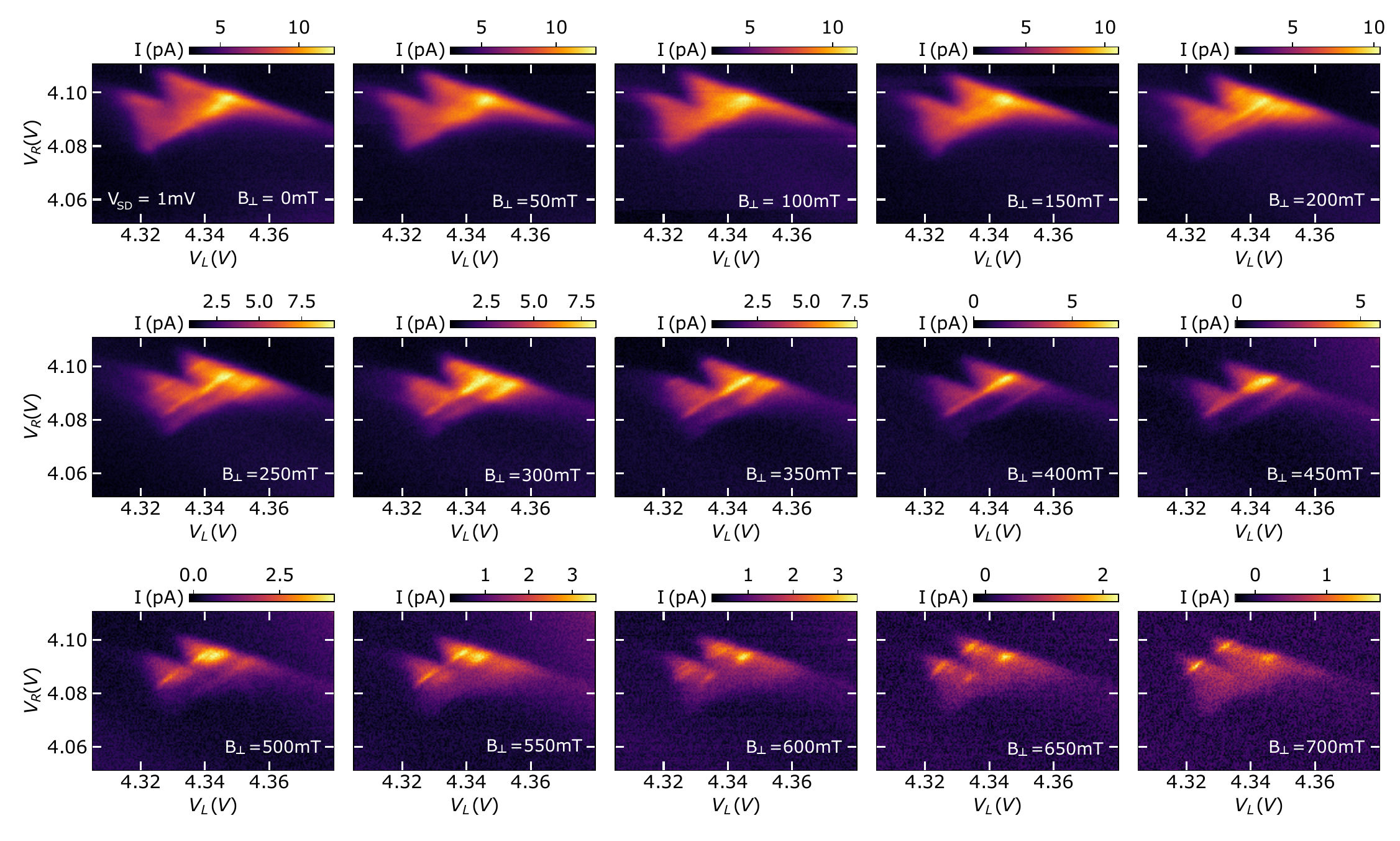}
\caption[]{Zoom-in on the $(1,1) \rightarrow (2,0)$ bias triangle for $V_\mathrm{SD} =  1$ mV and different perpendicular magnetic fields, complementary to Fig.~\ref{f2}.}
\label{fig:DetuningNonBlockadeA}
\end{figure*}

\begin{figure*}[]\centering
\includegraphics[draft=false,keepaspectratio=true,clip,width=\linewidth]{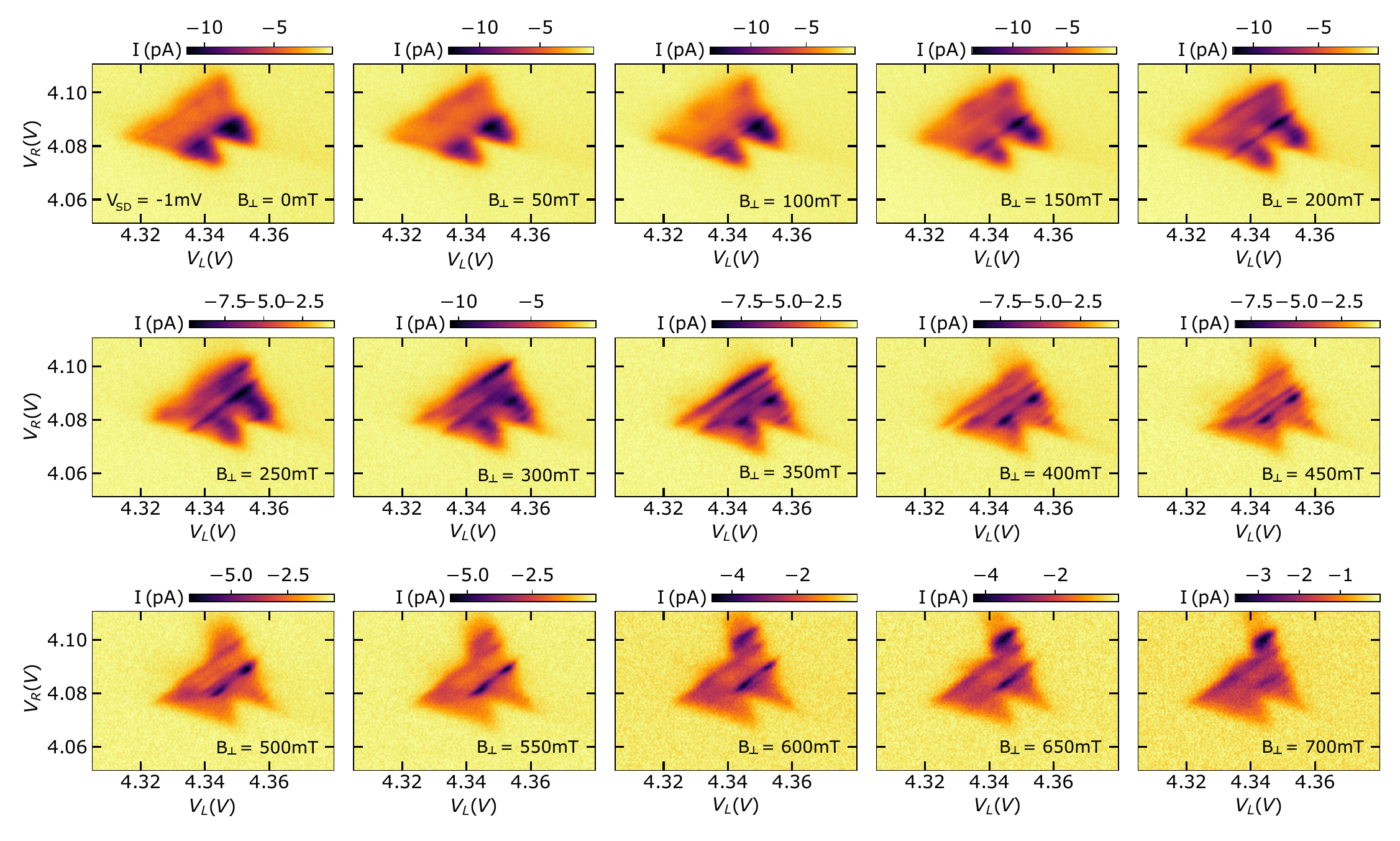}
\caption[]{Zoom-in on the $(1,1) \rightarrow (2,0)$ bias triangle for $V_\mathrm{SD} =  -1$ mV and different perpendicular magnetic fields, complementary to Fig.~\ref{f2}.}
\label{fig:DetuningBlockadeA}
\end{figure*}

\begin{figure*}[]\centering
\includegraphics[draft=false,keepaspectratio=true,clip,width=\linewidth]{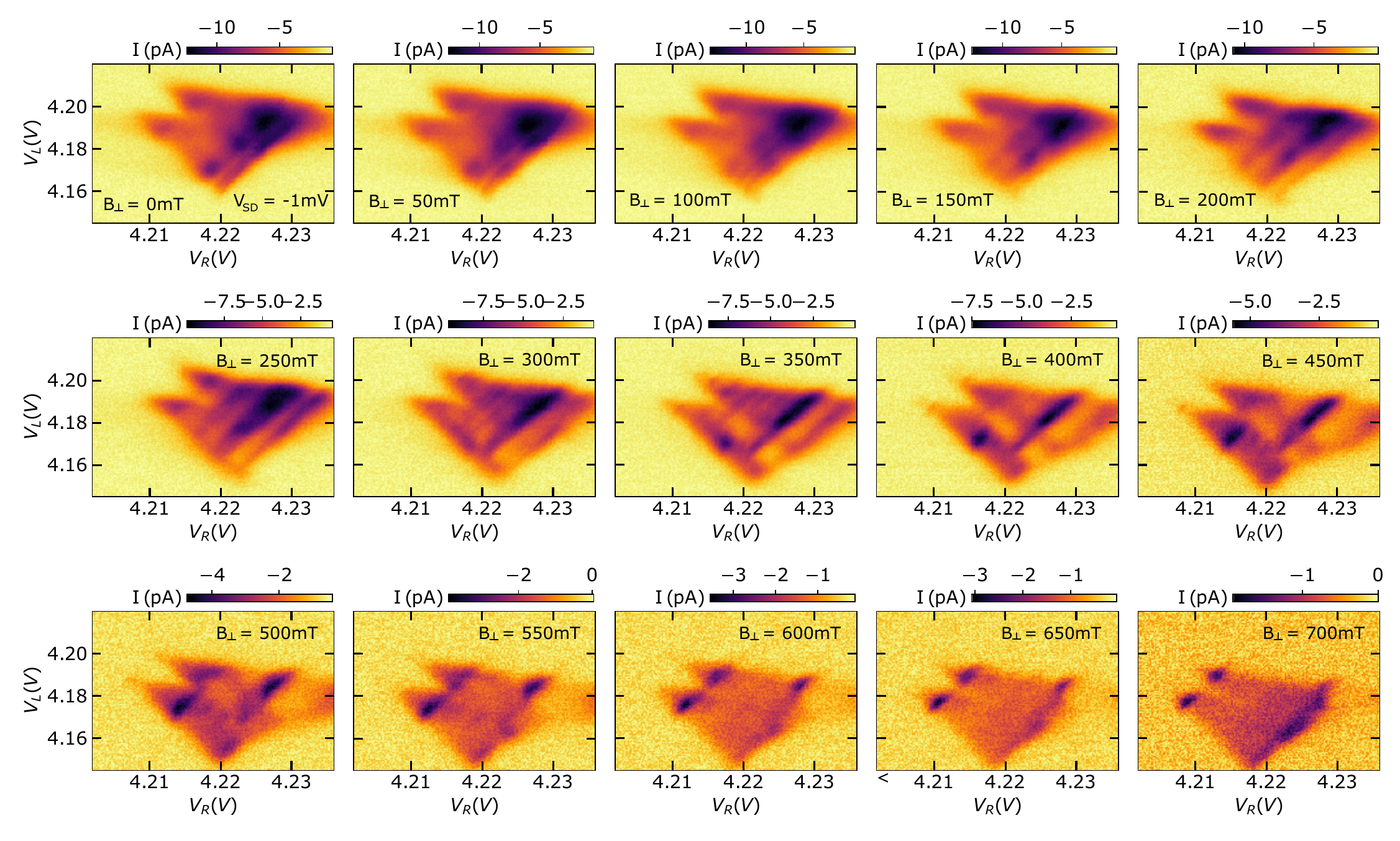}
\caption[Detuning cut of triple point A at $V_\mathrm{SD} = -1$ mV in $B_\perp$]{Zoom-in on the $(1,1) \rightarrow (0,2)$ bias triangle for $V_\mathrm{SD} = -1$ mV and different perpendicular magnetic fields.}
\label{fig:TPsNonBlockadeB}
\end{figure*}

\begin{figure}[]\centering
\includegraphics[draft=false,keepaspectratio=true,clip,width=\linewidth]{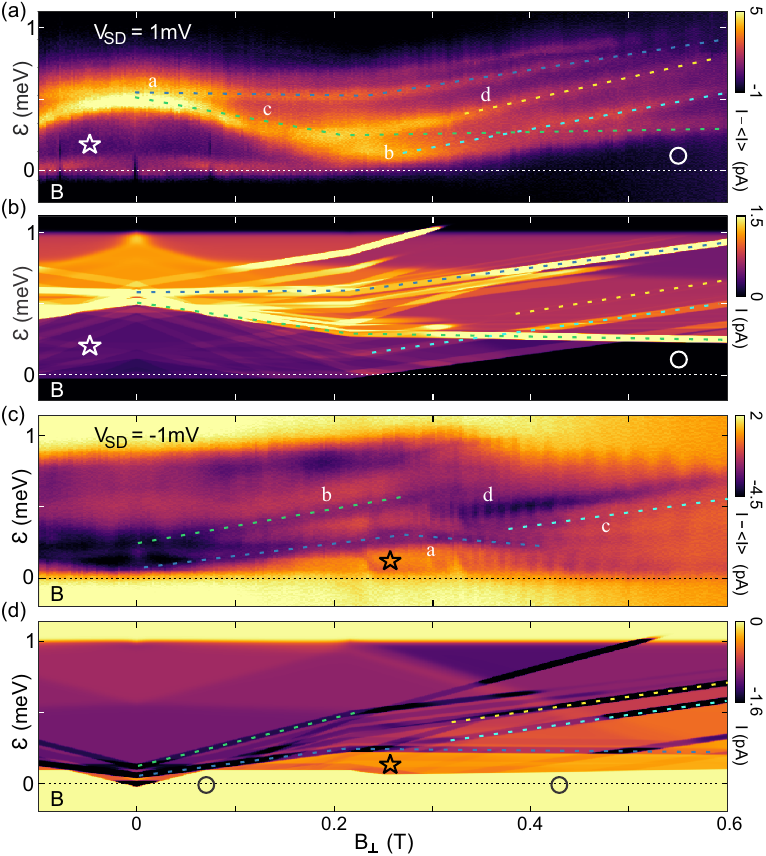}
\caption[]{\textbf{(a)}  Transport through the DQD measured along a detuning cut of triple point B at $V_\mathrm{SD} = 1$~mV as a function of perpendicular magnetic field. To compensate for the magnetic field induced reduction in tunnel rates, the average current was subtracted from each trace individually. The switch from valley to spin blockade is clearly visible, as indicated by the star and the circle, respectively. \textbf{(b)} Simulation of the measurement presented in (a) using the rate equation. Reasonable agreement with the data was achieved using $\Delta_\mathrm{Orb} = 0.575$~meV, $\delta_2 = 0.2$~meV, $g^{(1)}_v = 15$, $g^s_v = 18$, $g^a_v = 18$. Colored dashed lines highlight prominent resonances. Note that the valley blockade almost completely vanishes in the experimental data at $B_\perp \sim 0.22$~T, while it prevails in the simulation.
\textbf{(c)} As in (a) but for $V_\mathrm{SD} = -1$~mV. \textbf{(d)} Simulation of the measurement presented in (c) using the rate equation and the same paramters as in (a). Colored dashed lines highlight prominent features, while the star and circle indicate valley and spin blockade, respectively. Only the valley blocked region is observed in the experimental data.}
\label{fig:DetuningsTriplePointB}
\end{figure}

We present additional measurements of triple point A and B (as defined in Fig.~\ref{f1}) as a function of perpendicular magnetic field. Figs.~\ref{fig:DetuningNonBlockadeA} and~\ref{fig:DetuningBlockadeA} show the bias triangles of triple point A for $V_\mathrm{SD} = 1$~mV and $V_\mathrm{SD} = -1$~mV, respectively. The measurements are complementary to Fig.~\ref{f2}, adding intermediate magnetic fields, displayed in a linear color scale.
\begin{figure*}[]\centering
\includegraphics[draft=false,keepaspectratio=true,clip,width=\linewidth]{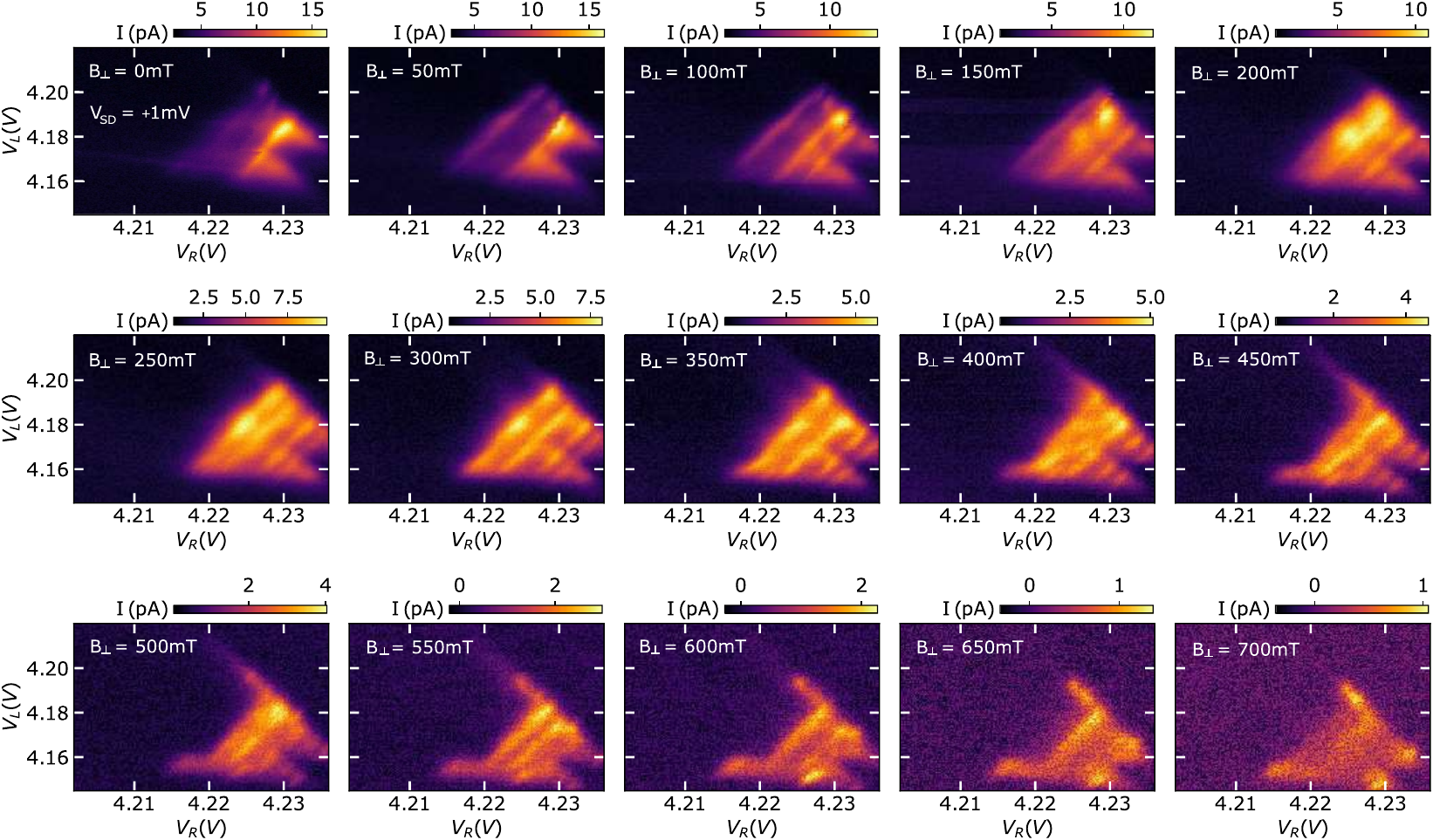}
\caption[]{Zoom-in on the $(1,1) \rightarrow (0,2)$ bias triangle for $V_\mathrm{SD} = 1$ mV and different perpendicular magnetic fields. At zero magnetic field, a distinct valley blocked region is visible. For increasing magnetic field, the extend of the valley blockade along the detuning axis reduces and eventually, a spin blocked region appears.}
\label{fig:TPsBlockadeB}
\end{figure*}
Figs.~\ref{fig:TPsBlockadeB} and~\ref{fig:TPsNonBlockadeB} show the bias triangles of triple point B for $V_\mathrm{SD} = 1$~mV and $V_\mathrm{SD} = -1$~mV, respectively. 
Similar to Fig.~\ref{f4} and Fig.~\ref{f5}(b,c), we investigate the magnetic field behavior of the resonances within bias triangle B by measuring along the detuning axis $\varepsilon$. We compare the measurement using our simulation, as shown in Fig.~\ref{fig:DetuningsTriplePointB} for  $V_\mathrm{SD} = 1$~mV (a,b) and $V_\mathrm{SD} = -1$~mV (c,d). As in the main text, we are able to reproduce the magnetic field tunable spin- and valley blockade (highlighted by the circle and star, respectively) as well as the most prominent resonances.


\subsection{Rate equation based transport simulation}
\label{sec:RateEquationAppendix}
\begin{figure*}[]\centering
\includegraphics[draft=false,keepaspectratio=true,clip,width=0.7\linewidth]{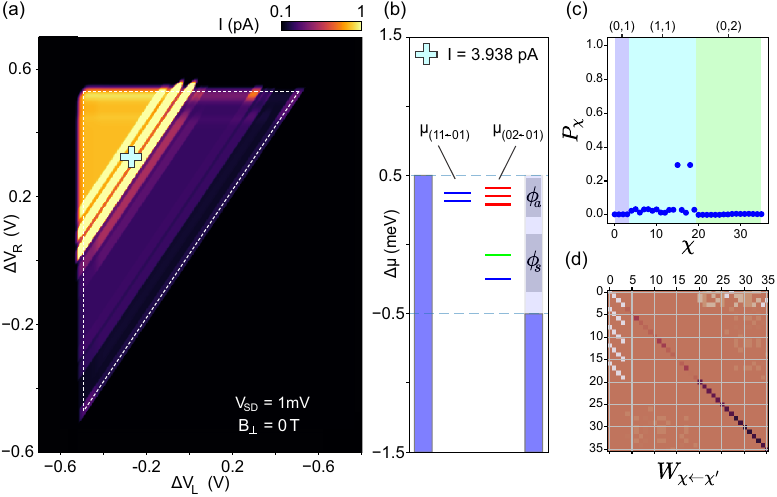}
\caption[Interactive simulation of the $(1,1) \rightarrow (0,2)$ triple point]{\textbf{(a)}  Simulated current through the DQD at the $(1,1) \rightarrow (0,2)$ triple point for $V_\mathrm{SD} = 1$ mV and $B_\perp = 0$. The white dashed line illustrates the outline of the triple point. \textbf{(b)} Schematics illustrating the configuration of source, drain and DQD chemical potentials at a chooseable position in gate space, e.g. at the position of the blue cross. The left horizontal lines depict the chemical potential of the four possible states of the single electron in the left QD. The right horizontal lines depict the chemical potentials of transitions from $ N = 1 \rightarrow 2$, into the orbitally symmetric states $\ket{S^v T^s_{0\pm}}$ (blue), $\ket{T^v_{0\pm} S^s}$ (green), and the orbitally antisymmetric states,  $\ket{T^v_{0\pm} T^s_{0\pm}}, \ket{S^v S^s}$ (red). 
\textbf{(c)} Probability vector $\mathrm{P}_{\chi}$, showing the probability of the DQD to be in a given state $\chi$ in the stationary limit, at the position of the blue cross. The DQD is mostly in $(1,1)$ states, transport is limited by the interdot transition. 
\textbf{(d)} Visualization of the entries of the rate matrix $W_{ \chi \leftarrow \chi'}$ at the position of the blue cross.}
\label{fig:1102_InteractiveSimulation}
\end{figure*}
We simulate the $(1,1) \leftrightarrow (0,2)$ charge transition using a rate equation, following the approach of Refs.~\cite{Bonet2002JanRateEQ, Timm2009AugRateEquationTheory, Knothe2022Apr}. We consider the $(0,1) \rightarrow (1,1) \rightarrow (0,2) \rightarrow (0,1)$ charge cycle, which needs to be possible in the stationary limit in order to allow for transport. The DQD has 36 possible configurations, 4 different $(0,1)$ states, 16 $(1,1)$ states and 16 $(0,2)$ states. Here, we neglect the next single particle orbital states in the $(0,1)$ and $(1,1)$ configuration, assuming they  are high enough in energy to not play a role. The energies\footnote{Charging energy and tuning by the gates is taken into account separately.} of the $(0,1)$ states, $\varepsilon_{01}$, in a perpendicular magnetic field are given by
\begin{align*}
    \varepsilon_{01}(v_R,s_R) &= \tfrac{1}{2} \Delta_\text{SO} v_R s_R + \frac{1}{2} g_\text{s} \mu_\text{B} B_\perp s_R + \frac{1}{2} g^{(1)}_\text{v} \mu_\text{B} \mathrm{B}_\perp v_R 
\end{align*}
with the spin and valley g-factors $g_\text{s}$ and $g_\text{v}$, the Bohr magneton $\mu_\text{B}$, the proximity enhanced (intrinsic) Kane-Mele spin-orbit coupling $\Delta_\text{SO} \approx 60\, \mu$eV and the spin and valley quantum numbers of the electron in the right QD, $(v_R, s_R) = (\pm 1, \pm 1)$. Correspondingly, the energies of the $(1,1)$ states, $\varepsilon_{11}$, are given by 
\begin{align*}
    \varepsilon_{11}(v_L,s_L, v_R,s_R) &= \varepsilon_{01}(v_L,s_L,) + \varepsilon_{01}(v_R,s_R).
\end{align*}
This approximation does not take inot account the mixing of the single particle states in the left and right QD. The strength of the mixing depends mostly on the magnitude of the interdot tunnel coupling but also on the shape and size of the confinement potential of the DQD, and the strength of the surrounding screening, making the resulting $(1,1)$ spectrum non-trivial~\cite{Knothe2024Jun}.
The energies of the $(0,2)$ states, $\varepsilon_{02}$, are given by~\cite{Moller2021Dec}
\begin{align*}
\varepsilon_{02}\,(\ket{S^v T^s_\pm}) &=  \pm g_s \mu_B B_\perp  ,\\
\varepsilon_{02}\,(\ket{S^v T^s_0}) &= 0 ,\\
\varepsilon_{02}\,(\ket{T^v_\pm S^s}) &=  \delta_2 \pm g^{s}_v \mu_B B_\perp  ,\\
\varepsilon_{02}\,(\ket{ T^v_0 S^s}) &=  \delta_2 +  \delta_1 ,\\
\notag\\
\varepsilon_{02}\,(\ket{ S^v S^s}) &= \Delta_\mathrm{Orb},\\
\varepsilon_{02}\,(\ket{ T^v_0 T^s_0}) &= \Delta_\mathrm{Orb},\\
\varepsilon_{02}\,(\ket{ T^v_0 T^s_\pm}) &= \Delta_\mathrm{Orb} \pm g_s \mu_B B_\perp ,\\
\varepsilon_{02}\,(\ket{ T^v_\pm T^s_0}) &= \Delta_\mathrm{Orb} \pm g^{a}_v \mu_B B_\perp,\\
\varepsilon_{02}\,(\ket{ T^v_\pm T^s_-}) &= \Delta_\mathrm{Orb} - \Delta_\mathrm{SO} \pm g^{a}_v \mu_B B_\perp  - g_s \mu_B B_\perp,\\
\varepsilon_{02}\,(\ket{ T^v_\pm T^s_+}) &= \Delta_\mathrm{Orb} + \Delta_\mathrm{SO} \pm g^{a}_v \mu_B B_\perp  + g_s \mu_B B_\perp,\\
\end{align*}
with the short-range splittings $\delta_1, \delta_2$, the valley g-factors for the symmetric, $g^s_v$, and antisymmetric orbital, $g^a_v$, the Kane-Mele SOC, $\Delta_\mathrm{SO}$, and the orbital splitting between symmetric and antisymmetric orbital, $\Delta_\mathrm{Orb}$. 
The electrostatic energy of each charge configuration is approximated by 
\begin{equation*}
    E_\text{electrostatic} (N_\text{R}, N_\text{L}) = e N_\text{R}  V_\text{R} + e N_\text{L}  V_\text{L} + N_\text{R} E_C\,,
\end{equation*}
with the absolute value of the elementary charge, $e$, the QD occupation number $N_\text{L} = 0,
1$,  $N_\text{R} = 1,2$, the charging energy in the right QD, $E_C$, and the gate voltages $V_\text{R}$ and $V_\text{L}$. The total energy of the DQD is then given by $E_{N_\text{R}, N_\text{L}} = \varepsilon_\mathrm{N_\text{R}, N_\text{L}} +  E_\text{electrostatic} (N_\text{R}, N_\text{L})$.

For solving the rate equation we define a 36 dimensional vector, describing the state of the DQD,
\begin{equation*}
    \chi = ((0,1), (1,1), (0,2))^\mathrm{T}.
\end{equation*}
The probabilities of the system to be in a given state $\mathrm{P}_{\chi}$, are related by
\begin{equation*}
\dot{\mathrm{P}}_{\chi} =\sum_{\chi'} (W_{ \chi \leftarrow \chi'} \, \mathrm{P}_{\chi'}  - W_{ \chi' \leftarrow \chi} \,\mathrm{P}_{\chi} ) ,
\label{eqn:rateeqn}
\end{equation*}
with the matrix $W_{\chi \leftarrow \chi'}$ describing the transition rates between states $\chi$ and $\chi'$. For computing the transition rates, we assume no mixing between lead and QD states and equal tunnel probabilities to and from the leads for all states, $\gamma^\text{L,R}$. Thus, we obtain the transition rates between QD states involving tunneling processes from the left lead by computing
\begin{align*}
W^\text{L}_{(0,1) \leftrightarrow (1,1)} &= \gamma^\text{L} \,  f(E_{01} - E_{11} - \mu^\text{L}),
\end{align*}
with the Fermi-function, $f$, at $T = 0.1\,$K and the chemical potential of the left lead, $\mu^\text{L}$. Similarly, tunneling to the right lead involves the  $(0,1)$ and $(0,2)$ states,  \begin{align*}
W^\text{R}_{(0,1) \leftrightarrow (0,2)} &= \gamma^\text{R} \,  f(E_{01} - E_{02} - \mu^\text{R})
\end{align*}
For interdot transitions, we compute
\begin{align}
W^\text{inter}_{(1,1) \leftrightarrow (0,2)} &= \gamma^\text{inter} \, |\bra{(1,1)} \ket{(0,2)}|^2  \nonumber \\
& \frac{1}{\sqrt{2 \pi \sigma}} \mathrm{exp}\left(  {-\frac{(E_{11} - E_{02})^2}{4 \sigma^2}} \right),
\label{align:interdot}
\end{align} 
with the interdot tunnel rate $\gamma^\text{inter}$. The Gaussian energy broadening, $\sigma$, allows to model the experimentally observed peak width, which is expected to originate from voltage fluctuations of the FGs. 

In our software, we may allow spin or valley flips to happen during tunneling, which is achieved by only computing the overlap in the valley or spin subspace, respectively. However, we introduce a valley/spin flip penalty parameter, $\zeta_{s,v}$, which reduces the overlap if a spin or valley flip was needed, i.e. $\bra{K^- \uparrow , K^+ \uparrow} \ket{ T^v_- T^s_-} = \bra{K^- \uparrow, K^+ \uparrow} \ket{K^- \uparrow K^- \uparrow}  = \zeta_v $.
With equation (\ref{align:interdot}) we implicitly assume that the states in the left and right QD have no coherent phase relation, i.e. no difference is made between tunneling into a singlet or triplet-0 state. Additionally, we add to eqution (10) an inelastic background term, 
\begin{align*}
\delta W^\text{inter}_{(1,1) \leftrightarrow (0,2)} &= \alpha_\text{inel} \, \gamma^\text{inter} \, |\bra{(1,1)} \ket{(0,2)}|^2  \\ &\Theta (E_\mathrm{inital} - E_\mathrm{final}),
\end{align*} 
with the dimensionless scaling parameter $\alpha_\text{inel}$ and the step function $\Theta$. Afterwards,  the columns of the transition matrix are normalized to 0,  $W_{ \chi\chi} = - \Sigma_{\chi' \neq \chi} W_{\chi' \chi}$ for computational reasons.
In the stationary limit, $\dot{\mathrm{P}}_{\chi} = 0$ and we can obtain $\mathrm{P}_{\chi}$ by computing the eigenvector of $W_{ \chi \leftarrow \chi'}$ for its vanishing eigenvalue, $\lambda_W = 0$ and normalizing the probabilities to $\sum_{\chi} \mathrm{P}_{\chi} = 1$~\cite{Timm2009AugRateEquationTheory}. We can compute the current through the DQD by computing the current flow from the right QD to lead R:
\begin{equation} 
 I^\text{R}= e \sum_{(0,1), (0,2)} \left( W^\text{R}_{(0,2) \leftarrow (0,1)} \mathrm{P}_{(0,1)}
-  W^\text{R}_{(0,1) \leftarrow (0,2) } \mathrm{P}_{(0,2)} \right).
 \label{eqn:IRateEQ}
\end{equation}

Our software follows these steps for every point in gate space, $V_\text{L}$, $V_\text{R}$, and different magnetic fields, creating charge stability diagrams or magnetic field dependent detuning cuts as presented in Figs.~\ref{f2}, \ref{f3} and \ref{f4}. In order to so, we fixed the following parameters for all calculations, $E_C = 5$~meV, $\sigma = 6\, \mu$eV, $g_s = 2$, $\Delta_\mathrm{SO} = 60\,\mu$eV, $\zeta_s = 0$, $\zeta_v = \frac{1}{\sqrt{2}}$, $\alpha_\mathrm{inel} = 0.01$, $\gamma^\mathrm{inter} = 2$~MHz, and $\gamma^R = \gamma^L = 5$~GHz. The remaining parameters, describing the DQD's energy dispersion, $\Delta_\mathrm{Orb}$, $\delta_{1,2}$, $g^{(1)}_v$, $g^s_v$, and $g^a_v$, are chosen for triple point A and B individually.

The software further allows to interactively inspect the charge stability diagrams, as illustrated by Fig.~\ref{fig:1102_InteractiveSimulation}, where the DQD parameters in the simulation were set to $\Delta_\mathrm{Orb} = 0.575$~meV, $\delta_2 = 0.2$~meV, $\delta_1 = 0$, $g^{(1)}_v = 15$, $g^s_v = 18$, $g^a_v = 18$. Fig.~\ref{fig:1102_InteractiveSimulation}(a) shows the resulting triple point for $B_\perp = 0$. After double-clicking a point in gate space, Fig.~\ref{fig:1102_InteractiveSimulation}(b) displays the corresponding current and configuration of the chemical potentials in the right and left QD. The left horizontal lines in (b) depict the chemical potential of the four possible states of the single electron in the left QD. The right horizontal lines depict the chemical potentials of transitions from $ N = 1 \rightarrow 2$, into the orbitally symmetric states $\ket{S^v T^s_{0\pm}}$ (blue), $\ket{T^v_{0\pm} S^s}$ (green), and the orbitally antisymmetric states,  $\ket{T^v_{0\pm} T^s_{0\pm}}, \ket{S^v S^s}$ (red), assuming the single electron in the right QD to be in the GS before the transition.
Fig.~\ref{fig:1102_InteractiveSimulation}(c) and (d) show the corresponding probability vector $\mathrm{P}_{\chi}$ and the rate matrix  $W_{ \chi \leftarrow \chi'}$. Having clicked at the position of the blue cross, we can indeed verify the interpretation of Fig.~\ref{f3}, that the blockade is lifted as soon as the detuning compensates the orbital splitting and the antisymmetric states become accessible. In the probability vector we can see that the DQD is in the $(1,1)$ configuration most of the time, i.e. we are limited by the interdot tunnel rate, just as in our experimental data. 

\FloatBarrier
\subsection{Data availability}
The data and evaluation scripts supporting the findings of this work are available in a Zenodo repository under XXX.
The code of the simulation can be found in another Zenodo repository under~\cite{ZenodoSimulation}.


\begin{thebibliography}{72}%
\makeatletter
\providecommand \@ifxundefined [1]{%
 \@ifx{#1\undefined}
}%
\providecommand \@ifnum [1]{%
 \ifnum #1\expandafter \@firstoftwo
 \else \expandafter \@secondoftwo
 \fi
}%
\providecommand \@ifx [1]{%
 \ifx #1\expandafter \@firstoftwo
 \else \expandafter \@secondoftwo
 \fi
}%
\providecommand \natexlab [1]{#1}%
\providecommand \enquote  [1]{``#1''}%
\providecommand \bibnamefont  [1]{#1}%
\providecommand \bibfnamefont [1]{#1}%
\providecommand \citenamefont [1]{#1}%
\providecommand \href@noop [0]{\@secondoftwo}%
\providecommand \href [0]{\begingroup \@sanitize@url \@href}%
\providecommand \@href[1]{\@@startlink{#1}\@@href}%
\providecommand \@@href[1]{\endgroup#1\@@endlink}%
\providecommand \@sanitize@url [0]{\catcode `\\12\catcode `\$12\catcode `\&12\catcode `\#12\catcode `\^12\catcode `\_12\catcode `\%12\relax}%
\providecommand \@@startlink[1]{}%
\providecommand \@@endlink[0]{}%
\providecommand \url  [0]{\begingroup\@sanitize@url \@url }%
\providecommand \@url [1]{\endgroup\@href {#1}{\urlprefix }}%
\providecommand \urlprefix  [0]{URL }%
\providecommand \Eprint [0]{\href }%
\providecommand \doibase [0]{https://doi.org/}%
\providecommand \selectlanguage [0]{\@gobble}%
\providecommand \bibinfo  [0]{\@secondoftwo}%
\providecommand \bibfield  [0]{\@secondoftwo}%
\providecommand \translation [1]{[#1]}%
\providecommand \BibitemOpen [0]{}%
\providecommand \bibitemStop [0]{}%
\providecommand \bibitemNoStop [0]{.\EOS\space}%
\providecommand \EOS [0]{\spacefactor3000\relax}%
\providecommand \BibitemShut  [1]{\csname bibitem#1\endcsname}%
\let\auto@bib@innerbib\@empty
\bibitem [{\citenamefont {Loss}\ and\ \citenamefont {DiVincenzo}(1998)}]{Loss1998Jan}%
  \BibitemOpen
  \bibfield  {author} {\bibinfo {author} {\bibfnamefont {D.}~\bibnamefont {Loss}}\ and\ \bibinfo {author} {\bibfnamefont {D.~P.}\ \bibnamefont {DiVincenzo}},\ }\bibfield  {title} {\bibinfo {title} {{Quantum computation with quantum dots}},\ }\href {https://doi.org/10.1103/PhysRevA.57.120} {\bibfield  {journal} {\bibinfo  {journal} {Phys. Rev. A}\ }\textbf {\bibinfo {volume} {57}},\ \bibinfo {pages} {120} (\bibinfo {year} {1998})}\BibitemShut {NoStop}%
\bibitem [{\citenamefont {Vajner}\ \emph {et~al.}(2022)\citenamefont {Vajner}, \citenamefont {Rickert}, \citenamefont {Gao}, \citenamefont {Kaymazlar},\ and\ \citenamefont {Heindel}}]{Vajner2022Jul}%
  \BibitemOpen
  \bibfield  {author} {\bibinfo {author} {\bibfnamefont {D.~A.}\ \bibnamefont {Vajner}}, \bibinfo {author} {\bibfnamefont {L.}~\bibnamefont {Rickert}}, \bibinfo {author} {\bibfnamefont {T.}~\bibnamefont {Gao}}, \bibinfo {author} {\bibfnamefont {K.}~\bibnamefont {Kaymazlar}},\ and\ \bibinfo {author} {\bibfnamefont {T.}~\bibnamefont {Heindel}},\ }\bibfield  {title} {\bibinfo {title} {{Quantum Communication Using Semiconductor Quantum Dots}},\ }\href {https://doi.org/10.1002/qute.202100116} {\bibfield  {journal} {\bibinfo  {journal} {Adv. Quantum Technol.}\ }\textbf {\bibinfo {volume} {5}},\ \bibinfo {pages} {2100116} (\bibinfo {year} {2022})}\BibitemShut {NoStop}%
\bibitem [{\citenamefont {Burkard}\ \emph {et~al.}(1999)\citenamefont {Burkard}, \citenamefont {Loss},\ and\ \citenamefont {DiVincenzo}}]{Burkard1999Jan}%
  \BibitemOpen
  \bibfield  {author} {\bibinfo {author} {\bibfnamefont {G.}~\bibnamefont {Burkard}}, \bibinfo {author} {\bibfnamefont {D.}~\bibnamefont {Loss}},\ and\ \bibinfo {author} {\bibfnamefont {D.~P.}\ \bibnamefont {DiVincenzo}},\ }\bibfield  {title} {\bibinfo {title} {{Coupled quantum dots as quantum gates}},\ }\href {https://doi.org/10.1103/PhysRevB.59.2070} {\bibfield  {journal} {\bibinfo  {journal} {Phys. Rev. B}\ }\textbf {\bibinfo {volume} {59}},\ \bibinfo {pages} {2070} (\bibinfo {year} {1999})}\BibitemShut {NoStop}%
\bibitem [{\citenamefont {Petta}\ \emph {et~al.}(2005)\citenamefont {Petta}, \citenamefont {Johnson}, \citenamefont {Taylor}, \citenamefont {Laird}, \citenamefont {Yacoby}, \citenamefont {Lukin}, \citenamefont {Marcus}, \citenamefont {Hanson},\ and\ \citenamefont {Gossard}}]{Petta2005Sep}%
  \BibitemOpen
  \bibfield  {author} {\bibinfo {author} {\bibfnamefont {J.~R.}\ \bibnamefont {Petta}}, \bibinfo {author} {\bibfnamefont {A.~C.}\ \bibnamefont {Johnson}}, \bibinfo {author} {\bibfnamefont {J.~M.}\ \bibnamefont {Taylor}}, \bibinfo {author} {\bibfnamefont {E.~A.}\ \bibnamefont {Laird}}, \bibinfo {author} {\bibfnamefont {A.}~\bibnamefont {Yacoby}}, \bibinfo {author} {\bibfnamefont {M.~D.}\ \bibnamefont {Lukin}}, \bibinfo {author} {\bibfnamefont {C.~M.}\ \bibnamefont {Marcus}}, \bibinfo {author} {\bibfnamefont {M.~P.}\ \bibnamefont {Hanson}},\ and\ \bibinfo {author} {\bibfnamefont {A.~C.}\ \bibnamefont {Gossard}},\ }\bibfield  {title} {\bibinfo {title} {{Coherent Manipulation of Coupled Electron Spins in Semiconductor Quantum Dots}},\ }\href {https://doi.org/10.1126/science.1116955} {\bibfield  {journal} {\bibinfo  {journal} {Science}\ }\textbf {\bibinfo {volume} {309}},\ \bibinfo {pages} {2180} (\bibinfo {year} {2005})}\BibitemShut {NoStop}%
\bibitem [{\citenamefont {Bluhm}\ \emph {et~al.}(2011)\citenamefont {Bluhm}, \citenamefont {Foletti}, \citenamefont {Neder}, \citenamefont {Rudner}, \citenamefont {Mahalu}, \citenamefont {Umansky},\ and\ \citenamefont {Yacoby}}]{Bluhm2011Feb}%
  \BibitemOpen
  \bibfield  {author} {\bibinfo {author} {\bibfnamefont {H.}~\bibnamefont {Bluhm}}, \bibinfo {author} {\bibfnamefont {S.}~\bibnamefont {Foletti}}, \bibinfo {author} {\bibfnamefont {I.}~\bibnamefont {Neder}}, \bibinfo {author} {\bibfnamefont {M.}~\bibnamefont {Rudner}}, \bibinfo {author} {\bibfnamefont {D.}~\bibnamefont {Mahalu}}, \bibinfo {author} {\bibfnamefont {V.}~\bibnamefont {Umansky}},\ and\ \bibinfo {author} {\bibfnamefont {A.}~\bibnamefont {Yacoby}},\ }\bibfield  {title} {\bibinfo {title} {{Dephasing time of GaAs electron-spin qubits coupled to a nuclear bath exceeding 200{\hspace{0.167em}}{$\mu$}s}},\ }\href {https://doi.org/10.1038/nphys1856} {\bibfield  {journal} {\bibinfo  {journal} {Nat. Phys.}\ }\textbf {\bibinfo {volume} {7}},\ \bibinfo {pages} {109} (\bibinfo {year} {2011})}\BibitemShut {NoStop}%
\bibitem [{\citenamefont {Cerfontaine}\ \emph {et~al.}(2020)\citenamefont {Cerfontaine}, \citenamefont {Botzem}, \citenamefont {Ritzmann}, \citenamefont {Humpohl}, \citenamefont {Ludwig}, \citenamefont {Schuh}, \citenamefont {Bougeard}, \citenamefont {Wieck},\ and\ \citenamefont {Bluhm}}]{Cerfontaine2020Aug}%
  \BibitemOpen
  \bibfield  {author} {\bibinfo {author} {\bibfnamefont {P.}~\bibnamefont {Cerfontaine}}, \bibinfo {author} {\bibfnamefont {T.}~\bibnamefont {Botzem}}, \bibinfo {author} {\bibfnamefont {J.}~\bibnamefont {Ritzmann}}, \bibinfo {author} {\bibfnamefont {S.~S.}\ \bibnamefont {Humpohl}}, \bibinfo {author} {\bibfnamefont {A.}~\bibnamefont {Ludwig}}, \bibinfo {author} {\bibfnamefont {D.}~\bibnamefont {Schuh}}, \bibinfo {author} {\bibfnamefont {D.}~\bibnamefont {Bougeard}}, \bibinfo {author} {\bibfnamefont {A.~D.}\ \bibnamefont {Wieck}},\ and\ \bibinfo {author} {\bibfnamefont {H.}~\bibnamefont {Bluhm}},\ }\bibfield  {title} {\bibinfo {title} {{Closed-loop control of a GaAs-based singlet-triplet spin qubit with 99.5{\%} gate fidelity and low leakage}},\ }\href {https://doi.org/10.1038/s41467-020-17865-3} {\bibfield  {journal} {\bibinfo  {journal} {Nat. Commun.}\ }\textbf {\bibinfo {volume} {11}},\ \bibinfo {pages} {4144} (\bibinfo {year} {2020})}\BibitemShut {NoStop}%
\bibitem [{\citenamefont {Sala}\ \emph {et~al.}(2020)\citenamefont {Sala}, \citenamefont {Qvist},\ and\ \citenamefont {Danon}}]{Sala2020Mar}%
  \BibitemOpen
  \bibfield  {author} {\bibinfo {author} {\bibfnamefont {A.}~\bibnamefont {Sala}}, \bibinfo {author} {\bibfnamefont {J.~H.}\ \bibnamefont {Qvist}},\ and\ \bibinfo {author} {\bibfnamefont {J.}~\bibnamefont {Danon}},\ }\bibfield  {title} {\bibinfo {title} {{Highly tunable exchange-only singlet-only qubit in a GaAs triple quantum dot}},\ }\href {https://doi.org/10.1103/PhysRevResearch.2.012062} {\bibfield  {journal} {\bibinfo  {journal} {Phys. Rev. Res.}\ }\textbf {\bibinfo {volume} {2}},\ \bibinfo {pages} {012062} (\bibinfo {year} {2020})}\BibitemShut {NoStop}%
\bibitem [{\citenamefont {Philips}\ \emph {et~al.}(2022)\citenamefont {Philips}, \citenamefont {Madzik}, \citenamefont {Amitonov}, \citenamefont {de~Snoo}, \citenamefont {Russ}, \citenamefont {Kalhor}, \citenamefont {Volk}, \citenamefont {Lawrie}, \citenamefont {Brousse}, \citenamefont {Tryputen}, \citenamefont {Wuetz}, \citenamefont {Sammak}, \citenamefont {Veldhorst}, \citenamefont {Scappucci},\ and\ \citenamefont {Vandersypen}}]{Philips2022Sep}%
  \BibitemOpen
  \bibfield  {author} {\bibinfo {author} {\bibfnamefont {S.~G.~J.}\ \bibnamefont {Philips}}, \bibinfo {author} {\bibfnamefont {M.~T.}\ \bibnamefont {Madzik}}, \bibinfo {author} {\bibfnamefont {S.~V.}\ \bibnamefont {Amitonov}}, \bibinfo {author} {\bibfnamefont {S.~L.}\ \bibnamefont {de~Snoo}}, \bibinfo {author} {\bibfnamefont {M.}~\bibnamefont {Russ}}, \bibinfo {author} {\bibfnamefont {N.}~\bibnamefont {Kalhor}}, \bibinfo {author} {\bibfnamefont {C.}~\bibnamefont {Volk}}, \bibinfo {author} {\bibfnamefont {W.~I.~L.}\ \bibnamefont {Lawrie}}, \bibinfo {author} {\bibfnamefont {D.}~\bibnamefont {Brousse}}, \bibinfo {author} {\bibfnamefont {L.}~\bibnamefont {Tryputen}}, \bibinfo {author} {\bibfnamefont {B.~P.}\ \bibnamefont {Wuetz}}, \bibinfo {author} {\bibfnamefont {A.}~\bibnamefont {Sammak}}, \bibinfo {author} {\bibfnamefont {M.}~\bibnamefont {Veldhorst}}, \bibinfo {author} {\bibfnamefont {G.}~\bibnamefont {Scappucci}},\ and\ \bibinfo {author} {\bibfnamefont {L.~M.~K.}\ \bibnamefont {Vandersypen}},\ }\bibfield
  {title} {\bibinfo {title} {{Universal control of a six-qubit quantum processor in silicon}},\ }\href {https://doi.org/10.1038/s41586-022-05117-x} {\bibfield  {journal} {\bibinfo  {journal} {Nature}\ }\textbf {\bibinfo {volume} {609}},\ \bibinfo {pages} {919} (\bibinfo {year} {2022})}\BibitemShut {NoStop}%
\bibitem [{\citenamefont {Dumoulin~Stuyck}\ \emph {et~al.}(2021)\citenamefont {Dumoulin~Stuyck}, \citenamefont {Mohiyaddin}, \citenamefont {Li}, \citenamefont {Heyns}, \citenamefont {Govoreanu},\ and\ \citenamefont {Radu}}]{DumoulinStuyck2021Aug}%
  \BibitemOpen
  \bibfield  {author} {\bibinfo {author} {\bibfnamefont {N.~I.}\ \bibnamefont {Dumoulin~Stuyck}}, \bibinfo {author} {\bibfnamefont {F.~A.}\ \bibnamefont {Mohiyaddin}}, \bibinfo {author} {\bibfnamefont {R.}~\bibnamefont {Li}}, \bibinfo {author} {\bibfnamefont {M.}~\bibnamefont {Heyns}}, \bibinfo {author} {\bibfnamefont {B.}~\bibnamefont {Govoreanu}},\ and\ \bibinfo {author} {\bibfnamefont {I.~P.}\ \bibnamefont {Radu}},\ }\bibfield  {title} {\bibinfo {title} {{Low dephasing and robust micromagnet designs for silicon spin qubits}},\ }\href {https://doi.org/10.1063/5.0059939} {\bibfield  {journal} {\bibinfo  {journal} {Appl. Phys. Lett.}\ }\textbf {\bibinfo {volume} {119}},\ \bibinfo {pages} {094001} (\bibinfo {year} {2021})}\BibitemShut {NoStop}%
\bibitem [{\citenamefont {Klemt}\ \emph {et~al.}(2023)\citenamefont {Klemt}, \citenamefont {Elhomsy}, \citenamefont {Nurizzo}, \citenamefont {Hamonic}, \citenamefont {Martinez}, \citenamefont {Cardoso~Paz}, \citenamefont {Spence}, \citenamefont {Dartiailh}, \citenamefont {Jadot}, \citenamefont {Chanrion}, \citenamefont {Thiney}, \citenamefont {Lethiecq}, \citenamefont {Bertrand}, \citenamefont {Niebojewski}, \citenamefont {B{\ifmmode\ddot{a}\else\"{a}\fi}uerle}, \citenamefont {Vinet}, \citenamefont {Niquet}, \citenamefont {Meunier},\ and\ \citenamefont {Urdampilleta}}]{Klemt2023Oct}%
  \BibitemOpen
  \bibfield  {author} {\bibinfo {author} {\bibfnamefont {B.}~\bibnamefont {Klemt}}, \bibinfo {author} {\bibfnamefont {V.}~\bibnamefont {Elhomsy}}, \bibinfo {author} {\bibfnamefont {M.}~\bibnamefont {Nurizzo}}, \bibinfo {author} {\bibfnamefont {P.}~\bibnamefont {Hamonic}}, \bibinfo {author} {\bibfnamefont {B.}~\bibnamefont {Martinez}}, \bibinfo {author} {\bibfnamefont {B.}~\bibnamefont {Cardoso~Paz}}, \bibinfo {author} {\bibfnamefont {C.}~\bibnamefont {Spence}}, \bibinfo {author} {\bibfnamefont {M.~C.}\ \bibnamefont {Dartiailh}}, \bibinfo {author} {\bibfnamefont {B.}~\bibnamefont {Jadot}}, \bibinfo {author} {\bibfnamefont {E.}~\bibnamefont {Chanrion}}, \bibinfo {author} {\bibfnamefont {V.}~\bibnamefont {Thiney}}, \bibinfo {author} {\bibfnamefont {R.}~\bibnamefont {Lethiecq}}, \bibinfo {author} {\bibfnamefont {B.}~\bibnamefont {Bertrand}}, \bibinfo {author} {\bibfnamefont {H.}~\bibnamefont {Niebojewski}}, \bibinfo {author} {\bibfnamefont {C.}~\bibnamefont {B{\ifmmode\ddot{a}\else\"{a}\fi}uerle}}, \bibinfo
  {author} {\bibfnamefont {M.}~\bibnamefont {Vinet}}, \bibinfo {author} {\bibfnamefont {Y.-M.}\ \bibnamefont {Niquet}}, \bibinfo {author} {\bibfnamefont {T.}~\bibnamefont {Meunier}},\ and\ \bibinfo {author} {\bibfnamefont {M.}~\bibnamefont {Urdampilleta}},\ }\bibfield  {title} {\bibinfo {title} {{Electrical manipulation of a single electron spin in CMOS using a micromagnet and spin-valley coupling}},\ }\href {https://doi.org/10.1038/s41534-023-00776-8} {\bibfield  {journal} {\bibinfo  {journal} {npj Quantum Inf.}\ }\textbf {\bibinfo {volume} {9}},\ \bibinfo {pages} {107} (\bibinfo {year} {2023})}\BibitemShut {NoStop}%
\bibitem [{\citenamefont {Huang}\ \emph {et~al.}(2024)\citenamefont {Huang}, \citenamefont {Su}, \citenamefont {Lim}, \citenamefont {Feng}, \citenamefont {van Straaten}, \citenamefont {Severin}, \citenamefont {Gilbert}, \citenamefont {Dumoulin~Stuyck}, \citenamefont {Tanttu}, \citenamefont {Serrano}, \citenamefont {Cifuentes}, \citenamefont {Hansen}, \citenamefont {Seedhouse}, \citenamefont {Vahapoglu}, \citenamefont {Leon}, \citenamefont {Abrosimov}, \citenamefont {Pohl}, \citenamefont {Thewalt}, \citenamefont {Hudson}, \citenamefont {Escott}, \citenamefont {Ares}, \citenamefont {Bartlett}, \citenamefont {Morello}, \citenamefont {Saraiva}, \citenamefont {Laucht}, \citenamefont {Dzurak},\ and\ \citenamefont {Yang}}]{Huang2024Mar}%
  \BibitemOpen
  \bibfield  {author} {\bibinfo {author} {\bibfnamefont {J.~Y.}\ \bibnamefont {Huang}}, \bibinfo {author} {\bibfnamefont {R.~Y.}\ \bibnamefont {Su}}, \bibinfo {author} {\bibfnamefont {W.~H.}\ \bibnamefont {Lim}}, \bibinfo {author} {\bibfnamefont {M.}~\bibnamefont {Feng}}, \bibinfo {author} {\bibfnamefont {B.}~\bibnamefont {van Straaten}}, \bibinfo {author} {\bibfnamefont {B.}~\bibnamefont {Severin}}, \bibinfo {author} {\bibfnamefont {W.}~\bibnamefont {Gilbert}}, \bibinfo {author} {\bibfnamefont {N.}~\bibnamefont {Dumoulin~Stuyck}}, \bibinfo {author} {\bibfnamefont {T.}~\bibnamefont {Tanttu}}, \bibinfo {author} {\bibfnamefont {S.}~\bibnamefont {Serrano}}, \bibinfo {author} {\bibfnamefont {J.~D.}\ \bibnamefont {Cifuentes}}, \bibinfo {author} {\bibfnamefont {I.}~\bibnamefont {Hansen}}, \bibinfo {author} {\bibfnamefont {A.~E.}\ \bibnamefont {Seedhouse}}, \bibinfo {author} {\bibfnamefont {E.}~\bibnamefont {Vahapoglu}}, \bibinfo {author} {\bibfnamefont {R.~C.~C.}\ \bibnamefont {Leon}}, \bibinfo {author}
  {\bibfnamefont {N.~V.}\ \bibnamefont {Abrosimov}}, \bibinfo {author} {\bibfnamefont {H.-J.}\ \bibnamefont {Pohl}}, \bibinfo {author} {\bibfnamefont {M.~L.~W.}\ \bibnamefont {Thewalt}}, \bibinfo {author} {\bibfnamefont {F.~E.}\ \bibnamefont {Hudson}}, \bibinfo {author} {\bibfnamefont {C.~C.}\ \bibnamefont {Escott}}, \bibinfo {author} {\bibfnamefont {N.}~\bibnamefont {Ares}}, \bibinfo {author} {\bibfnamefont {S.~D.}\ \bibnamefont {Bartlett}}, \bibinfo {author} {\bibfnamefont {A.}~\bibnamefont {Morello}}, \bibinfo {author} {\bibfnamefont {A.}~\bibnamefont {Saraiva}}, \bibinfo {author} {\bibfnamefont {A.}~\bibnamefont {Laucht}}, \bibinfo {author} {\bibfnamefont {A.~S.}\ \bibnamefont {Dzurak}},\ and\ \bibinfo {author} {\bibfnamefont {C.~H.}\ \bibnamefont {Yang}},\ }\bibfield  {title} {\bibinfo {title} {{High-fidelity spin qubit operation and algorithmic initialization above 1 K}},\ }\href {https://doi.org/10.1038/s41586-024-07160-2} {\bibfield  {journal} {\bibinfo  {journal} {Nature}\ }\textbf {\bibinfo {volume}
  {627}},\ \bibinfo {pages} {772} (\bibinfo {year} {2024})}\BibitemShut {NoStop}%
\bibitem [{\citenamefont {Cai}\ \emph {et~al.}(2023)\citenamefont {Cai}, \citenamefont {Connors}, \citenamefont {Edge},\ and\ \citenamefont {Nichol}}]{Cai2023Mar}%
  \BibitemOpen
  \bibfield  {author} {\bibinfo {author} {\bibfnamefont {X.}~\bibnamefont {Cai}}, \bibinfo {author} {\bibfnamefont {E.~J.}\ \bibnamefont {Connors}}, \bibinfo {author} {\bibfnamefont {L.~F.}\ \bibnamefont {Edge}},\ and\ \bibinfo {author} {\bibfnamefont {J.~M.}\ \bibnamefont {Nichol}},\ }\bibfield  {title} {\bibinfo {title} {{Coherent spin{\textendash}valley oscillations in silicon}},\ }\href {https://doi.org/10.1038/s41567-022-01870-y} {\bibfield  {journal} {\bibinfo  {journal} {Nat. Phys.}\ }\textbf {\bibinfo {volume} {19}},\ \bibinfo {pages} {386} (\bibinfo {year} {2023})}\BibitemShut {NoStop}%
\bibitem [{\citenamefont {Maurand}\ \emph {et~al.}(2016)\citenamefont {Maurand}, \citenamefont {Jehl}, \citenamefont {Kotekar-Patil}, \citenamefont {Corna}, \citenamefont {Bohuslavskyi}, \citenamefont {Lavi{\ifmmode\acute{e}\else\'{e}\fi}ville}, \citenamefont {Hutin}, \citenamefont {Barraud}, \citenamefont {Vinet}, \citenamefont {Sanquer},\ and\ \citenamefont {De~Franceschi}}]{Maurand2016Nov}%
  \BibitemOpen
  \bibfield  {author} {\bibinfo {author} {\bibfnamefont {R.}~\bibnamefont {Maurand}}, \bibinfo {author} {\bibfnamefont {X.}~\bibnamefont {Jehl}}, \bibinfo {author} {\bibfnamefont {D.}~\bibnamefont {Kotekar-Patil}}, \bibinfo {author} {\bibfnamefont {A.}~\bibnamefont {Corna}}, \bibinfo {author} {\bibfnamefont {H.}~\bibnamefont {Bohuslavskyi}}, \bibinfo {author} {\bibfnamefont {R.}~\bibnamefont {Lavi{\ifmmode\acute{e}\else\'{e}\fi}ville}}, \bibinfo {author} {\bibfnamefont {L.}~\bibnamefont {Hutin}}, \bibinfo {author} {\bibfnamefont {S.}~\bibnamefont {Barraud}}, \bibinfo {author} {\bibfnamefont {M.}~\bibnamefont {Vinet}}, \bibinfo {author} {\bibfnamefont {M.}~\bibnamefont {Sanquer}},\ and\ \bibinfo {author} {\bibfnamefont {S.}~\bibnamefont {De~Franceschi}},\ }\bibfield  {title} {\bibinfo {title} {{A CMOS silicon spin qubit}},\ }\href {https://doi.org/10.1038/ncomms13575} {\bibfield  {journal} {\bibinfo  {journal} {Nat. Commun.}\ }\textbf {\bibinfo {volume} {7}},\ \bibinfo {pages} {13575} (\bibinfo {year}
  {2016})}\BibitemShut {NoStop}%
\bibitem [{\citenamefont {Kim}\ \emph {et~al.}(2015)\citenamefont {Kim}, \citenamefont {Ward}, \citenamefont {Simmons}, \citenamefont {Savage}, \citenamefont {Lagally}, \citenamefont {Friesen}, \citenamefont {Coppersmith},\ and\ \citenamefont {Eriksson}}]{Kim2015Oct}%
  \BibitemOpen
  \bibfield  {author} {\bibinfo {author} {\bibfnamefont {D.}~\bibnamefont {Kim}}, \bibinfo {author} {\bibfnamefont {D.~R.}\ \bibnamefont {Ward}}, \bibinfo {author} {\bibfnamefont {C.~B.}\ \bibnamefont {Simmons}}, \bibinfo {author} {\bibfnamefont {D.~E.}\ \bibnamefont {Savage}}, \bibinfo {author} {\bibfnamefont {M.~G.}\ \bibnamefont {Lagally}}, \bibinfo {author} {\bibfnamefont {M.}~\bibnamefont {Friesen}}, \bibinfo {author} {\bibfnamefont {S.~N.}\ \bibnamefont {Coppersmith}},\ and\ \bibinfo {author} {\bibfnamefont {M.~A.}\ \bibnamefont {Eriksson}},\ }\bibfield  {title} {\bibinfo {title} {{High-fidelity resonant gating of a silicon-based quantum dot hybrid qubit}},\ }\href {https://doi.org/10.1038/npjqi.2015.4} {\bibfield  {journal} {\bibinfo  {journal} {npj Quantum Inf.}\ }\textbf {\bibinfo {volume} {1}},\ \bibinfo {pages} {15004} (\bibinfo {year} {2015})}\BibitemShut {NoStop}%
\bibitem [{\citenamefont {Shi}\ \emph {et~al.}(2012)\citenamefont {Shi}, \citenamefont {Simmons}, \citenamefont {Prance}, \citenamefont {Gamble}, \citenamefont {Koh}, \citenamefont {Shim}, \citenamefont {Hu}, \citenamefont {Savage}, \citenamefont {Lagally}, \citenamefont {Eriksson}, \citenamefont {Friesen},\ and\ \citenamefont {Coppersmith}}]{Shi2012Apr}%
  \BibitemOpen
  \bibfield  {author} {\bibinfo {author} {\bibfnamefont {Z.}~\bibnamefont {Shi}}, \bibinfo {author} {\bibfnamefont {C.~B.}\ \bibnamefont {Simmons}}, \bibinfo {author} {\bibfnamefont {J.~R.}\ \bibnamefont {Prance}}, \bibinfo {author} {\bibfnamefont {J.~K.}\ \bibnamefont {Gamble}}, \bibinfo {author} {\bibfnamefont {T.~S.}\ \bibnamefont {Koh}}, \bibinfo {author} {\bibfnamefont {Y.-P.}\ \bibnamefont {Shim}}, \bibinfo {author} {\bibfnamefont {X.}~\bibnamefont {Hu}}, \bibinfo {author} {\bibfnamefont {D.~E.}\ \bibnamefont {Savage}}, \bibinfo {author} {\bibfnamefont {M.~G.}\ \bibnamefont {Lagally}}, \bibinfo {author} {\bibfnamefont {M.~A.}\ \bibnamefont {Eriksson}}, \bibinfo {author} {\bibfnamefont {M.}~\bibnamefont {Friesen}},\ and\ \bibinfo {author} {\bibfnamefont {S.~N.}\ \bibnamefont {Coppersmith}},\ }\bibfield  {title} {\bibinfo {title} {{Fast Hybrid Silicon Double-Quantum-Dot Qubit}},\ }\href {https://doi.org/10.1103/PhysRevLett.108.140503} {\bibfield  {journal} {\bibinfo  {journal} {Phys. Rev. Lett.}\ }\textbf
  {\bibinfo {volume} {108}},\ \bibinfo {pages} {140503} (\bibinfo {year} {2012})}\BibitemShut {NoStop}%
\bibitem [{\citenamefont {Jirovec}\ \emph {et~al.}(2021)\citenamefont {Jirovec}, \citenamefont {Hofmann}, \citenamefont {Ballabio}, \citenamefont {Mutter}, \citenamefont {Tavani}, \citenamefont {Botifoll}, \citenamefont {Crippa}, \citenamefont {Kukucka}, \citenamefont {Sagi}, \citenamefont {Martins}, \citenamefont {Saez-Mollejo}, \citenamefont {Prieto}, \citenamefont {Borovkov}, \citenamefont {Arbiol}, \citenamefont {Chrastina}, \citenamefont {Isella},\ and\ \citenamefont {Katsaros}}]{Jirovec2021Aug}%
  \BibitemOpen
  \bibfield  {author} {\bibinfo {author} {\bibfnamefont {D.}~\bibnamefont {Jirovec}}, \bibinfo {author} {\bibfnamefont {A.}~\bibnamefont {Hofmann}}, \bibinfo {author} {\bibfnamefont {A.}~\bibnamefont {Ballabio}}, \bibinfo {author} {\bibfnamefont {P.~M.}\ \bibnamefont {Mutter}}, \bibinfo {author} {\bibfnamefont {G.}~\bibnamefont {Tavani}}, \bibinfo {author} {\bibfnamefont {M.}~\bibnamefont {Botifoll}}, \bibinfo {author} {\bibfnamefont {A.}~\bibnamefont {Crippa}}, \bibinfo {author} {\bibfnamefont {J.}~\bibnamefont {Kukucka}}, \bibinfo {author} {\bibfnamefont {O.}~\bibnamefont {Sagi}}, \bibinfo {author} {\bibfnamefont {F.}~\bibnamefont {Martins}}, \bibinfo {author} {\bibfnamefont {J.}~\bibnamefont {Saez-Mollejo}}, \bibinfo {author} {\bibfnamefont {I.}~\bibnamefont {Prieto}}, \bibinfo {author} {\bibfnamefont {M.}~\bibnamefont {Borovkov}}, \bibinfo {author} {\bibfnamefont {J.}~\bibnamefont {Arbiol}}, \bibinfo {author} {\bibfnamefont {D.}~\bibnamefont {Chrastina}}, \bibinfo {author} {\bibfnamefont {G.}~\bibnamefont
  {Isella}},\ and\ \bibinfo {author} {\bibfnamefont {G.}~\bibnamefont {Katsaros}},\ }\bibfield  {title} {\bibinfo {title} {{A singlet-triplet hole spin qubit in planar Ge}},\ }\href {https://doi.org/10.1038/s41563-021-01022-2} {\bibfield  {journal} {\bibinfo  {journal} {Nat. Mater.}\ }\textbf {\bibinfo {volume} {20}},\ \bibinfo {pages} {1106} (\bibinfo {year} {2021})}\BibitemShut {NoStop}%
\bibitem [{\citenamefont {Watzinger}\ \emph {et~al.}(2018)\citenamefont {Watzinger}, \citenamefont {Kuku{\ifmmode\check{c}\else\v{c}\fi}ka}, \citenamefont {Vuku{\ifmmode\check{s}\else\v{s}\fi}i{\ifmmode\acute{c}\else\'{c}\fi}}, \citenamefont {Gao}, \citenamefont {Wang}, \citenamefont {Sch{\ifmmode\ddot{a}\else\"{a}\fi}ffler}, \citenamefont {Zhang},\ and\ \citenamefont {Katsaros}}]{Watzinger2018Sep}%
  \BibitemOpen
  \bibfield  {author} {\bibinfo {author} {\bibfnamefont {H.}~\bibnamefont {Watzinger}}, \bibinfo {author} {\bibfnamefont {J.}~\bibnamefont {Kuku{\ifmmode\check{c}\else\v{c}\fi}ka}}, \bibinfo {author} {\bibfnamefont {L.}~\bibnamefont {Vuku{\ifmmode\check{s}\else\v{s}\fi}i{\ifmmode\acute{c}\else\'{c}\fi}}}, \bibinfo {author} {\bibfnamefont {F.}~\bibnamefont {Gao}}, \bibinfo {author} {\bibfnamefont {T.}~\bibnamefont {Wang}}, \bibinfo {author} {\bibfnamefont {F.}~\bibnamefont {Sch{\ifmmode\ddot{a}\else\"{a}\fi}ffler}}, \bibinfo {author} {\bibfnamefont {J.-J.}\ \bibnamefont {Zhang}},\ and\ \bibinfo {author} {\bibfnamefont {G.}~\bibnamefont {Katsaros}},\ }\bibfield  {title} {\bibinfo {title} {{A germanium hole spin qubit}},\ }\href {https://doi.org/10.1038/s41467-018-06418-4} {\bibfield  {journal} {\bibinfo  {journal} {Nat. Commun.}\ }\textbf {\bibinfo {volume} {9}},\ \bibinfo {pages} {3902} (\bibinfo {year} {2018})}\BibitemShut {NoStop}%
\bibitem [{\citenamefont {Wang}\ \emph {et~al.}(2022)\citenamefont {Wang}, \citenamefont {Xu}, \citenamefont {Gao}, \citenamefont {Liu}, \citenamefont {Ma}, \citenamefont {Zhang}, \citenamefont {Wang}, \citenamefont {Cao}, \citenamefont {Wang}, \citenamefont {Zhang}, \citenamefont {Culcer}, \citenamefont {Hu}, \citenamefont {Jiang}, \citenamefont {Li}, \citenamefont {Guo},\ and\ \citenamefont {Guo}}]{Wang2022Jan}%
  \BibitemOpen
  \bibfield  {author} {\bibinfo {author} {\bibfnamefont {K.}~\bibnamefont {Wang}}, \bibinfo {author} {\bibfnamefont {G.}~\bibnamefont {Xu}}, \bibinfo {author} {\bibfnamefont {F.}~\bibnamefont {Gao}}, \bibinfo {author} {\bibfnamefont {H.}~\bibnamefont {Liu}}, \bibinfo {author} {\bibfnamefont {R.-L.}\ \bibnamefont {Ma}}, \bibinfo {author} {\bibfnamefont {X.}~\bibnamefont {Zhang}}, \bibinfo {author} {\bibfnamefont {Z.}~\bibnamefont {Wang}}, \bibinfo {author} {\bibfnamefont {G.}~\bibnamefont {Cao}}, \bibinfo {author} {\bibfnamefont {T.}~\bibnamefont {Wang}}, \bibinfo {author} {\bibfnamefont {J.-J.}\ \bibnamefont {Zhang}}, \bibinfo {author} {\bibfnamefont {D.}~\bibnamefont {Culcer}}, \bibinfo {author} {\bibfnamefont {X.}~\bibnamefont {Hu}}, \bibinfo {author} {\bibfnamefont {H.-W.}\ \bibnamefont {Jiang}}, \bibinfo {author} {\bibfnamefont {H.-O.}\ \bibnamefont {Li}}, \bibinfo {author} {\bibfnamefont {G.-C.}\ \bibnamefont {Guo}},\ and\ \bibinfo {author} {\bibfnamefont {G.-P.}\ \bibnamefont {Guo}},\ }\bibfield
  {title} {\bibinfo {title} {{Ultrafast coherent control of a hole spin qubit in a germanium quantum dot}},\ }\href {https://doi.org/10.1038/s41467-021-27880-7} {\bibfield  {journal} {\bibinfo  {journal} {Nat. Commun.}\ }\textbf {\bibinfo {volume} {13}},\ \bibinfo {pages} {206} (\bibinfo {year} {2022})}\BibitemShut {NoStop}%
\bibitem [{\citenamefont {Hendrickx}\ \emph {et~al.}(2020)\citenamefont {Hendrickx}, \citenamefont {Franke}, \citenamefont {Sammak}, \citenamefont {Scappucci},\ and\ \citenamefont {Veldhorst}}]{Hendrickx2020Jan}%
  \BibitemOpen
  \bibfield  {author} {\bibinfo {author} {\bibfnamefont {N.~W.}\ \bibnamefont {Hendrickx}}, \bibinfo {author} {\bibfnamefont {D.~P.}\ \bibnamefont {Franke}}, \bibinfo {author} {\bibfnamefont {A.}~\bibnamefont {Sammak}}, \bibinfo {author} {\bibfnamefont {G.}~\bibnamefont {Scappucci}},\ and\ \bibinfo {author} {\bibfnamefont {M.}~\bibnamefont {Veldhorst}},\ }\bibfield  {title} {\bibinfo {title} {{Fast two-qubit logic with holes in germanium}},\ }\href {https://doi.org/10.1038/s41586-019-1919-3} {\bibfield  {journal} {\bibinfo  {journal} {Nature}\ }\textbf {\bibinfo {volume} {577}},\ \bibinfo {pages} {487} (\bibinfo {year} {2020})}\BibitemShut {NoStop}%
\bibitem [{\citenamefont {Laird}\ \emph {et~al.}(2013)\citenamefont {Laird}, \citenamefont {Pei},\ and\ \citenamefont {Kouwenhoven}}]{Laird2013Jul}%
  \BibitemOpen
  \bibfield  {author} {\bibinfo {author} {\bibfnamefont {E.~A.}\ \bibnamefont {Laird}}, \bibinfo {author} {\bibfnamefont {F.}~\bibnamefont {Pei}},\ and\ \bibinfo {author} {\bibfnamefont {L.~P.}\ \bibnamefont {Kouwenhoven}},\ }\bibfield  {title} {\bibinfo {title} {{A valley{\textendash}spin qubit in a carbon nanotube}},\ }\href {https://doi.org/10.1038/nnano.2013.140} {\bibfield  {journal} {\bibinfo  {journal} {Nat. Nanotechnol.}\ }\textbf {\bibinfo {volume} {8}},\ \bibinfo {pages} {565} (\bibinfo {year} {2013})}\BibitemShut {NoStop}%
\bibitem [{\citenamefont {{\ifmmode\dot{Z}\else\.{Z}\fi}ebrowski}\ \emph {et~al.}(2017)\citenamefont {{\ifmmode\dot{Z}\else\.{Z}\fi}ebrowski}, \citenamefont {Peeters},\ and\ \citenamefont {Szafran}}]{Zebrowski2017Jul}%
  \BibitemOpen
  \bibfield  {author} {\bibinfo {author} {\bibfnamefont {D.~P.}\ \bibnamefont {{\ifmmode\dot{Z}\else\.{Z}\fi}ebrowski}}, \bibinfo {author} {\bibfnamefont {F.~M.}\ \bibnamefont {Peeters}},\ and\ \bibinfo {author} {\bibfnamefont {B.}~\bibnamefont {Szafran}},\ }\bibfield  {title} {\bibinfo {title} {{Double quantum dots defined in bilayer graphene}},\ }\href {https://doi.org/10.1103/PhysRevB.96.035434} {\bibfield  {journal} {\bibinfo  {journal} {Phys. Rev. B}\ }\textbf {\bibinfo {volume} {96}},\ \bibinfo {pages} {035434} (\bibinfo {year} {2017})}\BibitemShut {NoStop}%
\bibitem [{\citenamefont {Eich}\ \emph {et~al.}(2018{\natexlab{a}})\citenamefont {Eich}, \citenamefont {Pisoni}, \citenamefont {Pally}, \citenamefont {Overweg}, \citenamefont {Kurzmann}, \citenamefont {Lee}, \citenamefont {Rickhaus}, \citenamefont {Watanabe}, \citenamefont {Taniguchi}, \citenamefont {Ensslin},\ and\ \citenamefont {Ihn}}]{Eich2018Aug}%
  \BibitemOpen
  \bibfield  {author} {\bibinfo {author} {\bibfnamefont {M.}~\bibnamefont {Eich}}, \bibinfo {author} {\bibfnamefont {R.}~\bibnamefont {Pisoni}}, \bibinfo {author} {\bibfnamefont {A.}~\bibnamefont {Pally}}, \bibinfo {author} {\bibfnamefont {H.}~\bibnamefont {Overweg}}, \bibinfo {author} {\bibfnamefont {A.}~\bibnamefont {Kurzmann}}, \bibinfo {author} {\bibfnamefont {Y.}~\bibnamefont {Lee}}, \bibinfo {author} {\bibfnamefont {P.}~\bibnamefont {Rickhaus}}, \bibinfo {author} {\bibfnamefont {K.}~\bibnamefont {Watanabe}}, \bibinfo {author} {\bibfnamefont {T.}~\bibnamefont {Taniguchi}}, \bibinfo {author} {\bibfnamefont {K.}~\bibnamefont {Ensslin}},\ and\ \bibinfo {author} {\bibfnamefont {T.}~\bibnamefont {Ihn}},\ }\bibfield  {title} {\bibinfo {title} {{Coupled Quantum Dots in Bilayer Graphene}},\ }\href {https://doi.org/10.1021/acs.nanolett.8b01859} {\bibfield  {journal} {\bibinfo  {journal} {Nano Lett.}\ }\textbf {\bibinfo {volume} {18}},\ \bibinfo {pages} {5042} (\bibinfo {year} {2018}{\natexlab{a}})}\BibitemShut
  {NoStop}%
\bibitem [{\citenamefont {Banszerus}\ \emph {et~al.}(2018)\citenamefont {Banszerus}, \citenamefont {Frohn}, \citenamefont {Epping}, \citenamefont {Neumaier}, \citenamefont {Watanabe}, \citenamefont {Taniguchi},\ and\ \citenamefont {Stampfer}}]{Banszerus2018AugFirstQD}%
  \BibitemOpen
  \bibfield  {author} {\bibinfo {author} {\bibfnamefont {L.}~\bibnamefont {Banszerus}}, \bibinfo {author} {\bibfnamefont {B.}~\bibnamefont {Frohn}}, \bibinfo {author} {\bibfnamefont {A.}~\bibnamefont {Epping}}, \bibinfo {author} {\bibfnamefont {D.}~\bibnamefont {Neumaier}}, \bibinfo {author} {\bibfnamefont {K.}~\bibnamefont {Watanabe}}, \bibinfo {author} {\bibfnamefont {T.}~\bibnamefont {Taniguchi}},\ and\ \bibinfo {author} {\bibfnamefont {C.}~\bibnamefont {Stampfer}},\ }\bibfield  {title} {\bibinfo {title} {{Gate-Defined Electron{\textendash}Hole Double Dots in Bilayer Graphene}},\ }\href {https://doi.org/10.1021/acs.nanolett.8b01303} {\bibfield  {journal} {\bibinfo  {journal} {Nano Lett.}\ }\textbf {\bibinfo {volume} {18}},\ \bibinfo {pages} {4785} (\bibinfo {year} {2018})}\BibitemShut {NoStop}%
\bibitem [{\citenamefont {Tong}\ \emph {et~al.}(2021)\citenamefont {Tong}, \citenamefont {Garreis}, \citenamefont {Knothe}, \citenamefont {Eich}, \citenamefont {Sacchi}, \citenamefont {Watanabe}, \citenamefont {Taniguchi}, \citenamefont {Fal{'}ko}, \citenamefont {Ihn}, \citenamefont {Ensslin},\ and\ \citenamefont {Kurzmann}}]{Tong2021JanTunableValleySplitting}%
  \BibitemOpen
  \bibfield  {author} {\bibinfo {author} {\bibfnamefont {C.}~\bibnamefont {Tong}}, \bibinfo {author} {\bibfnamefont {R.}~\bibnamefont {Garreis}}, \bibinfo {author} {\bibfnamefont {A.}~\bibnamefont {Knothe}}, \bibinfo {author} {\bibfnamefont {M.}~\bibnamefont {Eich}}, \bibinfo {author} {\bibfnamefont {A.}~\bibnamefont {Sacchi}}, \bibinfo {author} {\bibfnamefont {K.}~\bibnamefont {Watanabe}}, \bibinfo {author} {\bibfnamefont {T.}~\bibnamefont {Taniguchi}}, \bibinfo {author} {\bibfnamefont {V.}~\bibnamefont {Fal{'}ko}}, \bibinfo {author} {\bibfnamefont {T.}~\bibnamefont {Ihn}}, \bibinfo {author} {\bibfnamefont {K.}~\bibnamefont {Ensslin}},\ and\ \bibinfo {author} {\bibfnamefont {A.}~\bibnamefont {Kurzmann}},\ }\bibfield  {title} {\bibinfo {title} {{Tunable Valley Splitting and Bipolar Operation in Graphene Quantum Dots}},\ }\href {https://doi.org/10.1021/acs.nanolett.0c04343} {\bibfield  {journal} {\bibinfo  {journal} {Nano Lett.}\ }\textbf {\bibinfo {volume} {21}},\ \bibinfo {pages} {1068} (\bibinfo {year}
  {2021})}\BibitemShut {NoStop}%
\bibitem [{\citenamefont {Banszerus}\ \emph {et~al.}(2023)\citenamefont {Banszerus}, \citenamefont {M{\ifmmode\ddot{o}\else\"{o}\fi}ller}, \citenamefont {Hecker}, \citenamefont {Icking}, \citenamefont {Watanabe}, \citenamefont {Taniguchi}, \citenamefont {Hassler}, \citenamefont {Volk},\ and\ \citenamefont {Stampfer}}]{Banszerus2023MayNature}%
  \BibitemOpen
  \bibfield  {author} {\bibinfo {author} {\bibfnamefont {L.}~\bibnamefont {Banszerus}}, \bibinfo {author} {\bibfnamefont {S.}~\bibnamefont {M{\ifmmode\ddot{o}\else\"{o}\fi}ller}}, \bibinfo {author} {\bibfnamefont {K.}~\bibnamefont {Hecker}}, \bibinfo {author} {\bibfnamefont {E.}~\bibnamefont {Icking}}, \bibinfo {author} {\bibfnamefont {K.}~\bibnamefont {Watanabe}}, \bibinfo {author} {\bibfnamefont {T.}~\bibnamefont {Taniguchi}}, \bibinfo {author} {\bibfnamefont {F.}~\bibnamefont {Hassler}}, \bibinfo {author} {\bibfnamefont {C.}~\bibnamefont {Volk}},\ and\ \bibinfo {author} {\bibfnamefont {C.}~\bibnamefont {Stampfer}},\ }\bibfield  {title} {\bibinfo {title} {{Particle{\textendash}hole symmetry protects spin-valley blockade in graphene quantum dots}},\ }\href {https://doi.org/10.1038/s41586-023-05953-5} {\bibfield  {journal} {\bibinfo  {journal} {Nature}\ }\textbf {\bibinfo {volume} {618}},\ \bibinfo {pages} {51} (\bibinfo {year} {2023})}\BibitemShut {NoStop}%
\bibitem [{\citenamefont {Banszerus}\ \emph {et~al.}(2021{\natexlab{a}})\citenamefont {Banszerus}, \citenamefont {M{\ifmmode\ddot{o}\else\"{o}\fi}ller}, \citenamefont {Steiner}, \citenamefont {Icking}, \citenamefont {Trellenkamp}, \citenamefont {Lentz}, \citenamefont {Watanabe}, \citenamefont {Taniguchi}, \citenamefont {Volk},\ and\ \citenamefont {Stampfer}}]{Banszerus2021SepSpinOrbit}%
  \BibitemOpen
  \bibfield  {author} {\bibinfo {author} {\bibfnamefont {L.}~\bibnamefont {Banszerus}}, \bibinfo {author} {\bibfnamefont {S.}~\bibnamefont {M{\ifmmode\ddot{o}\else\"{o}\fi}ller}}, \bibinfo {author} {\bibfnamefont {C.}~\bibnamefont {Steiner}}, \bibinfo {author} {\bibfnamefont {E.}~\bibnamefont {Icking}}, \bibinfo {author} {\bibfnamefont {S.}~\bibnamefont {Trellenkamp}}, \bibinfo {author} {\bibfnamefont {F.}~\bibnamefont {Lentz}}, \bibinfo {author} {\bibfnamefont {K.}~\bibnamefont {Watanabe}}, \bibinfo {author} {\bibfnamefont {T.}~\bibnamefont {Taniguchi}}, \bibinfo {author} {\bibfnamefont {C.}~\bibnamefont {Volk}},\ and\ \bibinfo {author} {\bibfnamefont {C.}~\bibnamefont {Stampfer}},\ }\bibfield  {title} {\bibinfo {title} {{Spin-valley coupling in single-electron bilayer graphene quantum dots}},\ }\href {https://doi.org/10.1038/s41467-021-25498-3} {\bibfield  {journal} {\bibinfo  {journal} {Nat. Commun.}\ }\textbf {\bibinfo {volume} {12}},\ \bibinfo {pages} {5250} (\bibinfo {year}
  {2021}{\natexlab{a}})}\BibitemShut {NoStop}%
\bibitem [{\citenamefont {Banszerus}\ \emph {et~al.}(2020{\natexlab{a}})\citenamefont {Banszerus}, \citenamefont {M{\ifmmode\ddot{o}\else\"{o}\fi}ller}, \citenamefont {Icking}, \citenamefont {Watanabe}, \citenamefont {Taniguchi}, \citenamefont {Volk},\ and\ \citenamefont {Stampfer}}]{Banszerus2020MarSingleElectronDQD}%
  \BibitemOpen
  \bibfield  {author} {\bibinfo {author} {\bibfnamefont {L.}~\bibnamefont {Banszerus}}, \bibinfo {author} {\bibfnamefont {S.}~\bibnamefont {M{\ifmmode\ddot{o}\else\"{o}\fi}ller}}, \bibinfo {author} {\bibfnamefont {E.}~\bibnamefont {Icking}}, \bibinfo {author} {\bibfnamefont {K.}~\bibnamefont {Watanabe}}, \bibinfo {author} {\bibfnamefont {T.}~\bibnamefont {Taniguchi}}, \bibinfo {author} {\bibfnamefont {C.}~\bibnamefont {Volk}},\ and\ \bibinfo {author} {\bibfnamefont {C.}~\bibnamefont {Stampfer}},\ }\bibfield  {title} {\bibinfo {title} {{Single-Electron Double Quantum Dots in Bilayer Graphene}},\ }\href {https://doi.org/10.1021/acs.nanolett.9b05295} {\bibfield  {journal} {\bibinfo  {journal} {Nano Lett.}\ }\textbf {\bibinfo {volume} {20}},\ \bibinfo {pages} {2005} (\bibinfo {year} {2020}{\natexlab{a}})}\BibitemShut {NoStop}%
\bibitem [{\citenamefont {Banszerus}\ \emph {et~al.}(2021{\natexlab{b}})\citenamefont {Banszerus}, \citenamefont {Rothstein}, \citenamefont {Icking}, \citenamefont {M{\ifmmode\ddot{o}\else\"{o}\fi}ller}, \citenamefont {Watanabe}, \citenamefont {Taniguchi}, \citenamefont {Stampfer},\ and\ \citenamefont {Volk}}]{Banszerus2021MarTunableInterdot}%
  \BibitemOpen
  \bibfield  {author} {\bibinfo {author} {\bibfnamefont {L.}~\bibnamefont {Banszerus}}, \bibinfo {author} {\bibfnamefont {A.}~\bibnamefont {Rothstein}}, \bibinfo {author} {\bibfnamefont {E.}~\bibnamefont {Icking}}, \bibinfo {author} {\bibfnamefont {S.}~\bibnamefont {M{\ifmmode\ddot{o}\else\"{o}\fi}ller}}, \bibinfo {author} {\bibfnamefont {K.}~\bibnamefont {Watanabe}}, \bibinfo {author} {\bibfnamefont {T.}~\bibnamefont {Taniguchi}}, \bibinfo {author} {\bibfnamefont {C.}~\bibnamefont {Stampfer}},\ and\ \bibinfo {author} {\bibfnamefont {C.}~\bibnamefont {Volk}},\ }\bibfield  {title} {\bibinfo {title} {{Tunable interdot coupling in few-electron bilayer graphene double quantum dots}},\ }\href {https://doi.org/10.1063/5.0035300} {\bibfield  {journal} {\bibinfo  {journal} {Appl. Phys. Lett.}\ }\textbf {\bibinfo {volume} {118}},\ \bibinfo {pages} {103101} (\bibinfo {year} {2021}{\natexlab{b}})}\BibitemShut {NoStop}%
\bibitem [{\citenamefont {Tong}\ \emph {et~al.}(2022)\citenamefont {Tong}, \citenamefont {Kurzmann}, \citenamefont {Garreis}, \citenamefont {Huang}, \citenamefont {Jele}, \citenamefont {Eich}, \citenamefont {Ginzburg}, \citenamefont {Mittag}, \citenamefont {Watanabe}, \citenamefont {Taniguchi}, \citenamefont {Ensslin},\ and\ \citenamefont {Ihn}}]{Tong2022FebBlockade}%
  \BibitemOpen
  \bibfield  {author} {\bibinfo {author} {\bibfnamefont {C.}~\bibnamefont {Tong}}, \bibinfo {author} {\bibfnamefont {A.}~\bibnamefont {Kurzmann}}, \bibinfo {author} {\bibfnamefont {R.}~\bibnamefont {Garreis}}, \bibinfo {author} {\bibfnamefont {W.~W.}\ \bibnamefont {Huang}}, \bibinfo {author} {\bibfnamefont {S.}~\bibnamefont {Jele}}, \bibinfo {author} {\bibfnamefont {M.}~\bibnamefont {Eich}}, \bibinfo {author} {\bibfnamefont {L.}~\bibnamefont {Ginzburg}}, \bibinfo {author} {\bibfnamefont {C.}~\bibnamefont {Mittag}}, \bibinfo {author} {\bibfnamefont {K.}~\bibnamefont {Watanabe}}, \bibinfo {author} {\bibfnamefont {T.}~\bibnamefont {Taniguchi}}, \bibinfo {author} {\bibfnamefont {K.}~\bibnamefont {Ensslin}},\ and\ \bibinfo {author} {\bibfnamefont {T.}~\bibnamefont {Ihn}},\ }\bibfield  {title} {\bibinfo {title} {{Pauli Blockade of Tunable Two-Electron Spin and Valley States in Graphene Quantum Dots}},\ }\href {https://doi.org/10.1103/PhysRevLett.128.067702} {\bibfield  {journal} {\bibinfo  {journal} {Phys. Rev.
  Lett.}\ }\textbf {\bibinfo {volume} {128}},\ \bibinfo {pages} {067702} (\bibinfo {year} {2022})}\BibitemShut {NoStop}%
\bibitem [{\citenamefont {Hecker}\ \emph {et~al.}(2023)\citenamefont {Hecker}, \citenamefont {Banszerus}, \citenamefont {Sch{\ifmmode\ddot{a}\else\"{a}\fi}pers}, \citenamefont {M{\ifmmode\ddot{o}\else\"{o}\fi}ller}, \citenamefont {Peters}, \citenamefont {Icking}, \citenamefont {Watanabe}, \citenamefont {Taniguchi}, \citenamefont {Volk},\ and\ \citenamefont {Stampfer}}]{Hecker2023Nov}%
  \BibitemOpen
  \bibfield  {author} {\bibinfo {author} {\bibfnamefont {K.}~\bibnamefont {Hecker}}, \bibinfo {author} {\bibfnamefont {L.}~\bibnamefont {Banszerus}}, \bibinfo {author} {\bibfnamefont {A.}~\bibnamefont {Sch{\ifmmode\ddot{a}\else\"{a}\fi}pers}}, \bibinfo {author} {\bibfnamefont {S.}~\bibnamefont {M{\ifmmode\ddot{o}\else\"{o}\fi}ller}}, \bibinfo {author} {\bibfnamefont {A.}~\bibnamefont {Peters}}, \bibinfo {author} {\bibfnamefont {E.}~\bibnamefont {Icking}}, \bibinfo {author} {\bibfnamefont {K.}~\bibnamefont {Watanabe}}, \bibinfo {author} {\bibfnamefont {T.}~\bibnamefont {Taniguchi}}, \bibinfo {author} {\bibfnamefont {C.}~\bibnamefont {Volk}},\ and\ \bibinfo {author} {\bibfnamefont {C.}~\bibnamefont {Stampfer}},\ }\bibfield  {title} {\bibinfo {title} {{Coherent charge oscillations in a bilayer graphene double quantum dot}},\ }\href {https://doi.org/10.1038/s41467-023-43541-3} {\bibfield  {journal} {\bibinfo  {journal} {Nat. Commun.}\ }\textbf {\bibinfo {volume} {14}},\ \bibinfo {pages} {7911} (\bibinfo {year}
  {2023})}\BibitemShut {NoStop}%
\bibitem [{\citenamefont {Tong}\ \emph {et~al.}(2024{\natexlab{a}})\citenamefont {Tong}, \citenamefont {Kurzmann}, \citenamefont {Garreis}, \citenamefont {Watanabe}, \citenamefont {Taniguchi}, \citenamefont {Ihn},\ and\ \citenamefont {Ensslin}}]{Tong2024JanCatalogue}%
  \BibitemOpen
  \bibfield  {author} {\bibinfo {author} {\bibfnamefont {C.}~\bibnamefont {Tong}}, \bibinfo {author} {\bibfnamefont {A.}~\bibnamefont {Kurzmann}}, \bibinfo {author} {\bibfnamefont {R.}~\bibnamefont {Garreis}}, \bibinfo {author} {\bibfnamefont {K.}~\bibnamefont {Watanabe}}, \bibinfo {author} {\bibfnamefont {T.}~\bibnamefont {Taniguchi}}, \bibinfo {author} {\bibfnamefont {T.}~\bibnamefont {Ihn}},\ and\ \bibinfo {author} {\bibfnamefont {K.}~\bibnamefont {Ensslin}},\ }\bibfield  {title} {\bibinfo {title} {{Pauli blockade catalogue and three- and four-particle Kondo effect in bilayer graphene quantum dots}},\ }\href {https://doi.org/10.1103/PhysRevResearch.6.L012006} {\bibfield  {journal} {\bibinfo  {journal} {Phys. Rev. Res.}\ }\textbf {\bibinfo {volume} {6}},\ \bibinfo {pages} {L012006} (\bibinfo {year} {2024}{\natexlab{a}})}\BibitemShut {NoStop}%
\bibitem [{\citenamefont {Garreis}\ \emph {et~al.}(2024)\citenamefont {Garreis}, \citenamefont {Tong}, \citenamefont {Terle}, \citenamefont {Ruckriegel}, \citenamefont {Gerber}, \citenamefont {G{\ifmmode\ddot{a}\else\"{a}\fi}chter}, \citenamefont {Watanabe}, \citenamefont {Taniguchi}, \citenamefont {Ihn}, \citenamefont {Ensslin},\ and\ \citenamefont {Huang}}]{Garreis2024MarLongLived}%
  \BibitemOpen
  \bibfield  {author} {\bibinfo {author} {\bibfnamefont {R.}~\bibnamefont {Garreis}}, \bibinfo {author} {\bibfnamefont {C.}~\bibnamefont {Tong}}, \bibinfo {author} {\bibfnamefont {J.}~\bibnamefont {Terle}}, \bibinfo {author} {\bibfnamefont {M.~J.}\ \bibnamefont {Ruckriegel}}, \bibinfo {author} {\bibfnamefont {J.~D.}\ \bibnamefont {Gerber}}, \bibinfo {author} {\bibfnamefont {L.~M.}\ \bibnamefont {G{\ifmmode\ddot{a}\else\"{a}\fi}chter}}, \bibinfo {author} {\bibfnamefont {K.}~\bibnamefont {Watanabe}}, \bibinfo {author} {\bibfnamefont {T.}~\bibnamefont {Taniguchi}}, \bibinfo {author} {\bibfnamefont {T.}~\bibnamefont {Ihn}}, \bibinfo {author} {\bibfnamefont {K.}~\bibnamefont {Ensslin}},\ and\ \bibinfo {author} {\bibfnamefont {W.~W.}\ \bibnamefont {Huang}},\ }\bibfield  {title} {\bibinfo {title} {{Long-lived valley states in bilayer graphene quantum dots}},\ }\href {https://doi.org/10.1038/s41567-023-02334-7} {\bibfield  {journal} {\bibinfo  {journal} {Nat. Phys.}\ }\textbf {\bibinfo {volume} {20}},\ \bibinfo
  {pages} {428} (\bibinfo {year} {2024})}\BibitemShut {NoStop}%
\bibitem [{\citenamefont {Knothe}\ and\ \citenamefont {Burkard}(2024)}]{Knothe2024Jun}%
  \BibitemOpen
  \bibfield  {author} {\bibinfo {author} {\bibfnamefont {A.}~\bibnamefont {Knothe}}\ and\ \bibinfo {author} {\bibfnamefont {G.}~\bibnamefont {Burkard}},\ }\bibfield  {title} {\bibinfo {title} {{Extended Hubbard model describing small multidot arrays in bilayer graphene}},\ }\href {https://doi.org/10.1103/PhysRevB.109.245401} {\bibfield  {journal} {\bibinfo  {journal} {Phys. Rev. B}\ }\textbf {\bibinfo {volume} {109}},\ \bibinfo {pages} {245401} (\bibinfo {year} {2024})}\BibitemShut {NoStop}%
\bibitem [{\citenamefont {Tong}\ \emph {et~al.}(2024{\natexlab{b}})\citenamefont {Tong}, \citenamefont {Ginzel}, \citenamefont {Kurzmann}, \citenamefont {Garreis}, \citenamefont {Ostertag}, \citenamefont {Gerber}, \citenamefont {Huang}, \citenamefont {Watanabe}, \citenamefont {Taniguchi}, \citenamefont {Burkard}, \citenamefont {Danon}, \citenamefont {Ihn},\ and\ \citenamefont {Ensslin}}]{Tong2024Jul}%
  \BibitemOpen
  \bibfield  {author} {\bibinfo {author} {\bibfnamefont {C.}~\bibnamefont {Tong}}, \bibinfo {author} {\bibfnamefont {F.}~\bibnamefont {Ginzel}}, \bibinfo {author} {\bibfnamefont {A.}~\bibnamefont {Kurzmann}}, \bibinfo {author} {\bibfnamefont {R.}~\bibnamefont {Garreis}}, \bibinfo {author} {\bibfnamefont {L.}~\bibnamefont {Ostertag}}, \bibinfo {author} {\bibfnamefont {J.~D.}\ \bibnamefont {Gerber}}, \bibinfo {author} {\bibfnamefont {W.~W.}\ \bibnamefont {Huang}}, \bibinfo {author} {\bibfnamefont {K.}~\bibnamefont {Watanabe}}, \bibinfo {author} {\bibfnamefont {T.}~\bibnamefont {Taniguchi}}, \bibinfo {author} {\bibfnamefont {G.}~\bibnamefont {Burkard}}, \bibinfo {author} {\bibfnamefont {J.}~\bibnamefont {Danon}}, \bibinfo {author} {\bibfnamefont {T.}~\bibnamefont {Ihn}},\ and\ \bibinfo {author} {\bibfnamefont {K.}~\bibnamefont {Ensslin}},\ }\bibfield  {title} {\bibinfo {title} {{Three-Carrier Spin Blockade and Coupling in Bilayer Graphene Double Quantum Dots}},\ }\href
  {https://doi.org/10.1103/PhysRevLett.133.017001} {\bibfield  {journal} {\bibinfo  {journal} {Phys. Rev. Lett.}\ }\textbf {\bibinfo {volume} {133}},\ \bibinfo {pages} {017001} (\bibinfo {year} {2024}{\natexlab{b}})}\BibitemShut {NoStop}%
\bibitem [{\citenamefont {McCann}\ and\ \citenamefont {Koshino}(2013)}]{McCann2013Apr}%
  \BibitemOpen
  \bibfield  {author} {\bibinfo {author} {\bibfnamefont {E.}~\bibnamefont {McCann}}\ and\ \bibinfo {author} {\bibfnamefont {M.}~\bibnamefont {Koshino}},\ }\bibfield  {title} {\bibinfo {title} {{The electronic properties of bilayer graphene}},\ }\href {https://doi.org/10.1088/0034-4885/76/5/056503} {\bibfield  {journal} {\bibinfo  {journal} {Rep. Prog. Phys.}\ }\textbf {\bibinfo {volume} {76}},\ \bibinfo {pages} {056503} (\bibinfo {year} {2013})}\BibitemShut {NoStop}%
\bibitem [{\citenamefont {Trauzettel}\ \emph {et~al.}(2007)\citenamefont {Trauzettel}, \citenamefont {Bulaev}, \citenamefont {Loss},\ and\ \citenamefont {Burkard}}]{Trauzettel2007Feb}%
  \BibitemOpen
  \bibfield  {author} {\bibinfo {author} {\bibfnamefont {B.}~\bibnamefont {Trauzettel}}, \bibinfo {author} {\bibfnamefont {D.~V.}\ \bibnamefont {Bulaev}}, \bibinfo {author} {\bibfnamefont {D.}~\bibnamefont {Loss}},\ and\ \bibinfo {author} {\bibfnamefont {G.}~\bibnamefont {Burkard}},\ }\bibfield  {title} {\bibinfo {title} {{Spin qubits in graphene quantum dots}},\ }\href {https://doi.org/10.1038/nphys544} {\bibfield  {journal} {\bibinfo  {journal} {Nat. Phys.}\ }\textbf {\bibinfo {volume} {3}},\ \bibinfo {pages} {192} (\bibinfo {year} {2007})}\BibitemShut {NoStop}%
\bibitem [{\citenamefont {M{\ifmmode\ddot{o}\else\"{o}\fi}ller}\ \emph {et~al.}(2023)\citenamefont {M{\ifmmode\ddot{o}\else\"{o}\fi}ller}, \citenamefont {Banszerus}, \citenamefont {Knothe}, \citenamefont {Valerius}, \citenamefont {Hecker}, \citenamefont {Icking}, \citenamefont {Watanabe}, \citenamefont {Taniguchi}, \citenamefont {Volk},\ and\ \citenamefont {Stampfer}}]{Moller2023SepShellFilling}%
  \BibitemOpen
  \bibfield  {author} {\bibinfo {author} {\bibfnamefont {S.}~\bibnamefont {M{\ifmmode\ddot{o}\else\"{o}\fi}ller}}, \bibinfo {author} {\bibfnamefont {L.}~\bibnamefont {Banszerus}}, \bibinfo {author} {\bibfnamefont {A.}~\bibnamefont {Knothe}}, \bibinfo {author} {\bibfnamefont {L.}~\bibnamefont {Valerius}}, \bibinfo {author} {\bibfnamefont {K.}~\bibnamefont {Hecker}}, \bibinfo {author} {\bibfnamefont {E.}~\bibnamefont {Icking}}, \bibinfo {author} {\bibfnamefont {K.}~\bibnamefont {Watanabe}}, \bibinfo {author} {\bibfnamefont {T.}~\bibnamefont {Taniguchi}}, \bibinfo {author} {\bibfnamefont {C.}~\bibnamefont {Volk}},\ and\ \bibinfo {author} {\bibfnamefont {C.}~\bibnamefont {Stampfer}},\ }\bibfield  {title} {\bibinfo {title} {{Impact of competing energy scales on the shell-filling sequence in elliptic bilayer graphene quantum dots}},\ }\href {https://doi.org/10.1103/PhysRevB.108.125128} {\bibfield  {journal} {\bibinfo  {journal} {Phys. Rev. B}\ }\textbf {\bibinfo {volume} {108}},\ \bibinfo {pages} {125128} (\bibinfo
  {year} {2023})}\BibitemShut {NoStop}%
\bibitem [{\citenamefont {Rohling}\ \emph {et~al.}(2014)\citenamefont {Rohling}, \citenamefont {Russ},\ and\ \citenamefont {Burkard}}]{Rohling2014Oct}%
  \BibitemOpen
  \bibfield  {author} {\bibinfo {author} {\bibfnamefont {N.}~\bibnamefont {Rohling}}, \bibinfo {author} {\bibfnamefont {M.}~\bibnamefont {Russ}},\ and\ \bibinfo {author} {\bibfnamefont {G.}~\bibnamefont {Burkard}},\ }\bibfield  {title} {\bibinfo {title} {{Hybrid Spin and Valley Quantum Computing with Singlet-Triplet Qubits}},\ }\href {https://doi.org/10.1103/PhysRevLett.113.176801} {\bibfield  {journal} {\bibinfo  {journal} {Phys. Rev. Lett.}\ }\textbf {\bibinfo {volume} {113}},\ \bibinfo {pages} {176801} (\bibinfo {year} {2014})}\BibitemShut {NoStop}%
\bibitem [{\citenamefont {Rohling}\ and\ \citenamefont {Burkard}(2012)}]{Rohling2012Aug}%
  \BibitemOpen
  \bibfield  {author} {\bibinfo {author} {\bibfnamefont {N.}~\bibnamefont {Rohling}}\ and\ \bibinfo {author} {\bibfnamefont {G.}~\bibnamefont {Burkard}},\ }\bibfield  {title} {\bibinfo {title} {{Universal quantum computing with spin and valley states}},\ }\href {https://doi.org/10.1088/1367-2630/14/8/083008} {\bibfield  {journal} {\bibinfo  {journal} {New J. Phys.}\ }\textbf {\bibinfo {volume} {14}},\ \bibinfo {pages} {083008} (\bibinfo {year} {2012})}\BibitemShut {NoStop}%
\bibitem [{\citenamefont {Wu}\ \emph {et~al.}(2013)\citenamefont {Wu}, \citenamefont {Lue},\ and\ \citenamefont {Chen}}]{Wu2013Sep}%
  \BibitemOpen
  \bibfield  {author} {\bibinfo {author} {\bibfnamefont {G.~Y.}\ \bibnamefont {Wu}}, \bibinfo {author} {\bibfnamefont {N.-Y.}\ \bibnamefont {Lue}},\ and\ \bibinfo {author} {\bibfnamefont {Y.-C.}\ \bibnamefont {Chen}},\ }\bibfield  {title} {\bibinfo {title} {{Quantum manipulation of valleys in bilayer graphene}},\ }\href {https://doi.org/10.1103/PhysRevB.88.125422} {\bibfield  {journal} {\bibinfo  {journal} {Phys. Rev. B}\ }\textbf {\bibinfo {volume} {88}},\ \bibinfo {pages} {125422} (\bibinfo {year} {2013})}\BibitemShut {NoStop}%
\bibitem [{\citenamefont {Noiri}\ \emph {et~al.}(2020)\citenamefont {Noiri}, \citenamefont {Takeda}, \citenamefont {Yoneda}, \citenamefont {Nakajima}, \citenamefont {Kodera},\ and\ \citenamefont {Tarucha}}]{Noiri2020Feb}%
  \BibitemOpen
  \bibfield  {author} {\bibinfo {author} {\bibfnamefont {A.}~\bibnamefont {Noiri}}, \bibinfo {author} {\bibfnamefont {K.}~\bibnamefont {Takeda}}, \bibinfo {author} {\bibfnamefont {J.}~\bibnamefont {Yoneda}}, \bibinfo {author} {\bibfnamefont {T.}~\bibnamefont {Nakajima}}, \bibinfo {author} {\bibfnamefont {T.}~\bibnamefont {Kodera}},\ and\ \bibinfo {author} {\bibfnamefont {S.}~\bibnamefont {Tarucha}},\ }\bibfield  {title} {\bibinfo {title} {{Radio-Frequency-Detected Fast Charge Sensing in Undoped Silicon Quantum Dots}},\ }\href {https://doi.org/10.1021/acs.nanolett.9b03847} {\bibfield  {journal} {\bibinfo  {journal} {Nano Lett.}\ }\textbf {\bibinfo {volume} {20}},\ \bibinfo {pages} {947} (\bibinfo {year} {2020})}\BibitemShut {NoStop}%
\bibitem [{\citenamefont {Volk}\ \emph {et~al.}(2019)\citenamefont {Volk}, \citenamefont {Chatterjee}, \citenamefont {Ansaloni}, \citenamefont {Marcus},\ and\ \citenamefont {Kuemmeth}}]{Volk2019Aug}%
  \BibitemOpen
  \bibfield  {author} {\bibinfo {author} {\bibfnamefont {C.}~\bibnamefont {Volk}}, \bibinfo {author} {\bibfnamefont {A.}~\bibnamefont {Chatterjee}}, \bibinfo {author} {\bibfnamefont {F.}~\bibnamefont {Ansaloni}}, \bibinfo {author} {\bibfnamefont {C.~M.}\ \bibnamefont {Marcus}},\ and\ \bibinfo {author} {\bibfnamefont {F.}~\bibnamefont {Kuemmeth}},\ }\bibfield  {title} {\bibinfo {title} {{Fast Charge Sensing of Si/SiGe Quantum Dots via a High-Frequency Accumulation Gate}},\ }\href {https://doi.org/10.1021/acs.nanolett.9b02149} {\bibfield  {journal} {\bibinfo  {journal} {Nano Lett.}\ }\textbf {\bibinfo {volume} {19}},\ \bibinfo {pages} {5628} (\bibinfo {year} {2019})}\BibitemShut {NoStop}%
\bibitem [{\citenamefont {Cassidy}\ \emph {et~al.}(2007)\citenamefont {Cassidy}, \citenamefont {Dzurak}, \citenamefont {Clark}, \citenamefont {Petersson}, \citenamefont {Farrer}, \citenamefont {Ritchie},\ and\ \citenamefont {Smith}}]{Cassidy2007Nov}%
  \BibitemOpen
  \bibfield  {author} {\bibinfo {author} {\bibfnamefont {M.~C.}\ \bibnamefont {Cassidy}}, \bibinfo {author} {\bibfnamefont {A.~S.}\ \bibnamefont {Dzurak}}, \bibinfo {author} {\bibfnamefont {R.~G.}\ \bibnamefont {Clark}}, \bibinfo {author} {\bibfnamefont {K.~D.}\ \bibnamefont {Petersson}}, \bibinfo {author} {\bibfnamefont {I.}~\bibnamefont {Farrer}}, \bibinfo {author} {\bibfnamefont {D.~A.}\ \bibnamefont {Ritchie}},\ and\ \bibinfo {author} {\bibfnamefont {C.~G.}\ \bibnamefont {Smith}},\ }\bibfield  {title} {\bibinfo {title} {{Single shot charge detection using a radio-frequency quantum point contact}},\ }\href {https://doi.org/10.1063/1.2809370} {\bibfield  {journal} {\bibinfo  {journal} {Appl. Phys. Lett.}\ }\textbf {\bibinfo {volume} {91}},\ \bibinfo {pages} {222104} (\bibinfo {year} {2007})}\BibitemShut {NoStop}%
\bibitem [{\citenamefont {Buitelaar}\ \emph {et~al.}(2008)\citenamefont {Buitelaar}, \citenamefont {Fransson}, \citenamefont {Cantone}, \citenamefont {Smith}, \citenamefont {Anderson}, \citenamefont {Jones}, \citenamefont {Ardavan}, \citenamefont {Khlobystov}, \citenamefont {Watt}, \citenamefont {Porfyrakis},\ and\ \citenamefont {Briggs}}]{Buitelaar2008Jun}%
  \BibitemOpen
  \bibfield  {author} {\bibinfo {author} {\bibfnamefont {M.~R.}\ \bibnamefont {Buitelaar}}, \bibinfo {author} {\bibfnamefont {J.}~\bibnamefont {Fransson}}, \bibinfo {author} {\bibfnamefont {A.~L.}\ \bibnamefont {Cantone}}, \bibinfo {author} {\bibfnamefont {C.~G.}\ \bibnamefont {Smith}}, \bibinfo {author} {\bibfnamefont {D.}~\bibnamefont {Anderson}}, \bibinfo {author} {\bibfnamefont {G.~A.~C.}\ \bibnamefont {Jones}}, \bibinfo {author} {\bibfnamefont {A.}~\bibnamefont {Ardavan}}, \bibinfo {author} {\bibfnamefont {A.~N.}\ \bibnamefont {Khlobystov}}, \bibinfo {author} {\bibfnamefont {A.~A.~R.}\ \bibnamefont {Watt}}, \bibinfo {author} {\bibfnamefont {K.}~\bibnamefont {Porfyrakis}},\ and\ \bibinfo {author} {\bibfnamefont {G.~A.~D.}\ \bibnamefont {Briggs}},\ }\bibfield  {title} {\bibinfo {title} {{Pauli spin blockade in carbon nanotube double quantum dots}},\ }\href {https://doi.org/10.1103/PhysRevB.77.245439} {\bibfield  {journal} {\bibinfo  {journal} {Phys. Rev. B}\ }\textbf {\bibinfo {volume} {77}},\ \bibinfo
  {pages} {245439} (\bibinfo {year} {2008})}\BibitemShut {NoStop}%
\bibitem [{\citenamefont {Churchill}\ \emph {et~al.}(2009{\natexlab{a}})\citenamefont {Churchill}, \citenamefont {Bestwick}, \citenamefont {Harlow}, \citenamefont {Kuemmeth}, \citenamefont {Marcos}, \citenamefont {Stwertka}, \citenamefont {Watson},\ and\ \citenamefont {Marcus}}]{Churchill2009Apr}%
  \BibitemOpen
  \bibfield  {author} {\bibinfo {author} {\bibfnamefont {H.~O.~H.}\ \bibnamefont {Churchill}}, \bibinfo {author} {\bibfnamefont {A.~J.}\ \bibnamefont {Bestwick}}, \bibinfo {author} {\bibfnamefont {J.~W.}\ \bibnamefont {Harlow}}, \bibinfo {author} {\bibfnamefont {F.}~\bibnamefont {Kuemmeth}}, \bibinfo {author} {\bibfnamefont {D.}~\bibnamefont {Marcos}}, \bibinfo {author} {\bibfnamefont {C.~H.}\ \bibnamefont {Stwertka}}, \bibinfo {author} {\bibfnamefont {S.~K.}\ \bibnamefont {Watson}},\ and\ \bibinfo {author} {\bibfnamefont {C.~M.}\ \bibnamefont {Marcus}},\ }\bibfield  {title} {\bibinfo {title} {{Electron{\textendash}nuclear interaction in 13C nanotube double quantum dots}},\ }\href {https://doi.org/10.1038/nphys1247} {\bibfield  {journal} {\bibinfo  {journal} {Nat. Phys.}\ }\textbf {\bibinfo {volume} {5}},\ \bibinfo {pages} {321} (\bibinfo {year} {2009}{\natexlab{a}})}\BibitemShut {NoStop}%
\bibitem [{\citenamefont {Churchill}\ \emph {et~al.}(2009{\natexlab{b}})\citenamefont {Churchill}, \citenamefont {Kuemmeth}, \citenamefont {Harlow}, \citenamefont {Bestwick}, \citenamefont {Rashba}, \citenamefont {Flensberg}, \citenamefont {Stwertka}, \citenamefont {Taychatanapat}, \citenamefont {Watson},\ and\ \citenamefont {Marcus}}]{Churchill2009AprA}%
  \BibitemOpen
  \bibfield  {author} {\bibinfo {author} {\bibfnamefont {H.~O.~H.}\ \bibnamefont {Churchill}}, \bibinfo {author} {\bibfnamefont {F.}~\bibnamefont {Kuemmeth}}, \bibinfo {author} {\bibfnamefont {J.~W.}\ \bibnamefont {Harlow}}, \bibinfo {author} {\bibfnamefont {A.~J.}\ \bibnamefont {Bestwick}}, \bibinfo {author} {\bibfnamefont {E.~I.}\ \bibnamefont {Rashba}}, \bibinfo {author} {\bibfnamefont {K.}~\bibnamefont {Flensberg}}, \bibinfo {author} {\bibfnamefont {C.~H.}\ \bibnamefont {Stwertka}}, \bibinfo {author} {\bibfnamefont {T.}~\bibnamefont {Taychatanapat}}, \bibinfo {author} {\bibfnamefont {S.~K.}\ \bibnamefont {Watson}},\ and\ \bibinfo {author} {\bibfnamefont {C.~M.}\ \bibnamefont {Marcus}},\ }\bibfield  {title} {\bibinfo {title} {{Relaxation and Dephasing in a Two-Electron $^{13}\mathbf{C}$ Nanotube Double Quantum Dot}},\ }\href {https://doi.org/10.1103/PhysRevLett.102.166802} {\bibfield  {journal} {\bibinfo  {journal} {Phys. Rev. Lett.}\ }\textbf {\bibinfo {volume} {102}},\ \bibinfo {pages} {166802} (\bibinfo
  {year} {2009}{\natexlab{b}})}\BibitemShut {NoStop}%
\bibitem [{\citenamefont {Lai}\ \emph {et~al.}(2011)\citenamefont {Lai}, \citenamefont {Lim}, \citenamefont {Yang}, \citenamefont {Zwanenburg}, \citenamefont {Coish}, \citenamefont {Qassemi}, \citenamefont {Morello},\ and\ \citenamefont {Dzurak}}]{Lai2011Oct}%
  \BibitemOpen
  \bibfield  {author} {\bibinfo {author} {\bibfnamefont {N.~S.}\ \bibnamefont {Lai}}, \bibinfo {author} {\bibfnamefont {W.~H.}\ \bibnamefont {Lim}}, \bibinfo {author} {\bibfnamefont {C.~H.}\ \bibnamefont {Yang}}, \bibinfo {author} {\bibfnamefont {F.~A.}\ \bibnamefont {Zwanenburg}}, \bibinfo {author} {\bibfnamefont {W.~A.}\ \bibnamefont {Coish}}, \bibinfo {author} {\bibfnamefont {F.}~\bibnamefont {Qassemi}}, \bibinfo {author} {\bibfnamefont {A.}~\bibnamefont {Morello}},\ and\ \bibinfo {author} {\bibfnamefont {A.~S.}\ \bibnamefont {Dzurak}},\ }\bibfield  {title} {\bibinfo {title} {{Pauli Spin Blockade in a Highly Tunable Silicon Double Quantum Dot}},\ }\href {https://doi.org/10.1038/srep00110} {\bibfield  {journal} {\bibinfo  {journal} {Sci. Rep.}\ }\textbf {\bibinfo {volume} {1}},\ \bibinfo {pages} {1} (\bibinfo {year} {2011})}\BibitemShut {NoStop}%
\bibitem [{\citenamefont {Pakkiam}\ \emph {et~al.}(2018)\citenamefont {Pakkiam}, \citenamefont {Timofeev}, \citenamefont {House}, \citenamefont {Hogg}, \citenamefont {Kobayashi}, \citenamefont {Koch}, \citenamefont {Rogge},\ and\ \citenamefont {Simmons}}]{Pakkiam2018Nov}%
  \BibitemOpen
  \bibfield  {author} {\bibinfo {author} {\bibfnamefont {P.}~\bibnamefont {Pakkiam}}, \bibinfo {author} {\bibfnamefont {A.~V.}\ \bibnamefont {Timofeev}}, \bibinfo {author} {\bibfnamefont {M.~G.}\ \bibnamefont {House}}, \bibinfo {author} {\bibfnamefont {M.~R.}\ \bibnamefont {Hogg}}, \bibinfo {author} {\bibfnamefont {T.}~\bibnamefont {Kobayashi}}, \bibinfo {author} {\bibfnamefont {M.}~\bibnamefont {Koch}}, \bibinfo {author} {\bibfnamefont {S.}~\bibnamefont {Rogge}},\ and\ \bibinfo {author} {\bibfnamefont {M.~Y.}\ \bibnamefont {Simmons}},\ }\bibfield  {title} {\bibinfo {title} {{Single-Shot Single-Gate rf Spin Readout in Silicon}},\ }\href {https://doi.org/10.1103/PhysRevX.8.041032} {\bibfield  {journal} {\bibinfo  {journal} {Phys. Rev. X}\ }\textbf {\bibinfo {volume} {8}},\ \bibinfo {pages} {041032} (\bibinfo {year} {2018})}\BibitemShut {NoStop}%
\bibitem [{\citenamefont {Seedhouse}\ \emph {et~al.}(2021)\citenamefont {Seedhouse}, \citenamefont {Tanttu}, \citenamefont {Leon}, \citenamefont {Zhao}, \citenamefont {Tan}, \citenamefont {Hensen}, \citenamefont {Hudson}, \citenamefont {Itoh}, \citenamefont {Yoneda}, \citenamefont {Yang}, \citenamefont {Morello}, \citenamefont {Laucht}, \citenamefont {Coppersmith}, \citenamefont {Saraiva},\ and\ \citenamefont {Dzurak}}]{Seedhouse2021Jan}%
  \BibitemOpen
  \bibfield  {author} {\bibinfo {author} {\bibfnamefont {A.~E.}\ \bibnamefont {Seedhouse}}, \bibinfo {author} {\bibfnamefont {T.}~\bibnamefont {Tanttu}}, \bibinfo {author} {\bibfnamefont {R.~C.~C.}\ \bibnamefont {Leon}}, \bibinfo {author} {\bibfnamefont {R.}~\bibnamefont {Zhao}}, \bibinfo {author} {\bibfnamefont {K.~Y.}\ \bibnamefont {Tan}}, \bibinfo {author} {\bibfnamefont {B.}~\bibnamefont {Hensen}}, \bibinfo {author} {\bibfnamefont {F.~E.}\ \bibnamefont {Hudson}}, \bibinfo {author} {\bibfnamefont {K.~M.}\ \bibnamefont {Itoh}}, \bibinfo {author} {\bibfnamefont {J.}~\bibnamefont {Yoneda}}, \bibinfo {author} {\bibfnamefont {C.~H.}\ \bibnamefont {Yang}}, \bibinfo {author} {\bibfnamefont {A.}~\bibnamefont {Morello}}, \bibinfo {author} {\bibfnamefont {A.}~\bibnamefont {Laucht}}, \bibinfo {author} {\bibfnamefont {S.~N.}\ \bibnamefont {Coppersmith}}, \bibinfo {author} {\bibfnamefont {A.}~\bibnamefont {Saraiva}},\ and\ \bibinfo {author} {\bibfnamefont {A.~S.}\ \bibnamefont {Dzurak}},\ }\bibfield  {title} {\bibinfo
  {title} {{Pauli Blockade in Silicon Quantum Dots with Spin-Orbit Control}},\ }\href {https://doi.org/10.1103/PRXQuantum.2.010303} {\bibfield  {journal} {\bibinfo  {journal} {PRX Quantum}\ }\textbf {\bibinfo {volume} {2}},\ \bibinfo {pages} {010303} (\bibinfo {year} {2021})}\BibitemShut {NoStop}%
\bibitem [{\citenamefont {Eich}\ \emph {et~al.}(2018{\natexlab{b}})\citenamefont {Eich}, \citenamefont {Pisoni}, \citenamefont {Overweg}, \citenamefont {Kurzmann}, \citenamefont {Lee}, \citenamefont {Rickhaus}, \citenamefont {Ihn}, \citenamefont {Ensslin}, \citenamefont {Herman}, \citenamefont {Sigrist}, \citenamefont {Watanabe},\ and\ \citenamefont {Taniguchi}}]{Eich2018Jul}%
  \BibitemOpen
  \bibfield  {author} {\bibinfo {author} {\bibfnamefont {M.}~\bibnamefont {Eich}}, \bibinfo {author} {\bibfnamefont {R.}~\bibnamefont {Pisoni}}, \bibinfo {author} {\bibfnamefont {H.}~\bibnamefont {Overweg}}, \bibinfo {author} {\bibfnamefont {A.}~\bibnamefont {Kurzmann}}, \bibinfo {author} {\bibfnamefont {Y.}~\bibnamefont {Lee}}, \bibinfo {author} {\bibfnamefont {P.}~\bibnamefont {Rickhaus}}, \bibinfo {author} {\bibfnamefont {T.}~\bibnamefont {Ihn}}, \bibinfo {author} {\bibfnamefont {K.}~\bibnamefont {Ensslin}}, \bibinfo {author} {\bibfnamefont {F.}~\bibnamefont {Herman}}, \bibinfo {author} {\bibfnamefont {M.}~\bibnamefont {Sigrist}}, \bibinfo {author} {\bibfnamefont {K.}~\bibnamefont {Watanabe}},\ and\ \bibinfo {author} {\bibfnamefont {T.}~\bibnamefont {Taniguchi}},\ }\bibfield  {title} {\bibinfo {title} {{Spin and Valley States in Gate-Defined Bilayer Graphene Quantum Dots}},\ }\href {https://doi.org/10.1103/PhysRevX.8.031023} {\bibfield  {journal} {\bibinfo  {journal} {Phys. Rev. X}\ }\textbf {\bibinfo
  {volume} {8}},\ \bibinfo {pages} {031023} (\bibinfo {year} {2018}{\natexlab{b}})}\BibitemShut {NoStop}%
\bibitem [{\citenamefont {Garreis}\ \emph {et~al.}(2021)\citenamefont {Garreis}, \citenamefont {Knothe}, \citenamefont {Tong}, \citenamefont {Eich}, \citenamefont {Gold}, \citenamefont {Watanabe}, \citenamefont {Taniguchi}, \citenamefont {Fal{'}ko}, \citenamefont {Ihn}, \citenamefont {Ensslin},\ and\ \citenamefont {Kurzmann}}]{Garreis2021AprShellfillingWarping}%
  \BibitemOpen
  \bibfield  {author} {\bibinfo {author} {\bibfnamefont {R.}~\bibnamefont {Garreis}}, \bibinfo {author} {\bibfnamefont {A.}~\bibnamefont {Knothe}}, \bibinfo {author} {\bibfnamefont {C.}~\bibnamefont {Tong}}, \bibinfo {author} {\bibfnamefont {M.}~\bibnamefont {Eich}}, \bibinfo {author} {\bibfnamefont {C.}~\bibnamefont {Gold}}, \bibinfo {author} {\bibfnamefont {K.}~\bibnamefont {Watanabe}}, \bibinfo {author} {\bibfnamefont {T.}~\bibnamefont {Taniguchi}}, \bibinfo {author} {\bibfnamefont {V.}~\bibnamefont {Fal{'}ko}}, \bibinfo {author} {\bibfnamefont {T.}~\bibnamefont {Ihn}}, \bibinfo {author} {\bibfnamefont {K.}~\bibnamefont {Ensslin}},\ and\ \bibinfo {author} {\bibfnamefont {A.}~\bibnamefont {Kurzmann}},\ }\bibfield  {title} {\bibinfo {title} {{Shell Filling and Trigonal Warping in Graphene Quantum Dots}},\ }\href {https://doi.org/10.1103/PhysRevLett.126.147703} {\bibfield  {journal} {\bibinfo  {journal} {Phys. Rev. Lett.}\ }\textbf {\bibinfo {volume} {126}},\ \bibinfo {pages} {147703} (\bibinfo {year}
  {2021})}\BibitemShut {NoStop}%
\bibitem [{\citenamefont {Engels}\ \emph {et~al.}(2014)\citenamefont {Engels}, \citenamefont {Terr{\ifmmode \acute{e} \else \'{e}\fi}s}, \citenamefont {Epping}, \citenamefont {Khodkov}, \citenamefont {Watanabe}, \citenamefont {Taniguchi}, \citenamefont {Beschoten},\ and\ \citenamefont {Stampfer}}]{Engels2014Sep}%
  \BibitemOpen
  \bibfield  {author} {\bibinfo {author} {\bibfnamefont {S.}~\bibnamefont {Engels}}, \bibinfo {author} {\bibfnamefont {B.}~\bibnamefont {Terr{\ifmmode \acute{e} \else \'{e}\fi}s}}, \bibinfo {author} {\bibfnamefont {A.}~\bibnamefont {Epping}}, \bibinfo {author} {\bibfnamefont {T.}~\bibnamefont {Khodkov}}, \bibinfo {author} {\bibfnamefont {K.}~\bibnamefont {Watanabe}}, \bibinfo {author} {\bibfnamefont {T.}~\bibnamefont {Taniguchi}}, \bibinfo {author} {\bibfnamefont {B.}~\bibnamefont {Beschoten}},\ and\ \bibinfo {author} {\bibfnamefont {C.}~\bibnamefont {Stampfer}},\ }\bibfield  {title} {\bibinfo {title} {{Limitations to Carrier Mobility and Phase-Coherent Transport in Bilayer Graphene}},\ }\href {https://doi.org/10.1103/PhysRevLett.113.126801} {\bibfield  {journal} {\bibinfo  {journal} {Phys. Rev. Lett.}\ }\textbf {\bibinfo {volume} {113}},\ \bibinfo {pages} {126801} (\bibinfo {year} {2014})}\BibitemShut {NoStop}%
\bibitem [{\citenamefont {Wang}\ \emph {et~al.}(2013)\citenamefont {Wang}, \citenamefont {Meric}, \citenamefont {Huang}, \citenamefont {Gao}, \citenamefont {Gao}, \citenamefont {Tran}, \citenamefont {Taniguchi}, \citenamefont {Watanabe}, \citenamefont {Campos}, \citenamefont {Muller}, \citenamefont {Guo}, \citenamefont {Kim}, \citenamefont {Hone}, \citenamefont {Shepard},\ and\ \citenamefont {Dean}}]{Wang2013Nov}%
  \BibitemOpen
  \bibfield  {author} {\bibinfo {author} {\bibfnamefont {L.}~\bibnamefont {Wang}}, \bibinfo {author} {\bibfnamefont {I.}~\bibnamefont {Meric}}, \bibinfo {author} {\bibfnamefont {P.~Y.}\ \bibnamefont {Huang}}, \bibinfo {author} {\bibfnamefont {Q.}~\bibnamefont {Gao}}, \bibinfo {author} {\bibfnamefont {Y.}~\bibnamefont {Gao}}, \bibinfo {author} {\bibfnamefont {H.}~\bibnamefont {Tran}}, \bibinfo {author} {\bibfnamefont {T.}~\bibnamefont {Taniguchi}}, \bibinfo {author} {\bibfnamefont {K.}~\bibnamefont {Watanabe}}, \bibinfo {author} {\bibfnamefont {L.~M.}\ \bibnamefont {Campos}}, \bibinfo {author} {\bibfnamefont {D.~A.}\ \bibnamefont {Muller}}, \bibinfo {author} {\bibfnamefont {J.}~\bibnamefont {Guo}}, \bibinfo {author} {\bibfnamefont {P.}~\bibnamefont {Kim}}, \bibinfo {author} {\bibfnamefont {J.}~\bibnamefont {Hone}}, \bibinfo {author} {\bibfnamefont {K.~L.}\ \bibnamefont {Shepard}},\ and\ \bibinfo {author} {\bibfnamefont {C.~R.}\ \bibnamefont {Dean}},\ }\bibfield  {title} {\bibinfo {title} {{One-Dimensional
  Electrical Contact to a Two-Dimensional Material}},\ }\href {https://doi.org/10.1126/science.1244358} {\bibfield  {journal} {\bibinfo  {journal} {Science}\ }\textbf {\bibinfo {volume} {342}},\ \bibinfo {pages} {614} (\bibinfo {year} {2013})}\BibitemShut {NoStop}%
\bibitem [{\citenamefont {Icking}\ \emph {et~al.}(2022)\citenamefont {Icking}, \citenamefont {Banszerus}, \citenamefont {W{\ifmmode\ddot{o}\else\"{o}\fi}rtche}, \citenamefont {Volmer},\ and\ \citenamefont {Stampfer}}]{Icking2022Jul}%
  \BibitemOpen
  \bibfield  {author} {\bibinfo {author} {\bibfnamefont {E.}~\bibnamefont {Icking}}, \bibinfo {author} {\bibfnamefont {L.}~\bibnamefont {Banszerus}}, \bibinfo {author} {\bibfnamefont {F.}~\bibnamefont {W{\ifmmode\ddot{o}\else\"{o}\fi}rtche}}, \bibinfo {author} {\bibfnamefont {F.}~\bibnamefont {Volmer}},\ and\ \bibinfo {author} {\bibfnamefont {C.}~\bibnamefont {Stampfer}},\ }\bibfield  {title} {\bibinfo {title} {{Transport Spectroscopy of Ultraclean Tunable Band Gaps in Bilayer Graphene}},\ }\href {https://doi.org/10.1002/aelm.202200510} {\bibfield  {journal} {\bibinfo  {journal} {Adv. Electron. Mater.}\ }\textbf {\bibinfo {volume} {8}},\ \bibinfo {pages} {2200510} (\bibinfo {year} {2022})}\BibitemShut {NoStop}%
\bibitem [{\citenamefont {Banszerus}\ \emph {et~al.}(2020{\natexlab{b}})\citenamefont {Banszerus}, \citenamefont {Rothstein}, \citenamefont {Fabian}, \citenamefont {M{\ifmmode\ddot{o}\else\"{o}\fi}ller}, \citenamefont {Icking}, \citenamefont {Trellenkamp}, \citenamefont {Lentz}, \citenamefont {Neumaier}, \citenamefont {Watanabe}, \citenamefont {Taniguchi}, \citenamefont {Libisch}, \citenamefont {Volk},\ and\ \citenamefont {Stampfer}}]{Banszerus2020OctEHcrossover}%
  \BibitemOpen
  \bibfield  {author} {\bibinfo {author} {\bibfnamefont {L.}~\bibnamefont {Banszerus}}, \bibinfo {author} {\bibfnamefont {A.}~\bibnamefont {Rothstein}}, \bibinfo {author} {\bibfnamefont {T.}~\bibnamefont {Fabian}}, \bibinfo {author} {\bibfnamefont {S.}~\bibnamefont {M{\ifmmode\ddot{o}\else\"{o}\fi}ller}}, \bibinfo {author} {\bibfnamefont {E.}~\bibnamefont {Icking}}, \bibinfo {author} {\bibfnamefont {S.}~\bibnamefont {Trellenkamp}}, \bibinfo {author} {\bibfnamefont {F.}~\bibnamefont {Lentz}}, \bibinfo {author} {\bibfnamefont {D.}~\bibnamefont {Neumaier}}, \bibinfo {author} {\bibfnamefont {K.}~\bibnamefont {Watanabe}}, \bibinfo {author} {\bibfnamefont {T.}~\bibnamefont {Taniguchi}}, \bibinfo {author} {\bibfnamefont {F.}~\bibnamefont {Libisch}}, \bibinfo {author} {\bibfnamefont {C.}~\bibnamefont {Volk}},\ and\ \bibinfo {author} {\bibfnamefont {C.}~\bibnamefont {Stampfer}},\ }\bibfield  {title} {\bibinfo {title} {{Electron{\textendash}Hole Crossover in Gate-Controlled Bilayer Graphene Quantum Dots}},\ }\href
  {https://doi.org/10.1021/acs.nanolett.0c03227} {\bibfield  {journal} {\bibinfo  {journal} {Nano Lett.}\ }\textbf {\bibinfo {volume} {20}},\ \bibinfo {pages} {7709} (\bibinfo {year} {2020}{\natexlab{b}})}\BibitemShut {NoStop}%
\bibitem [{\citenamefont {Banszerus}\ \emph {et~al.}(2020{\natexlab{c}})\citenamefont {Banszerus}, \citenamefont {Fabian}, \citenamefont {M{\ifmmode\ddot{o}\else\"{o}\fi}ller}, \citenamefont {Icking}, \citenamefont {Heiming}, \citenamefont {Trellenkamp}, \citenamefont {Lentz}, \citenamefont {Neumaier}, \citenamefont {Otto}, \citenamefont {Watanabe}, \citenamefont {Taniguchi}, \citenamefont {Libisch}, \citenamefont {Volk},\ and\ \citenamefont {Stampfer}}]{Banszerus2020DecPSSB}%
  \BibitemOpen
  \bibfield  {author} {\bibinfo {author} {\bibfnamefont {L.}~\bibnamefont {Banszerus}}, \bibinfo {author} {\bibfnamefont {T.}~\bibnamefont {Fabian}}, \bibinfo {author} {\bibfnamefont {S.}~\bibnamefont {M{\ifmmode\ddot{o}\else\"{o}\fi}ller}}, \bibinfo {author} {\bibfnamefont {E.}~\bibnamefont {Icking}}, \bibinfo {author} {\bibfnamefont {H.}~\bibnamefont {Heiming}}, \bibinfo {author} {\bibfnamefont {S.}~\bibnamefont {Trellenkamp}}, \bibinfo {author} {\bibfnamefont {F.}~\bibnamefont {Lentz}}, \bibinfo {author} {\bibfnamefont {D.}~\bibnamefont {Neumaier}}, \bibinfo {author} {\bibfnamefont {M.}~\bibnamefont {Otto}}, \bibinfo {author} {\bibfnamefont {K.}~\bibnamefont {Watanabe}}, \bibinfo {author} {\bibfnamefont {T.}~\bibnamefont {Taniguchi}}, \bibinfo {author} {\bibfnamefont {F.}~\bibnamefont {Libisch}}, \bibinfo {author} {\bibfnamefont {C.}~\bibnamefont {Volk}},\ and\ \bibinfo {author} {\bibfnamefont {C.}~\bibnamefont {Stampfer}},\ }\bibfield  {title} {\bibinfo {title} {{Electrostatic Detection of
  Shubnikov{\textendash}de Haas Oscillations in Bilayer Graphene by Coulomb Resonances in Gate-Defined Quantum Dots}},\ }\href {https://doi.org/10.1002/pssb.202000333} {\bibfield  {journal} {\bibinfo  {journal} {Phys. Status Solidi B}\ }\textbf {\bibinfo {volume} {257}},\ \bibinfo {pages} {2000333} (\bibinfo {year} {2020}{\natexlab{c}})}\BibitemShut {NoStop}%
\bibitem [{\citenamefont {M{\ifmmode\ddot{o}\else\"{o}\fi}ller}\ \emph {et~al.}(2021)\citenamefont {M{\ifmmode\ddot{o}\else\"{o}\fi}ller}, \citenamefont {Banszerus}, \citenamefont {Knothe}, \citenamefont {Steiner}, \citenamefont {Icking}, \citenamefont {Trellenkamp}, \citenamefont {Lentz}, \citenamefont {Watanabe}, \citenamefont {Taniguchi}, \citenamefont {Glazman}, \citenamefont {Fal{'}ko}, \citenamefont {Volk},\ and\ \citenamefont {Stampfer}}]{Moller2021Dec}%
  \BibitemOpen
  \bibfield  {author} {\bibinfo {author} {\bibfnamefont {S.}~\bibnamefont {M{\ifmmode\ddot{o}\else\"{o}\fi}ller}}, \bibinfo {author} {\bibfnamefont {L.}~\bibnamefont {Banszerus}}, \bibinfo {author} {\bibfnamefont {A.}~\bibnamefont {Knothe}}, \bibinfo {author} {\bibfnamefont {C.}~\bibnamefont {Steiner}}, \bibinfo {author} {\bibfnamefont {E.}~\bibnamefont {Icking}}, \bibinfo {author} {\bibfnamefont {S.}~\bibnamefont {Trellenkamp}}, \bibinfo {author} {\bibfnamefont {F.}~\bibnamefont {Lentz}}, \bibinfo {author} {\bibfnamefont {K.}~\bibnamefont {Watanabe}}, \bibinfo {author} {\bibfnamefont {T.}~\bibnamefont {Taniguchi}}, \bibinfo {author} {\bibfnamefont {L.~I.}\ \bibnamefont {Glazman}}, \bibinfo {author} {\bibfnamefont {V.~I.}\ \bibnamefont {Fal{'}ko}}, \bibinfo {author} {\bibfnamefont {C.}~\bibnamefont {Volk}},\ and\ \bibinfo {author} {\bibfnamefont {C.}~\bibnamefont {Stampfer}},\ }\bibfield  {title} {\bibinfo {title} {{Probing Two-Electron Multiplets in Bilayer Graphene Quantum Dots}},\ }\href
  {https://doi.org/10.1103/PhysRevLett.127.256802} {\bibfield  {journal} {\bibinfo  {journal} {Phys. Rev. Lett.}\ }\textbf {\bibinfo {volume} {127}},\ \bibinfo {pages} {256802} (\bibinfo {year} {2021})}\BibitemShut {NoStop}%
\bibitem [{\citenamefont {Knothe}\ \emph {et~al.}(2022)\citenamefont {Knothe}, \citenamefont {Glazman},\ and\ \citenamefont {Fal{'}ko}}]{Knothe2022Apr}%
  \BibitemOpen
  \bibfield  {author} {\bibinfo {author} {\bibfnamefont {A.}~\bibnamefont {Knothe}}, \bibinfo {author} {\bibfnamefont {L.~I.}\ \bibnamefont {Glazman}},\ and\ \bibinfo {author} {\bibfnamefont {V.~I.}\ \bibnamefont {Fal{'}ko}},\ }\bibfield  {title} {\bibinfo {title} {{Tunneling theory for a bilayer graphene quantum dot{'}s single- and two-electron states}},\ }\href {https://doi.org/10.1088/1367-2630/ac5d00} {\bibfield  {journal} {\bibinfo  {journal} {New J. Phys.}\ }\textbf {\bibinfo {volume} {24}},\ \bibinfo {pages} {043003} (\bibinfo {year} {2022})}\BibitemShut {NoStop}%
\bibitem [{\citenamefont {Knothe}\ and\ \citenamefont {Fal'ko}(2020)}]{Knothe2020JunQuartetStates}%
  \BibitemOpen
  \bibfield  {author} {\bibinfo {author} {\bibfnamefont {A.}~\bibnamefont {Knothe}}\ and\ \bibinfo {author} {\bibfnamefont {V.}~\bibnamefont {Fal'ko}},\ }\bibfield  {title} {\bibinfo {title} {{Quartet states in two-electron quantum dots in bilayer graphene}},\ }\href {https://doi.org/10.1103/PhysRevB.101.235423} {\bibfield  {journal} {\bibinfo  {journal} {Phys. Rev. B}\ }\textbf {\bibinfo {volume} {101}},\ \bibinfo {pages} {235423} (\bibinfo {year} {2020})}\BibitemShut {NoStop}%
\bibitem [{\citenamefont {Lemonik}\ \emph {et~al.}(2010)\citenamefont {Lemonik}, \citenamefont {Aleiner}, \citenamefont {Toke},\ and\ \citenamefont {Fal{'}ko}}]{Lemonik2010NovLifshitzSymmBreak}%
  \BibitemOpen
  \bibfield  {author} {\bibinfo {author} {\bibfnamefont {Y.}~\bibnamefont {Lemonik}}, \bibinfo {author} {\bibfnamefont {I.~L.}\ \bibnamefont {Aleiner}}, \bibinfo {author} {\bibfnamefont {C.}~\bibnamefont {Toke}},\ and\ \bibinfo {author} {\bibfnamefont {V.~I.}\ \bibnamefont {Fal{'}ko}},\ }\bibfield  {title} {\bibinfo {title} {{Spontaneous symmetry breaking and Lifshitz transition in bilayer graphene}},\ }\href {https://doi.org/10.1103/PhysRevB.82.201408} {\bibfield  {journal} {\bibinfo  {journal} {Phys. Rev. B}\ }\textbf {\bibinfo {volume} {82}},\ \bibinfo {pages} {201408} (\bibinfo {year} {2010})}\BibitemShut {NoStop}%
\bibitem [{\citenamefont {Lemonik}\ \emph {et~al.}(2012)\citenamefont {Lemonik}, \citenamefont {Aleiner},\ and\ \citenamefont {Fal'ko}}]{Lemonik2012JunCompetingBLGOrders}%
  \BibitemOpen
  \bibfield  {author} {\bibinfo {author} {\bibfnamefont {Y.}~\bibnamefont {Lemonik}}, \bibinfo {author} {\bibfnamefont {I.}~\bibnamefont {Aleiner}},\ and\ \bibinfo {author} {\bibfnamefont {V.~I.}\ \bibnamefont {Fal'ko}},\ }\bibfield  {title} {\bibinfo {title} {{Competing nematic, antiferromagnetic, and spin-flux orders in the ground state of bilayer graphene}},\ }\href {https://doi.org/10.1103/PhysRevB.85.245451} {\bibfield  {journal} {\bibinfo  {journal} {Phys. Rev. B}\ }\textbf {\bibinfo {volume} {85}},\ \bibinfo {pages} {245451} (\bibinfo {year} {2012})}\BibitemShut {NoStop}%
\bibitem [{\citenamefont {Kurzmann}\ \emph {et~al.}(2021)\citenamefont {Kurzmann}, \citenamefont {Kleeorin}, \citenamefont {Tong}, \citenamefont {Garreis}, \citenamefont {Knothe}, \citenamefont {Eich}, \citenamefont {Mittag}, \citenamefont {Gold}, \citenamefont {de~Vries}, \citenamefont {Watanabe}, \citenamefont {Taniguchi}, \citenamefont {Fal{'}ko}, \citenamefont {Meir}, \citenamefont {Ihn},\ and\ \citenamefont {Ensslin}}]{Kurzmann2021OctKondo}%
  \BibitemOpen
  \bibfield  {author} {\bibinfo {author} {\bibfnamefont {A.}~\bibnamefont {Kurzmann}}, \bibinfo {author} {\bibfnamefont {Y.}~\bibnamefont {Kleeorin}}, \bibinfo {author} {\bibfnamefont {C.}~\bibnamefont {Tong}}, \bibinfo {author} {\bibfnamefont {R.}~\bibnamefont {Garreis}}, \bibinfo {author} {\bibfnamefont {A.}~\bibnamefont {Knothe}}, \bibinfo {author} {\bibfnamefont {M.}~\bibnamefont {Eich}}, \bibinfo {author} {\bibfnamefont {C.}~\bibnamefont {Mittag}}, \bibinfo {author} {\bibfnamefont {C.}~\bibnamefont {Gold}}, \bibinfo {author} {\bibfnamefont {F.~K.}\ \bibnamefont {de~Vries}}, \bibinfo {author} {\bibfnamefont {K.}~\bibnamefont {Watanabe}}, \bibinfo {author} {\bibfnamefont {T.}~\bibnamefont {Taniguchi}}, \bibinfo {author} {\bibfnamefont {V.}~\bibnamefont {Fal{'}ko}}, \bibinfo {author} {\bibfnamefont {Y.}~\bibnamefont {Meir}}, \bibinfo {author} {\bibfnamefont {T.}~\bibnamefont {Ihn}},\ and\ \bibinfo {author} {\bibfnamefont {K.}~\bibnamefont {Ensslin}},\ }\bibfield  {title} {\bibinfo {title} {{Kondo effect and
  spin{\textendash}orbit coupling in graphene quantum dots}},\ }\href {https://doi.org/10.1038/s41467-021-26149-3} {\bibfield  {journal} {\bibinfo  {journal} {Nat. Commun.}\ }\textbf {\bibinfo {volume} {12}},\ \bibinfo {pages} {6004} (\bibinfo {year} {2021})}\BibitemShut {NoStop}%
\bibitem [{\citenamefont {Banszerus}\ \emph {et~al.}(2022)\citenamefont {Banszerus}, \citenamefont {Hecker}, \citenamefont {M{\ifmmode\ddot{o}\else\"{o}\fi}ller}, \citenamefont {Icking}, \citenamefont {Watanabe}, \citenamefont {Taniguchi}, \citenamefont {Volk},\ and\ \citenamefont {Stampfer}}]{Banszerus2022JunSpinRelax}%
  \BibitemOpen
  \bibfield  {author} {\bibinfo {author} {\bibfnamefont {L.}~\bibnamefont {Banszerus}}, \bibinfo {author} {\bibfnamefont {K.}~\bibnamefont {Hecker}}, \bibinfo {author} {\bibfnamefont {S.}~\bibnamefont {M{\ifmmode\ddot{o}\else\"{o}\fi}ller}}, \bibinfo {author} {\bibfnamefont {E.}~\bibnamefont {Icking}}, \bibinfo {author} {\bibfnamefont {K.}~\bibnamefont {Watanabe}}, \bibinfo {author} {\bibfnamefont {T.}~\bibnamefont {Taniguchi}}, \bibinfo {author} {\bibfnamefont {C.}~\bibnamefont {Volk}},\ and\ \bibinfo {author} {\bibfnamefont {C.}~\bibnamefont {Stampfer}},\ }\bibfield  {title} {\bibinfo {title} {{Spin relaxation in a single-electron graphene quantum dot}},\ }\href {https://doi.org/10.1038/s41467-022-31231-5} {\bibfield  {journal} {\bibinfo  {journal} {Nat. Commun.}\ }\textbf {\bibinfo {volume} {13}},\ \bibinfo {pages} {3637} (\bibinfo {year} {2022})}\BibitemShut {NoStop}%
\bibitem [{\citenamefont {Banszerus}\ \emph {et~al.}(2024)\citenamefont {Banszerus}, \citenamefont {Hecker}, \citenamefont {Wang}, \citenamefont {M{\ifmmode\ddot{o}\else\"{o}\fi}ller}, \citenamefont {Watanabe}, \citenamefont {Taniguchi}, \citenamefont {Burkard}, \citenamefont {Volk},\ and\ \citenamefont {Stampfer}}]{Banszerus2024FebValleyLive}%
  \BibitemOpen
  \bibfield  {author} {\bibinfo {author} {\bibfnamefont {L.}~\bibnamefont {Banszerus}}, \bibinfo {author} {\bibfnamefont {K.}~\bibnamefont {Hecker}}, \bibinfo {author} {\bibfnamefont {L.}~\bibnamefont {Wang}}, \bibinfo {author} {\bibfnamefont {S.}~\bibnamefont {M{\ifmmode\ddot{o}\else\"{o}\fi}ller}}, \bibinfo {author} {\bibfnamefont {K.}~\bibnamefont {Watanabe}}, \bibinfo {author} {\bibfnamefont {T.}~\bibnamefont {Taniguchi}}, \bibinfo {author} {\bibfnamefont {G.}~\bibnamefont {Burkard}}, \bibinfo {author} {\bibfnamefont {C.}~\bibnamefont {Volk}},\ and\ \bibinfo {author} {\bibfnamefont {C.}~\bibnamefont {Stampfer}},\ }\bibfield  {title} {\bibinfo {title} {{Phonon-limited valley life times in single-particle bilayer graphene quantum dots}},\ }\bibfield  {journal} {\bibinfo  {journal} {arXiv}\ }\href {https://doi.org/10.48550/arXiv.2402.16691} {10.48550/arXiv.2402.16691} (\bibinfo {year} {2024}),\ \Eprint {https://arxiv.org/abs/2402.16691} {2402.16691} \BibitemShut {NoStop}%
\bibitem [{Note1()}]{Note1}%
  \BibitemOpen
  \bibinfo {note} {The software of the simulation allows to interactively inspect any position in gate space, where it illustrates the alignment of the chemical potentials in the left and right QD.}\BibitemShut {Stop}%
\bibitem [{Note2()}]{Note2}%
  \BibitemOpen
  \bibinfo {note} {We assume the tunnel current to be limited by the interdot tunnel rate.}\BibitemShut {Stop}%
\bibitem [{\citenamefont {Denisov}\ \emph {et~al.}(2025)\citenamefont {Denisov}, \citenamefont {Reckova}, \citenamefont {Cances}, \citenamefont {Ruckriegel}, \citenamefont {Masseroni}, \citenamefont {Adam}, \citenamefont {Tong}, \citenamefont {Gerber}, \citenamefont {Huang}, \citenamefont {Watanabe}, \citenamefont {Taniguchi}, \citenamefont {Ihn}, \citenamefont {Ensslin},\ and\ \citenamefont {Duprez}}]{Denisov2025Feb}%
  \BibitemOpen
  \bibfield  {author} {\bibinfo {author} {\bibfnamefont {A.~O.}\ \bibnamefont {Denisov}}, \bibinfo {author} {\bibfnamefont {V.}~\bibnamefont {Reckova}}, \bibinfo {author} {\bibfnamefont {S.}~\bibnamefont {Cances}}, \bibinfo {author} {\bibfnamefont {M.~J.}\ \bibnamefont {Ruckriegel}}, \bibinfo {author} {\bibfnamefont {M.}~\bibnamefont {Masseroni}}, \bibinfo {author} {\bibfnamefont {C.}~\bibnamefont {Adam}}, \bibinfo {author} {\bibfnamefont {C.}~\bibnamefont {Tong}}, \bibinfo {author} {\bibfnamefont {J.~D.}\ \bibnamefont {Gerber}}, \bibinfo {author} {\bibfnamefont {W.~W.}\ \bibnamefont {Huang}}, \bibinfo {author} {\bibfnamefont {K.}~\bibnamefont {Watanabe}}, \bibinfo {author} {\bibfnamefont {T.}~\bibnamefont {Taniguchi}}, \bibinfo {author} {\bibfnamefont {T.}~\bibnamefont {Ihn}}, \bibinfo {author} {\bibfnamefont {K.}~\bibnamefont {Ensslin}},\ and\ \bibinfo {author} {\bibfnamefont {H.}~\bibnamefont {Duprez}},\ }\bibfield  {title} {\bibinfo {title} {{Spin{\textendash}valley protected Kramers pair in bilayer
  graphene}},\ }\href {https://doi.org/10.1038/s41565-025-01858-8} {\bibfield  {journal} {\bibinfo  {journal} {Nat. Nanotechnol.}\ ,\ \bibinfo {pages} {1}} (\bibinfo {year} {2025})}\BibitemShut {NoStop}%
\bibitem [{\citenamefont {Albrecht}\ \emph {et~al.}(2017)\citenamefont {Albrecht}, \citenamefont {Moers},\ and\ \citenamefont {Hermanns}}]{Albrecht2017May}%
  \BibitemOpen
  \bibfield  {author} {\bibinfo {author} {\bibfnamefont {W.}~\bibnamefont {Albrecht}}, \bibinfo {author} {\bibfnamefont {J.}~\bibnamefont {Moers}},\ and\ \bibinfo {author} {\bibfnamefont {B.}~\bibnamefont {Hermanns}},\ }\bibfield  {title} {\bibinfo {title} {{HNF - Helmholtz Nano Facility}},\ }\href {https://doi.org/10.17815/jlsrf-3-158} {\bibfield  {journal} {\bibinfo  {journal} {Journal of Large-Scale Research Facilities}\ }\textbf {\bibinfo {volume} {3}},\ \bibinfo {pages} {112} (\bibinfo {year} {2017})}\BibitemShut {NoStop}%
\bibitem [{\citenamefont {Bonet}\ \emph {et~al.}(2002)\citenamefont {Bonet}, \citenamefont {Deshmukh},\ and\ \citenamefont {Ralph}}]{Bonet2002JanRateEQ}%
  \BibitemOpen
  \bibfield  {author} {\bibinfo {author} {\bibfnamefont {E.}~\bibnamefont {Bonet}}, \bibinfo {author} {\bibfnamefont {M.~M.}\ \bibnamefont {Deshmukh}},\ and\ \bibinfo {author} {\bibfnamefont {D.~C.}\ \bibnamefont {Ralph}},\ }\bibfield  {title} {\bibinfo {title} {{Solving rate equations for electron tunneling via discrete quantum states}},\ }\href {https://doi.org/10.1103/PhysRevB.65.045317} {\bibfield  {journal} {\bibinfo  {journal} {Phys. Rev. B}\ }\textbf {\bibinfo {volume} {65}},\ \bibinfo {pages} {045317} (\bibinfo {year} {2002})}\BibitemShut {NoStop}%
\bibitem [{\citenamefont {Timm}(2009)}]{Timm2009AugRateEquationTheory}%
  \BibitemOpen
  \bibfield  {author} {\bibinfo {author} {\bibfnamefont {C.}~\bibnamefont {Timm}},\ }\bibfield  {title} {\bibinfo {title} {{Random transition-rate matrices for the master equation}},\ }\href {https://doi.org/10.1103/PhysRevE.80.021140} {\bibfield  {journal} {\bibinfo  {journal} {Phys. Rev. E}\ }\textbf {\bibinfo {volume} {80}},\ \bibinfo {pages} {021140} (\bibinfo {year} {2009})}\BibitemShut {NoStop}%
\bibitem [{Note3()}]{Note3}%
  \BibitemOpen
  \bibinfo {note} {Charging energy and tuning by the gates is taken into account separately.}\BibitemShut {Stop}%
\bibitem [{Zen(2024)}]{ZenodoSimulation}%
  \BibitemOpen
  \bibfield  {title} {\bibinfo {title} {{Simulating transport via the (1,1) - (0,2) charge transition in a bilayer graphene double quantum dot}},\ }\bibfield  {journal} {\bibinfo  {journal} {Zenodo}\ }\href {https://doi.org/10.5281/zenodo.12759889} {10.5281/zenodo.12759889} (\bibinfo {year} {2024}),\ \bibinfo {note} {[Online; accessed 18. Jul. 2024]}\BibitemShut {NoStop}%
\end{thebibliography}
%

\end{document}